\documentclass[11pt]{article}
\usepackage[a4paper, total={6.5in, 9in}]{geometry}
\usepackage{amsmath}

\usepackage{amsfonts}
\usepackage{mathtools}
\usepackage{hyperref}
\usepackage{breqn} %for breaking equation in two lines
\usepackage{bm}
\usepackage{graphicx}
\usepackage{epstopdf}
\usepackage{subcaption}
\usepackage{authblk}
\usepackage[table]{xcolor}
\usepackage{color,soul}
\usepackage{tikz-cd} 
\usepackage{multirow}
\usepackage{comment}
\usepackage{array}
\usepackage[ruled,vlined]{algorithm2e}
\SetKwProg{Init}{Initialize:}{}{}
\usepackage[toc]{appendix}
\DeclarePairedDelimiterX{\norm}[1]{\lVert}{\rVert}{#1}
\def\equationautorefname~#1\null{Equation (#1)\null}

\title{Geometrically adapted Langevin dynamics for Markov chain Monte Carlo simulations}
\author[1]{Mariya Mamajiwala \thanks{mariya.mamajiwala.18@ucl.ac.uk}}
\affil[1]{Department of Statistical Science, University College London, United Kingdom}
\author[,2,3]{Debasish Roy} %\thanks{Corresponding author; Email: royd@iisc.ac.in}}
\affil[2]{Centre of Excellence in Advanced Mechanics of Materials, Indian
Institute of Science, Bangalore 560012, India }
\affil[3]{Computational Mechanics Lab, Department of Civil Engineering, Indian
Institute of Science, Bangalore 560012, India}
\author[1]{Serge Guillas}

\date{}

\newcolumntype{T}{>{\tiny}l} % define a new column type for \tiny
\newcolumntype{H}{>{\Huge}l} % define a new column type for \Huge
  
\begin{document}
\maketitle
\begin{abstract} 
Markov Chain Monte Carlo (MCMC) is one of the most powerful methods to sample from a given probability distribution, of which the Metropolis Adjusted Langevin Algorithm (MALA) is a variant wherein the gradient of the distribution is used towards faster convergence. However, being set up in the Euclidean framework, MALA might perform poorly in higher dimensional problems or in those involving anisotropic densities as the underlying non-Euclidean aspects of the geometry of the sample space remain unaccounted for. We make use of concepts from differential geometry and stochastic calculus on Riemannian manifolds to geometrically adapt a stochastic differential equation with a non-trivial drift term. This adaptation is also referred to as a stochastic development. We apply this method specifically to the Langevin diffusion equation and arrive at a geometrically adapted Langevin dynamics. This new approach far outperforms MALA, certain manifold variants of MALA, and other approaches such as Hamiltonian Monte Carlo (HMC), its adaptive variant the no-U-turn sampler (NUTS) implemented in Stan, especially as the dimension of the problem increases where often GALA is actually the only successful method. This is evidenced through several numerical examples that include parameter estimation of a broad class of probability distributions and a logistic regression problem.   
\vspace{10pt}

\textbf{Keywords}: differential geometry, HMC, MALA, MCMC, Riemannian manifold, stochastic development, stochastic differential equations
    \end{abstract} 
    
\section{Introduction}
 Markov Chain Monte Carlo (MCMC) is an active field of research with a rich body of literature that is fast growing. Significant applications of an MCMC algorithm include, among others, evaluating a complex integral and sampling from an unnormalized distribution. The latter is especially useful when it is difficult to obtain the normalizing constant of a distribution or when sampling from the density is quite non-trivial even though the density may itself have a simple form. MCMC is perhaps the only known general approach to find the volume enclosed by an $n$-dimensional convex body with a reasonable computational overhead \cite{convex}. It has also been used to sample from the posterior probability in stochastic filtering problems based on Sequential Monte Carlo (SMC). In a more general context, MCMC has been employed for optimization as well, see e.g.  \cite{MartinoL2016OpMm}. MCMC methods, in combination with existing machine learning algorithms, have been exploited in applications such as particle filtering \cite{mcmcfiltering}, robotics \cite{wang2011markov}, computational biology \cite{rieping2005inferential}, genetics \cite{geneticsmcmc} and machine learning \cite{mcmcmML}, to name only a few. 

We refer to \cite{robert2020markov} for a recent review of MCMC methods with an interesting discussion on a few popular misconceptions. The Metropolis Adjusted Langevin Algorithm (MALA) \cite{MALTA}, the  Hamiltonian Monte Carlo (HMC) approach \cite{DUANE1987216} and related methods that make use of the gradient of the target density to design a proposal distribution for the Markov chain may be considered as 'first-order'.  MALA uses Langevin dynamics in conjunction with the Metropolis accept-reject step. There are several MCMC algorithms that are based on Langevin dynamics, e.g. the Metropolis adjusted Langevin truncated algorithm or MALTA \cite{MALTA}, the unadjusted Langevin algorithm or ULA \cite{ULAconvergence} which is free from the Metropolis accept-reject step, the projected ULA \cite{projectedULA}, proximal MALA \cite{proximalMALA}, underdamped Langevin MCMC \cite{underdampedLangevin},  Moreau-Yosida Unadjusted Langevin Algorithm  (MYULA) and Moreau-Yosida Regularized Metropolis Adjusted Langevin Algorithm (MYMALA) \cite{MYMALA}. Note that \cite{MALTA}, \cite{roberts1998optimal}, \cite{breyer2004scalingRegression}, \cite{scalingLangevin}, \cite{ULAconvergence}, \cite{convergenceLMCnonconvex},  investigated the convergence properties of various Langevin diffusion based MCMC methods. There are also several studies that focus on the scaling, convergence and mixing properties of the Langevin-class of MCMC algorithms. For instance, in the context of sampling from a log-concave density using MALA, \cite{DwivediRaaz2018LsMa} prove a non-asymptotic upper bound on the mixing time to demonstrate the benefit of the accept-reject step, viz. an exponentially improved dependence on error tolerance. Similar bounds on the error of sampling from a target density based on three different schemes of discretized Langevin dynamics have been reported \cite{DalalyanArnak2019Bteo}. \cite{eberle2019couplings} propose a new approach to quantify convergence of underdamped Langevin dynamics to equilibirium.  

In this work, our focus is on MALA. It is a class of MCMC methods in which the Markov Chain evolves as per the overdamped Langevin dynamics. Specifically, the Langevin SDE (interpreted in the sense of Ito) is given by
\begin{equation}
dx_t = \frac{1}{2}\nabla L dt + dB_t .
\end{equation}
Here $L$ is the log-likelihood of the target density and $dB$ the standard Brownian increment such that $dB \sim \mathcal{N}(0,\sqrt{dt})$. Since the Langevin dynamics involves gradient information of the target distribution, the method is more likely to move towards regions of high probability which is a major advantage over the use of largely arbitrary proposal distributions.  

If the Langevin SDE could be solved exactly, all the particles would be accepted and there would be no need for a Metropolis adjust step. However this is rarely the case. The SDE may be solved by various numerical integrators -- the Euler-Maruyama method being often used --  that introduce integration errors and hence necessitates the Metropolis accept-reject step. This step also helps improving the convergence characteristics of the algorithm. It was shown \cite{roberts2001optimal} that the asymptotically optimal acceptance probability in MALA is 0.574 in contrast with 0.234 for MH. It means that MALA is significantly faster than MH. However, MALA does have its share of disadvantages, e.g. it is not the best choice where target distributions are heavy-tailed or in cases involving highly correlated multivariate distributions.

As we have just noted, MALA is based on the Euclidean Langevin SDE. Working with SDEs in the Euclidean setting however comes with its shortcomings. The two major inadequacies of working in the Euclidean setting are as follows. First, owing to the noise term in the SDE, there is a possibility of a gradual increase in the variance of numerical  solutions to SDEs. Second, the space-filling properties of Brownian motion may cause delayed convergence. These issues gain in importance as the dimension of the problem increases and form the motivation for our work that uses stochastic calculus in the geometric setting. Despite the spectrum of research areas in which MCMC finds application and the many flavours of it that have been explored, hardly an effort has been made at exploiting the differential geometric aspects to develop faster and more accurate algorithms. Whenever diffusions on Riemannian manifolds are considered, it is either directly in the language of frame bundles or exponential and log maps which is inaccessible to non-specialists, see e.g. \cite{sommer2015modelling}, or in the form of an SDE which simultaneously uses Amari's natural gradient \cite{amari1998natural} in conjunction with the equation for Brownian motion on a Riemannian manifold. \cite{GirolamiM2011RmLa} is a work on MCMC belonging to the latter category. To our understanding, this work is however beset with certain issues (discussed in detail in section \ref{sec:related}) which is indeed one of the motivating factors for this article. Since embedding within a higher dimensional Euclidean manifold is generally infeasible, the understanding of diffusion on a manifold that is intrinsically defined must be through the use of frame bundles \cite{elworthy1988geometric}. We present here, perhaps for the first time, a systematic derivation of the stochastic development of a general SDE on a Riemannian manifold, following \cite{HsuElton,hsu1995quasi} and use it specifically in the context of the Langevin diffusion equation to obtain the geometrically adapted version of MALA, which we will refer as Geometrically Adapted Langevin Algorithm (GALA) from here on. Stochastic development is the framework that is used for the derivation of the equation for Brownian motion on a manifold, leading to  the celebrated Laplace-Beltrami operator. We extend this approach for a general SDE, which is also applied to other interesting problems in \cite{MAMAJIWALA2022103179}. A brief review of the relevant literature is provided in section \ref{sec:prelim} followed by detailed derivation. The resulting algorithm for GALA is also given in section \ref{sec:prelim}, which may be considered a `second-order' method, as it makes use of derivatives up to the second order for the proposal step; this is unlike MALA which is a `first-order' method. 

Since MCMC methods are probabilistic wherein the objective is to sample from a given distribution, possibly under certain constraints, it naturally implies an underlying geometric structure. In the specific context of a Riemannian geometric worldview, this structure is adequately brought forth through an appropriate metric \cite{LeeJohn}, e.g. the Fisher-Information Matrix (FIM) \cite{CostaSueli} which is symmetric, positive definite and in conformity with the compatibility conditions \cite{LeeJohn}. This structure may therefore be exploited in principle to constrain the solution and hence improve certain features, including convergence, of the algorithm. 
 
The aim of this study is to provide, perhaps for the first time, a geometrically consistent and rigorously founded strategy to stochastically develop the Langevin SDE on the Riemannian manifold with a suitably constructed FIM and the associated connection. The key to our stochastic development is the notion of a horizontal frame bundle, a feature of a more general theory of fibre bundles \cite{HsuElton}, and we call the resulting Langevin SDE geometrically adapted. We show why it is important to work with the geometric adaptation of MCMC methods, and what potential it holds. Specifically, whilst working with the Langevin-diffusion based MCMC, a geometric adaptation of Langevin dynamics would enable us to restrict the evolving parameters on a hypersurface entirely consistent with the underlying constraints of motion. This in turn provides us with a handle to control the space-filling properties of Brownian motion that are physically meaningless and often the cause of delayed convergence. Moreover, the modified drift term that restricts the solution of the Langevin equation to remain on the Riemannian hypersurface provides for an additional means of faster convergence and higher accuracy. We also show, in addition to efficiency, that our new method is the only one that succeeds across a range of moderate and large dimensional problems.

The Whitney embedding theorem \cite{WhitneyEmbed}, guarantees an embedding of any Riemannian manifold within a sufficiently higher dimensional Euclidean space. Characterizing the embedding space is however no trivial task in general. As an alternative and with inspiration drawn from the work in \cite{HsuElton}, we introduce the additional construct of a frame bundle consisting of its vertical and horizontal components. This construct is used to geometrically adapt the Langevin dynamics originally posed in a Euclidean space (not the embedding space), which is isomorphic, though not isometric, with the tangent space to the Riemannian manifold. We illustrate the outcomes of this study with two sets of examples, all pertaining to parameter estimation. In the first, we determine the parameters of a broad range of probability distributions where the form of the distribution is known; this class of examples, though not covering a broad range, was also considered in \cite{GirolamiM2011RmLa}. Unfortunately, the geometric variant of the dynamics, as reported by these authors, was not properly developed. We demonstrate the specific advantages of our scheme vis-a-vis the limitations of that in the last reference. 
  
The rest of the paper is organized as follows. In section 2, we derive the stochastic development of an SDE on the Riemannian manifold starting with a brief review of differential geometry and stochastic calculus for completeness. Section 3 discusses work related to GALA. Section 4 contains an illustration of the method on the couple of problems discussed above. We conclude the article in section 5 with a  discussion on the outcomes and an appraisal of the future scope.

\section{Stochastic development of an SDE on Riemannian Manifold (RM) }\label{sec:prelim}

For the sake of completeness, brief reviews of a few concepts in stochastic calculus and differential geometry are provided in sections 2.1 and 2.2 respectively. In section 2.3, the notion of frame bundles on Riemannian manifolds is introduced to develop the Langevin SDE, and finally in section 2.4, the geometrically adapted MCMC procedure is described.

\subsection{A brief review of stochastic calculus}
Stochastic calculus affords a platform to analyze and simulate solutions of stochastic differential equations (SDEs). An SDE, in its typical form, may be given as:
\begin{equation} \label{SDE}
dx_t = \alpha(x_t,t)dt + \beta(x_t,t)dB_t
\end{equation}
Here, $x_t$ is a stochastic process, $\alpha(x_t,t)$ and $\beta(x_t,t)$ are vector valued functions and $dB_t$ is a Brownian vector (infinitesimal) increment, with all its scalar components being independent. $\alpha(x_t,t)dt$ is usually called the drift term and $\beta(x_t,t)dB_t$ the diffusion term. On removing the diffusion term from (\ref{SDE}), it reduces to an ordinary differential equation. Recall that the Brownian motion $B_t$ is everywhere continuous but nowhere differentiable and this calls for an approach different from standard calculus in $\mathbb{R}^N$ in solving an SDE. Much of the theory of stochastic calculus thus involves the interpretation of the diffusion (stochastic) integral $\int_{t_0}^{t}\beta(x_s,s)dB_s$. There are mainly two routes to this end, viz. Ito and Stratonovich, and it is possible to switch between the two. In Ito's calculus, the stochastic integral is interpreted as 
\begin{equation}
\int_{t_0}^t \beta(x_s,s) dB_s = \lim_{\max_i (t_{i+1} - t_i) \rightarrow 0} \sum_{i} \beta(x_{t_i},t_i) (B_{t_{i+1}} - B_{t_i})
\end{equation}
where $t_0 < t_1 <...<t_i<...$ is a discretization of the interval $[t_0,t]$. In Stratonovich calculus, on the other hand, this integral is interpreted as 
\begin{equation}
\int_{t_0}^t \beta(x_s,s) dB_s = \lim_{\max_i (t_{i+1} - t_i) \rightarrow 0} \sum_{i} \frac{1}{2}(\beta(x_{t_i},t_i)+\beta(x_{t_{i+1}},t_{i+1}))(B_{t_{i+1}} - B_{t_i})
\end{equation}
While the Ito version has the physical appeal of causality built into its construction, the Stratonovich version conforms better with the features of standard calculus in $\mathbb{R}^N$. We interpret the solution of SDEs in this work in Ito's sense. The basic ingredient of this calculus is Ito's formula which we now describe. Consider the stochastic process $x(t)$ which is the solution of the following SDE:
\begin{equation}
dx(t) = \mu(t) dt + \sigma(t) dB(t) 
\end{equation}
If $f(x)$ is a twice continuously differentiable function of $x$, then Ito's formula gives the following SDE for $f(x(t))$: 
\begin{eqnarray}
df(x(t)) &=& f'(x(t)) dx(t) + \frac{1}{2} f''(x(t)) d[x,x](t) \\
&=& (f'(x(t)) \mu(t) + \frac{1}{2} f''(x(t)) \sigma^2(t)) dt + f'(x(t)) \sigma(t) dB(t) \nonumber 
\end{eqnarray}
In the equation above, $[x,x](t)$ denotes the quadratic variation of $x(t)$ and is defined as 
\begin{equation}
[x,x](t) = \lim_{\delta_r \rightarrow 0} \sum_{i=1}^r ||(x(t_i^r)-x(t_{i-1}^r))||^2
\end{equation}
where this limit is taken over the set of all possible partitions: 
$$ 0=t_0^r < t_1^r < t_0^r <...< t_r^r = t \;\;\; \text{with } \;\;\;\  \delta_r = \max_{1 \leq i \leq r} (t_i^r - t_{i-1}^r)$$ 

One of the most remarkable results in the theory of stochastic calculus is that the quadratic variation of the Brownian motion is $[B,B](t) = t$ with probability $1$; the result is remarkable since, although Brownian motion is stochastic, its quadratic variation returns a strictly deterministic quantity \cite{KlebanerFimaC.2012Itsc}. Mainly owing to non-linearity in the drift and/or diffusion terms, an analytical solution to an SDE is generally not available. Solutions in general must therefore be obtained through various numerical integration schemes, such as the Euler-Maruyama \cite{RoyDebasish}. 

\subsection{Concepts from differential geometry: a brief review}

Differential geometry is the machinery for performing calculus over smooth hypersurfaces in any dimension, say $\mathbb{R}^d$, and can be seen as a non-trivial generalization of standard calculus. The departure from the Euclidean set-up is specifically captured through certain incompatibility tensors, e.g. the curvature tensor in Riemannian geometry. A small neighbourhood around every point in the hypersurface, which is referred to as a manifold, is represented by a local coordinate chart drawn from the embedding Euclidean space which is generally of a higher dimension, say $\mathbb{R}^n$. These local charts overlap smoothly to enable calculations on the manifold as a whole. An important concept in the theory of differential geometry is that of a tangent plane. As the name suggests, it is the unique plane tangent to the manifold at a given point. Formally, a manifold is called Riemannian if the tangent plane at every point $p$ is equipped with an inner product with respect to a given metric $g$ such that, if $X_p$ and $Y_p$ are two vectors on the tangent plane, then we have 
\begin{equation}
\langle{X_p,Y_p}\rangle = [g]_{ij}x^iy^j
\end{equation}
where $X_p=x^i e_i$, $Y_p=y^j e_j$, $\{e_i\}_{i=1}^d$ being the basis vectors in  $\mathbb{R}^d$ and $x^i,y^j$ the components of vectors $X_p$ and $Y_p$ respectively. Throughout the article, we make use of Einstein's summation convention unless otherwise specified.
 
In the Euclidean setting, $g_{ij} = \delta_{ij}$. Loosely speaking, $g$ encapsulates the notion of how distances and angles between two vectors are measured on a tangent plane. It is known that every Riemannian manifold (RM) is associated with a unique Riemannian metric. Now that we have seen that every point on the Riemannian manifold has a tangent plane attached to it and that every tangent plane in turn has a unique metric, one must also figure out a way to smoothly move from one tangent plane to another in a close neighbourhood of the former (parallel transport of vector and tensor fields). This is precisely where the concept of connection comes in. For a given Riemannian metric $g$, the connection is defined as 
\begin{equation}
\gamma^k_{ij} = \frac{1}{2}g^{kl}[\partial_ig_{jl} +\partial_jg_{il} -\partial_lg_{ij}]
\end{equation}
In the above equation,  $g^{kl} = g^{-1}_{kl}$, $\partial_p g_{qr}$ represents partial derivative of the $(q,r)^{th}$ component of $g$ with respect to the $i^{th}$ component of $x$ and the symbols $\gamma^k_{ij}$ are also referred to as Christoffel symbols. It must be noted that $\gamma$ is not a tensor, as it does not transform like one under a smooth change of coordinates. The usual concept of derivative in $\mathbb{R}^n$ does not apply on the RM, since any two vectors lying in different tangent planes are objects of different vector spaces, and hence cannot be added or subtracted in the usual way. The equivalent notion of derivative on the RM is known as covariant derivative and is defined in terms of the connection. The covariant derivative of a vector $Y$ along a vector $X$ in terms of the Christoffel symbols is defined as follows:
\begin{equation}\label{delXY}
\nabla_XY = [XY^k + X^iY^j\gamma^k_{ij}]e_k
\end{equation}
where $X = X^ie_i $, $Y=Y^je_j$, $e_i$ is the unit vector in the $i^{th}$ coordinate direction in terms of a local chart. We emphasize that (\ref{delXY}) is valid only within the cutlocus; roughly speaking the cutlocus at a point $p$ on the manifold is that neighbourhood (on the manifold) every point in which has a geodesic connecting the point $p$ (see below for the definition of a geodesic on the RM).

Now that we have a way of moving from one point on the manifold to another using the connection, we can define curves. An important example of a curve on the manifold, parametrized by $t$, is that of a geodesic. It is the shortest path joining two given points on the manifold. The equation of (the $k^{th}$ component of) a geodesic on a $d-$dimensional Riemannian manifold is as follows:
\begin{equation}
\overset{..}{x}^k(t) + \overset{.}{x}^i_t\overset{.}{x}^j_t\gamma^k_{ij}(x(t)) = 0 \;\; \mathrm{for} \;\; i,j,k \in [1,d]
\end{equation}
The Euclidean equivalent of the above equation is just $\overset{..}{x}^k(t)=0$, solutions to which are straight lines. \\

\subsection{The concept of stochastic development}
We now consider the notion of a frame bundle $F(M)$ of a manifold $M$ and reflect on how the connection $\nabla$ manifests itself on $F(M)$, which is the key in arriving at the stochastic development of an SDE. A frame at a point $x\in M$ is a linear isomorphism between the Euclidean space $\mathbb{R}^d$ where the solution of a standard SDE evolves and the $d$-dimensional tangent space $T_xM$ to $M$ on which the solution needs to be projected. Thus, it is through the frame bundle that we can track the paths on the manifold once we know how it evolves in $\mathbb{R}^d$. 

Let $E_1,...,E_d$ be the coordinate unit vectors of the $d$-dimensional Euclidean space. Consider a frame $q$ at $x$; so that the vectors $qE_1,...,qE_d$ make up a basis for $T_xM$. Let $F(M)_x$ denote the set of all frames at $x$ so that the elements of $F(M)_x$ may be acted upon by $GL(d,\mathbb{R})$, the general linear group, i.e. any linear transformation of $F(M)_x$ is also a valid frame at $x$. $F(M)_x$ is also called a fibre at $x$. Roughly speaking, a fibre $\mathcal{F}_x$ at a point $x$ on $M$ is defined as a space attached to that point such that there exists a surjective map $\pi : \mathcal{F}_x \longrightarrow M$. The frame or fibre bundle is then the collection of such sets of frames at different points on the manifold, i.e.  $F(M) = \bigcup_{x \in M} F(M)_x$. $ F(M)$ may itself be looked upon as a differentiable manifold of dimension $d + d^2$ and hence the canonical projection $\pi : F(M) \longrightarrow M$ is a smooth map.
Clearly, the tangent space of the frame bundle $T_qF(M)$ is a vector space of dimension $d+d^2$. A tangent vector $Y \in T_qF(M)$ is called vertical if it is tangent to the fibre $F(M)_{\pi q}$. The space of vertical vectors is denoted by $V_qF(M)$; it is a subspace of $T_qF(M)$ and of dimension $d^2$. Assuming that $M$ is equipped with a connection $\nabla$, a curve $ q_t$ in $F(M)$ is a smoothly varying field of frames such that the projected curve $x_t= \pi q_t$ on $M$ is smooth. $ q_t$ is called horizontal if for each $E \in \mathbb{R}^d$, the vector field $q_t E$ is parallel along $x_t$. We recall that a vector field V along a curve $x_t$ on $M$ is said to be parallel along the curve if $\nabla_{\overset{.}{x}}V = 0$ at every point of the curve and that the vector $V_{x_t}$ at $x_t$ is said to be the parallel transport of $V_{x_0}$ at $x_0$. 

\begin{comment}
        \includegraphics{explanatory_fig_1.PNG}
        \caption{q is an isomorphism between $\mathbb{R}^d$ and $T_x M$ - the tangent space at $x$ on $M$}
         \includegraphics{explanatory_fig_2.PNG}
        \caption{The horizontal motion of $q_t$ on $M$ : The tangent vectors ($q_0 E_1$, $q_0 E_2$) at $T_{x_0} M$ are parallelly transported according to the curvature of $M$ }
    \caption{Stochastic development : a schematic}
    \label{fig:stoch_dev}
\end{comment}

\begin{figure*}[h!]
    \centering
    \begin{subfigure}{0.9\textwidth}
        \centering
        \includegraphics[height=3in]{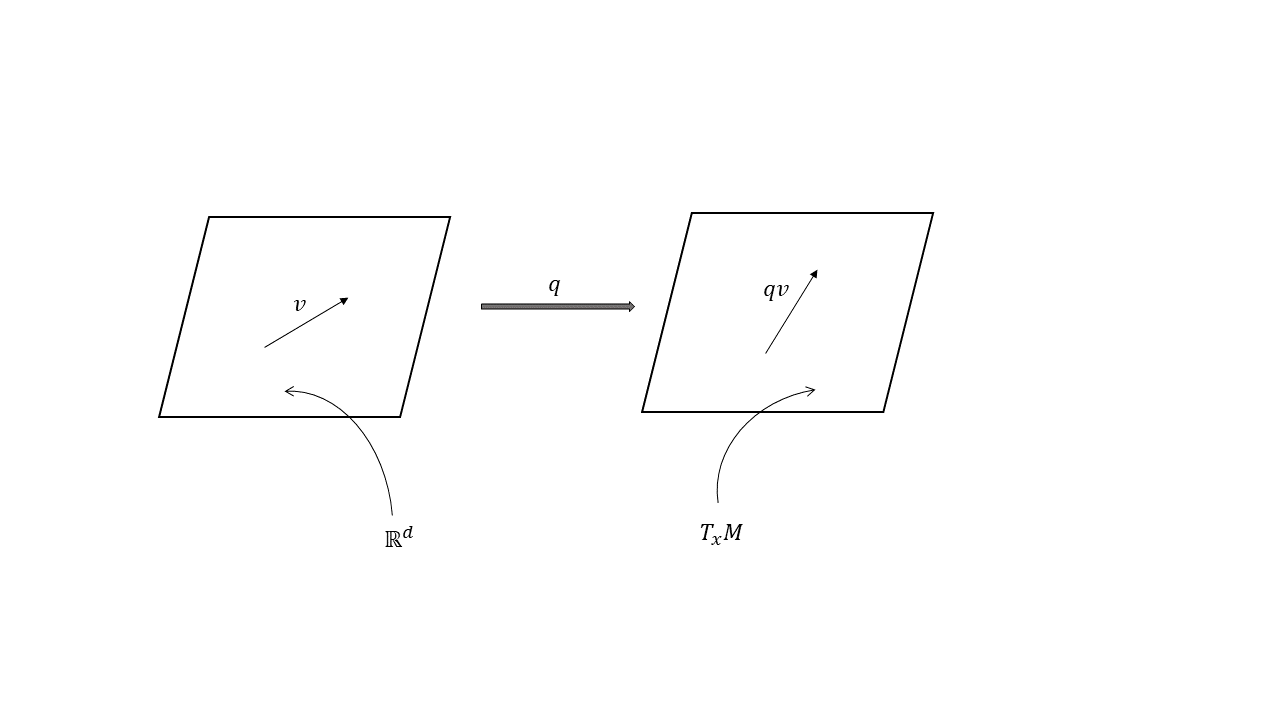}
        \caption{q is an isomorphism between $\mathbb{R}^d$ and $T_x M$ - the tangent space at $x$ on $M$}
    \end{subfigure}
    \begin{subfigure}{0.9\textwidth}
        \centering
        \includegraphics[height=3in]{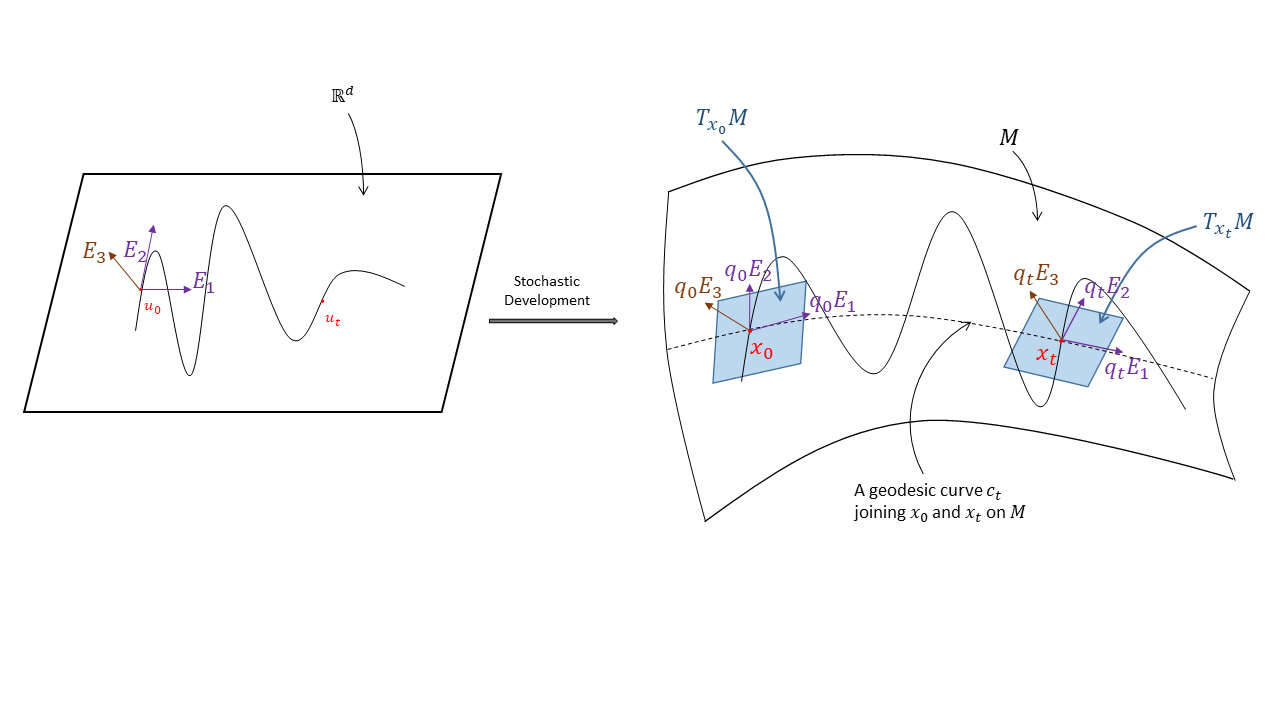}
        \caption{The horizontal motion of $q_t$ on $M$ : The tangent vectors ($q_0 E_1$, $q_0 E_2$) at $T_{x_0} M$ are parallelly transported according to the curvature of $M$ }
    \end{subfigure}
    \caption{Stochastic development : a schematic illustration}
\label{fig:stoch:dev}
\end{figure*}

A tangent vector $X \in T_qF(M)$ is called horizontal if it is the tangent vector of a horizontal curve $q_t$. The space of horizontal vectors at $q$ is denoted by $H_qF(M)$; it is a subspace of dimension $d$, and we have the decomposition $$ T_qF(M) = V_qF(M) \oplus H_qF(M)$$
Based on the projection $\pi : F(M) \longrightarrow M$, one may define a pushforward operation (an isomorphism) $\pi_* : H_qF(M) \longrightarrow T_{x}M$. Thus, for each $X \in T_xM$ and a frame $q$ at $x$, there is a unique horizontal vector $X^*$ , called the horizontal lift of $X$ to $q$ such that $\pi_* X^* = X$. For each $E \in \mathbb{R}^d$, the vector field $H_E$ at $q \in F(M)$ is defined  by the relation $H_E(q) = (qE)^* $. Hence, $(qE)^* $ which is the horizontal lift of  $qE \in T_{x}M$ to $q$ is a horizontal vector field on $F(M)$. Corresponding to the unit coordinate vectors $E_1,...,E_d$ in $\mathbb{R}^d$, $H_i := H_{E_i}, \;\; i=1,...,d$, are the corresponding horizontal fields of $F(M)$ that span $H_qF(M)$ at each $q \in F(M)$.

A local chart $x=\{x^i\}$ in a neighbourhood $O \subset M$ induces a local chart $\tilde{O} = \pi^{-1}(O)$ in $F(M)$. Specifically, let $X_i = \frac{\partial}{\partial x^i} , \;\; 1 \leq i \leq d $, be the associated moving frame.  For a frame $q \in \tilde{O} $ we have $qE_i = Q^j_i X_j$ for some matrix $Q = (Q^i_j) \in GL(d,\mathbb{R})$. This gives $(x,q) \in \mathbb{R}^{d+d^2}$ as the local chart for $\tilde{O}$. Then, the vertical subspace $V_qF(M)$ is spanned by $X_{kj} = \frac{\partial}{\partial Q^k_j}, \;\; 1 \leq j,k \leq d$. Moreover, the vector fields $\{X_i,X_{ij}, 1 \leq i,j \leq d\}$ span $T_qF(M)$, $q \in \tilde{O}$. A local expression  for the horizontal vector field $H_i$ is given as follows. We refer to \cite{HsuElton} for a proof. 
\begin{equation} \label{horz-vector}
H_i(q) = Q^j_iX_j - Q^j_i Q^l_m \gamma^k_{jl}(x) X_{km}
\end{equation}
From the definition of $\{q_t\}$, which is the horizontal lift of a differentiable curve $\{x_t\}$ on M, we have $q_t^{-1}\overset{.}{x}_t \in \mathbb{R}^d $ since $\overset{.}{x}_t \in T_{x_t}M$. We define the anti-development of $\{x_t\}$ on $M$ as a curve $\{ u_t\}$ in $\mathbb{R}^d$ to satisfy the equation 
$$u_t = \int_0^t q_s^{-1} \overset{.}{x}_s ds  .$$
In other words, $q_t \overset{.}{u}_t = \overset{.}{x}_t$ and by definition of horizontal vector fields, we have $H_{\overset{.}{u}_t}(q_t) = (q_t \overset{.}{u}_t)^* = (\overset{.}{x}_t)^* = \overset{.}{q_t} $, but we also have $H_{\overset{.}{u}_t}(q_t) = (q_t \overset{.}{u}_t)^* =  (q_t \overset{.}{u}^i_t e_i)^* = (q_t  e_i)^* \overset{.}{u}^i_t = H_i(q_t) \overset{.}{u}^i_t $. Thus, the anti-development $\{u_t\}$ and the horizontal lift $\{ q_t\}$ of a curve $\{x_t\}$ on $M$ are simply related by the following Ordinary Differential Equation (ODE). \begin{equation}\label{eqn:development}
\overset{.}{q}_t = H_i(q_t) \overset{.}{u}^i_t \end{equation}
If we start from a curve $\{ u_t\} \in \mathbb{R}^d$ and a frame $q_0$ at $x_0$, the unique solution of the above ODE is a horizontal curve $\{q_t\} \in F(M)$, which is referred to as the development of $\{u_t\}$ in $F(M)$. Equivalently, its projection on $M$ given by $\{\pi q_t\}$ is referred to as the development of $\{ u_t\}$ in $M$. The stochastic counterpart of (\ref{eqn:development}) is arrived at by interpreting it in the Stratonovich sense and then determining its Ito representation as in \cite{MAMAJIWALA2022103179}. In the present work, we adopt a slightly different route, in that we start with the equation for Brownian motion on a Riemannian manifold and find how an additional drift applied to the Euclidean SDE manifests itself on the Riemannian manifold.

\subsection{Stochastically developed SDE on RM}\label{sec:derivation}
In what follows, we discuss a method for intrinsically developing a stochastic differential equation from a $d$-dimensional Euclidean space to a Riemannian manifold $M$ of the same dimension. This  approach exploits the notion of an orthonormal frame bundle $F(M)$ on $M$. Here, every point $x$ in $M$ is furnished with an orthonormal frame $Q$ that serves as an isomorphism between the Euclidean space $\mathbb{R}^d$ and $T_x M$, the $d$-dimensional tangent space to $M$ at the point $x$. The procedure that we adopt largely follows the article by \cite{hsu1995quasi} and may be considered both an alternative and extension of the procedure explored in \cite{MAMAJIWALA2022103179} to reach the same result. To start with consider the following SDE in $\mathbb{R}^d$
\begin{equation}\label{eqn:orig:sde}
dW_t = \alpha(W_t) dt + dB_t
\end{equation}
where $W_t$ is an $\mathbb{R}^d$-valued stochastic process with $\alpha$ and $\beta$ being the drift and diffusion fields respectively. $B_t$ is an $\mathbb{R}^d$-valued Brownian motion with independently evolving scalar components. Before we proceed further, let us also recall from \cite{HsuElton} the standard equation for the Brownian motion $B_t$ on a Riemannian manifold $M$ in terms of the standard Euclidean Brownian motion $\bar{B}_t$ which is given by 
\begin{equation}\label{eqn:BMonRM}
dB^i_t = \left(\sqrt{G^{-1}} \right)_{ij} d\bar{B}^j_t - \frac{1}{2} G^{-1}_{jk} \gamma^i_{jk} dt
\end{equation}

 We now need a representation for the vector field $\alpha$ developed on $M$ and this is what we do next, based on the original work by \cite{driver1992cameron}  extended later by \cite{enchev1995towards} and \cite{hsu1995quasi}. This approach makes use of the Cartan's structure equations to arrive at an appropriate representation of an Euclidean vector field on $M$. Towards this, let us define a smooth path $x := x_{t_0:t}$ on $M$ and let $\alpha$ be a vector field in $\mathbb{R}^d$. The directional derivative at the point $x(t)$ along $\alpha$ defined as $D_{\alpha} (x(t))$ is given by $D_{\alpha} (x(t))= Q(x(t)) \alpha $. By $y_{\alpha}^x(s)$ we define the integral curve (flow) of the vector field $D_{\alpha} (x(t)))$, i.e.$\frac{\partial y_{\alpha}^{x(t)}(s)}{\partial s}\bigg|_{s=0} = D_{\alpha} (x(t))$ , where $ y_{\alpha}^{x(t)} (0) = x(t)$ . Since we are interested in parallel representations of curves in $\mathbb{R}^d$ and $M$, we also use the symbol $\mathcal{I}$ to relate the two representations. In other words,  $\bar{x} = \mathcal{I}^{-1} x $  is the representation in $\mathbb{R}^d$  of the curve $x$ in $M$. Accordingly, we have
\begin{equation}
\bar{y}^{\bar{x}(t)}_{\alpha} (s) = \mathcal{I}^{-1} \circ y^{x(t)}_{\alpha} (s) \circ \mathcal{I}
\end{equation}
This defines a corresponding pushforward map 
\begin{equation}
  \frac{\partial \bar{y}_{\alpha}^{\bar{x}(t)}(s) }{ \partial s} \bigg|_{s=0} =(\mathcal{I}^{-1}_* D_{\alpha})(\bar{x}(t)) = \mathcal{E}_{\alpha} (\bar{x}(t))
\end{equation}
The RHS of the last equation is clearly a vector field in $\mathbb{R}^d$ along $\bar{x}$ which we refer to as $\mathcal{E}_{\alpha} (\bar{x})$ and for which we wish to arrive at a representation. Towards this, define a canonical 1-form  $\phi$ in $T^* O(M)$ such that for any vector field $Z$ in $T(O(M))$, we have $\phi(Z) = Q^{-1} \pi_* (Z)$, where $\pi_*$ is the push-forward of the canonical projection $\pi : O(M) \rightarrow M$. To proceed further, we now make use of Cartan's structure equations given as follows. 
\begin{align}
d \phi ={}& -\omega \wedge \phi + \Theta \label{eqn:cartan1}\\
d \omega ={}& -\omega \wedge \omega + \Omega \label{eqn:cartan2}
\end{align}
%\begin{equation}
%d \omega^j_i = -\omega^j_h \wedge \omega^h_i  + \frac{1}{2} R^j_{ikl} \bar{x}^k \bar{x}^l
%\end{equation}
In the equations above, $\wedge$ is the skew wedge product of differential forms and $\omega$ denotes the $o(d)$-valued connection 1-forms, i.e. a $d \times d$ skew-symmetric matrix with each element being a 1-form. $\Theta$ is an $\mathbb{R}^d$-valued torsion 2-form which is identically zero for a Riemannian manifold.  $\Omega$ is the $o(d)$-valued curvature 2-form.  Since we are dealing with both the curves $x(t)$ which is parametrized in $t$ and $y_{\alpha}^{x(t)}(s)$ parametrized in $s$ and starting at $x(t)$, we may consider a frame $Q$ to be a function of both $t$ and $s$, i.e. $Q \equiv Q(t,s)$. This enables us to write the following velocities 
\begin{equation}
T = \frac{\partial Q(t,s)}{\partial t} ,\quad S=\frac{\partial Q(t,s)}{\partial s}, \quad N=\frac{\partial \bar{y}^{\bar{x}(t)}_{\alpha} (s)  }{\partial t}
\end{equation}
which yields the following identification
\begin{equation}
\mathcal{E}_{\alpha} (\bar{x}(t))= \int_0^t \left( \frac{ \partial N(\tau,s)}{ \partial s} \bigg|_{s=0} \right) d\tau \end{equation} 
Clearly, $ Q(t,s)$ for a fixed $t$ is the horizontal lift of  $\bar{y}^{\bar{x}(t) }_{\alpha} (s)$ on $O(M)$, so that we have $T=HN$, where $H$ is a horizontal vector field. This is also equivalent to $N = \phi(T)$ which leads to the following upon differentiation with respect to $s$. 
\begin{equation}
\frac{\partial N}{\partial s} = S \phi(T)
\end{equation}
At this stage we invoke the following formula for exterior differentiation. For two vector fields $T$ and $S$ , and the closed 2-form $d\phi$, we have 
\begin{equation}d \phi(T,S) = T \phi(S) - S \phi(T) - \phi([T,S]) \end{equation}
where, the Lie bracket $[T,S]$ is presently zero since the time like co-ordinates $s$ and $t$ are chosen independently. Hence, we have 
\begin{equation}
 S \phi(T) = T \phi(S) - d \phi(T,S)
  \end{equation}
Moreover, by observing that $\pi(Q(t,0)) = y_{\alpha}^{x(t)}(0)=x(t) $ , we directly have 
$\pi_* (S)= Q(t,0) \alpha$. In other words, from this we retrieve the horizontal component of $S$ as $H \alpha$, which is equivalent to the following
\begin{equation} 
\phi(S)=\alpha 
\end{equation}
This yields
\begin{equation}\label{eqn:delN}
\frac{\partial N(t,s) }{\partial s}  \bigg|_{s=0} = \overset{.}{\alpha} -d\phi(T,S)
\end{equation}
where overdot denotes derivative with respect to 't'. We need to simplify $d\phi(T,S)$ in the equation above using the two structure equations of Cartan. Using the first one, we immediately have
\begin{equation}\label{eqn:dphi} 
d\phi(T,S) = \omega(S) N 
\end{equation}
Note that, $S$ is not necessarily a purely horizontal vector field unlike $T$, i.e. we have $\omega(T)=0$ since the connection $\omega$ is a purely vertical 1-form. Now, using the second structure equation (\ref{eqn:cartan2}) and the formula for exterior differentiation \cite{edelen2005applied}, we get
\begin{equation}\label{eqn:domega}
d\omega(T,S) = T \omega(S) = \Omega(T,S)
\end{equation}
Integrating the last equation over $t$, we arrive at the required expression for the connection 1-form $\omega(S)$
\begin{equation}\label{eqn:omega}
\omega_{Q(t,0)}(S) = \int_0^t \Omega_{Q(\tau,0)}(T,S) d\tau
\end{equation}
Note that the curvature 2-form $\Omega$ is strictly horizontal. This, along with the fact that $T=H \frac{\partial \bar{y}^{\bar{x}(t)}_{\alpha} (s)  }{\partial t} \bigg|_{s=0} := H \overset{.}{ \bar{y}}^{\bar{x}(t)}_{\alpha} (0)$ leads to
\begin{equation}\label{eqn:omega2}
\omega_{Q(t,0)}(S) = \int_0^t \Omega_{Q(\tau,0)}(H \overset{.}{ \bar{y}}^{\bar{x}(t)}_{\alpha} (0),H \alpha) d\tau := \int_0^t \mathcal{K}_{\alpha}(\tau) d\tau 
\end{equation}
Substituting (\ref{eqn:omega2}) in (\ref{eqn:dphi}) and putting this back in (\ref{eqn:delN}), we have
\begin{equation}
\frac{\partial N(t,s) }{\partial s}  \bigg|_{s=0} = \overset{.}{\alpha} - \left(\int_0^t \mathcal{K}_{\alpha}(\tau) d\tau \right) \overset{.}{ \bar{y}}^{\bar{x}(t)}_{\alpha} (0)
\end{equation}
See the Appendix for the expression for $\mathcal{K}_{\alpha}$. Integrating once more with respect to $t$, we get 
\begin{equation}\label{eqn:Epsilon}
\mathcal{E}(\bar{x}(t)) = \alpha - \int_0^t \int_0^t \mathcal{K}_{\alpha}(\tau) d\tau d{ \bar{y}}^{\bar{x}(\tau)}_{\alpha} (0) =  \alpha - \int_0^t \int_0^t \mathcal{K}_{\alpha}(\tau) d\tau d\bar{x}(\tau)
\end{equation}

$\mathcal{K}_{\alpha}$ is clearly a matrix with scalar entries which are functions of $\tau$. Therefore, restricting the double integral in (\ref{eqn:Epsilon}) to $[t,t+\Delta t]$, we observe that the integral is of the order $(\Delta t)^{\frac{3}{2}}$, provided $\bar{x}(t)$ is a Brownian motion in $\mathbb{R}^d$. Hence, from the perspective of  numerical integration, it constitutes a higher order term which is ignored in this work. With this approximation in place, we may transfer the developed vector field $\mathcal{E}(\bar{x}(t))$ from $\mathbb{R}^d$ to $T_x M$ to get $Q_{(t,0)} \alpha$, which is the modified drift in the stochastically developed SDE that we shall make use of in this work. This additional drift when added to the equation for Brownian motion on an RM (\ref{eqn:BMonRM}), should lead to the equation for a general SDE on the RM. At this stage, we need a representation of $Q$ in terms of the Riemannian metric tensor $g$, which is given by $Q = \sqrt{g^{-1}}$, see chapter 3 of \cite{HsuElton}. The developed SDE corresponding to (\ref{eqn:orig:sde}) thus takes the form
\begin{equation}\label{GALE}
dx^i_t =  \left[\sqrt{g^{-1}(x_t)}\right]_{ij}\alpha^j(x_t)dt  -\frac{1}{2}  \left[g^{-1}(x_t) \right]_{kl} \gamma^i_{kl}(x_t) dt + \left[\sqrt{g^{-1}(x_t) } \right]_{im}dB^m_t
\end{equation}

Given our interest in MALA, the evolution of the parameter vector $\theta(t)$ is governed by the Langevin SDE, \begin{equation} \label{Lang-Classic}
d\theta(t) = \frac{1}{2} \nabla L(\theta(t)) dt + dB_t
\end{equation}
where $L$ is the log likelihood.
In accordance with (\ref{GALE}), the stochastically developed counterpart of (\ref{Lang-Classic}) is then given by 
\begin{equation} \label{GALA:eqn}
d\theta^i_t =  \frac{1}{2} [\sqrt{g^{-1}(\theta_t)}]_{ij}\nabla L(\theta_t)^j dt + [\sqrt{g^{-1}(\theta_t)}]_{im}dB^m_t -\frac{1}{2}  [g^{-1}(\theta_t)]_{kl} \gamma^i_{kl}(\theta_t) dt
\end{equation}

% Referring to the recent attempt at a geometrically inspired reformulation of MALA \cite{GirolamiM2011RmLa}, the authors construct the developed SDE by suitably altering the drift and diffusion terms, each based on a separate logic. While the drift is modified as per the natural gradient approach of (add Amari ref), the modified diffusion term is nothing but the well-known geometric B.M. i.e. stochastically developed B.M. on a RM (references). In our work however, we have extended the concept of stochastic development from a B.M.  to a general SDE. (elaborate??)

\noindent
\textbf{Algorithm for GALA}\\
The pseudo-code presented below is for estimating the parameter vector $\theta^{*}$ of a given distribution using GALA, when the observations $\{z\}_{i=1}^N$ are available from a known probability density function $p_x(x;\theta^{*})$, where $x\in \Omega$ and $(\Omega, \mathcal{F}, \mathcal{P})$ is a complete probability space. 

\begin{algorithm}[H]
\SetAlgoLined
\KwResult{MCMC chain of length K }
\KwData{N samples $\{z\}_{i=1}^N$ distributed as per  $p_x(x;\theta^{*})$}
\Init{$\theta_{\tau}$ for $\tau =1$}{ 
}
\For{$ \tau= 1:K-1$}{
Evaluate the log-likelihood $L(\theta_{\tau})$ as $L(\theta_{\tau})=\log(p_z(z|\theta_{\tau}) )$\\
Obtain the Fisher-Information matrix (Riemannian metric) $g(\theta_{\tau}) = E[(\nabla_{\theta} L(\theta_{\tau})) (\nabla_{\theta} L(\theta_{\tau}))^T]$\\
Determine the Riemannian connection $\Gamma^k_{ij}(\theta_{\tau}) = \frac{1}{2}g^{kl}(\theta_{\tau})[\partial_ig_{jl}(\theta_{\tau}) +\partial_jg_{il}(\theta_{\tau}) -\partial_lg_{ij}(\theta_{\tau})]$ \\
Integrating the SDE (\ref{GALA:eqn}) by Euler-Maruyama method, we have the following proposal \\
 $\theta_{\tau+1}^i = \theta_{\tau}^i + [\sqrt{g^{-1}(\theta_{\tau})}]_{ij}\nabla L(\theta_{\tau})^j \Delta t + [\sqrt{g^{-1}(\theta_{\tau})}]_{im}dB^m_t -\frac{1}{2}  [g^{-1}(\theta_{\tau})]_{kl} \Gamma^i_{kl}(\theta_{\tau}) \Delta t $\\
 Accept $\theta_{\tau +1}$ as per the Metropolis-Hastings acceptance probability \\
}
\Return $ \{ \theta_\tau \}_{\tau=1}^K$
\caption{GALA}
\end{algorithm}

\section{Related work}\label{sec:related}

In this section, we discuss work related to GALA. The closest by far is  \cite{GirolamiM2011RmLa}. The authors therein propose two major categories of MCMC methods on Riemannian manifolds, the first is based on Langevin dynamics and the second on Hamiltonian dynamics. Within the Langevin dynamics based methods, there are again two versions - the manifold Metropolis adjusted Langevin algorithm (MMALA) and the simplified MMALA. First, consider the MMALA which is closest to GALA. The equation for generating samples in  \cite{GirolamiM2011RmLa} is proposed as (after correcting for a  factor half in the last term):
\begin{equation}\label{eqn:GC}
\theta_{\tau+1}^i = \theta_{\tau}^i + [g^{-1}(\theta_{\tau})]_{ij}\nabla L(\theta_{\tau})^j \Delta t + [\sqrt{g^{-1}(\theta_{\tau})}]_{im}dB^m_t -\frac{1}{2}  [g^{-1}(\theta_{\tau})]_{kl} \Gamma^i_{kl}(\theta_{\tau}) \Delta t 
\end{equation}

The paper lacks a fully rigorous proof of this equation, the only justification provided is adding Amari's natural gradient ($ [g^{-1}(\theta_{\tau})]_{ij}\nabla L(\theta_{\tau})^j $) to the equation for Brownian motion on a Riemannian manifold (which is well-known in the literature). Amari's natural gradient is taken to be the equivalent of gradient on a Riemannian manifold (i.e. on a tangent plane of the RM), which is true in the case of deterministic curves, but not for diffusions. The article by Amari on natural gradient \cite{amari1998natural} has in fact used this gradient only for a deterministic method. Indeed, the origin of $G^{-1}$ as a multiplying factor to the drift vector field appearing in a differential equation on an RM can be traced to certain basic principles of geometric mechanics; e.g. see \cite{fiori2016nonlinear}. Specifically, a differential equation representing a balance law (e.g. of linear momentum) is  essentially a balance of forces which in turn are co-vectors. Representing such an equation in terms of vectors (e.g. velocity, acc. etc.) requires the sharpening operation using $G^{-1}$; see Chapter 3 of \cite{LeeJohn}. Unfortunately, for SDEs written in terms of incremental states (co-vectors), a vector representation is not meaningful, and hence $G^{-1}$ as a multiplier of the drift does not apply.  
 
Next, consider the simplified MMALA. In this approach, the connection term (last term in  (\ref{eqn:GC})) is dropped, apparently for the sake of simplification. Moreover, it is claimed that the invariant distribution remains unchanged despite dropping this term on account of the acceptance probability, which is clearly not true for two simple reasons. Since the SDE is different, so is the proposal density as well as the invariant distribution. The acceptance step which is claimed to be the reason for convergence to the invariant distribution, is in its own right not enough to achieve this. For instance, if the proposal SDE corresponds to a density whose invariant measure differs considerably from the target measure, it no longer behaves as an importance sampling scheme owing to a loss of absolute continuity of measures. Second, at least in the Euclidean setting, the invariant distribution of any SDE pertains to the stationary solution of the Fokker-Planck equation. Hence, the only case simplified MMALA may converge to the correct target distribution is when G is constant (see (\ref{eqn:smm}) in Appendix). This case simply corresponds to the preconditioned MALA, a well-established Euclidean MCMC method and not a Riemannian manifold method. In practice, for parameter estimation problems, when enough data is available so that the posterior is almost a Dirac measure and as parameters converge to one value, simplified MMALA may behave as a preconditioned MALA as iterations progress. However, this will no longer be available when considering sampling problems when $G(X_t)$ never converges to one value or even for parameter estimation problems when the data is scarce and the posterior has a large variance. An alternative could perhaps be to consider the multiplicative noise along with an appropriately added drift for the Langevin dynamics (see \cite{lau2007state}) which is known to converge to the correct distribution, even though it would still be a Euclidean method. This last version has exactly  been arrived at in a follow-up article by \cite{XifaraT2014Ldat} by a slightly different approach. It is done with the objective of correcting the proposal as per (\ref{eqn:GC}) so that it converges to the target distribution. In order to achieve this,  a couple of adjustments are made to (\ref{eqn:GC}) so that it becomes equal to the multiplicatively driven Langevin system \cite{lau2007state}. Even though it converges to the correct target distribution, it is indeed a Euclidean method as the derivation of this clearly indicates; see again \cite{lau2007state}. Specifically, therefore, it falls short of being characterizable as a Riemannian manifold method unlike claimed in \cite{XifaraT2014Ldat}.
 
Moving on to the RM-HMC method, the formalism in \cite{GirolamiM2011RmLa} seems to have been based on Theorem 6.4 of \cite{calin2006geometric} which corresponds to the case when the Hamiltonian consists of only kinetic energy.  This special case of Hamiltonian requires the evolution of the system state $\phi$ to follow a geodesic on the RM (so that $\nabla_{\overset{.}{\phi}} \overset{.}{\phi} =0 $). The Hamiltonian considered by \cite{GirolamiM2011RmLa}  however consists of both potential energy as well as kinetic energy, so the equations of motion used are not valid. Intuitively we can see that these are not valid from the following perspective. For an $n$-dimensional manifold with coordinates $x^1,x^2...x^n$, the time derivative (velocity $\overset{.}{x}^i$) is a tangent space object, see for instance Chapter 1 of \cite{marsden1994mathematical}. The definition of acceleration would therefore necessitate the underlying Riemannian connection in the expression since it requires an evolution across disjoint tangent spaces. In other words, the equation for time derivatives of momenta in Hamilton's equations must contain the Riemannian connection, see for instance the original work done in \cite{zlochin2001manifold} where the Riemannian connection was duly incorporated. A more general case of this (presence of external forcing) is also derived in \cite{fiori2011extended} and \cite{fiori2016nonlinear} by taking a variation of the action functional in terms of the Lagrangian, wherein the equations of motion are identical with those obtained in \cite{zlochin2001manifold} for the case of no external forcing. Moreover, even though the phase space volume interpreted in the Euclidean space is conserved in \cite{GirolamiM2011RmLa}, the Riemannian volume over the same phase space is not (e.g. see chapter 3 of \cite{LeeJohn} for the computation of the Riemannian volume).  

Unfortunately, in most of the available literature on Riemannian manifold based MCMC methods \cite{ma2015complete,XifaraT2014Ldat,livingstone2014information}, the invariant distribution of a $d-$ dimensional proposal on a Riemannian manifold is examined using the Fokker-Planck equation evolving in a $d-$dimensional Euclidean space. We think this is inappropriate because any path evolving on a Riemannian manifold of dimension $d$ is actually a $D-$dimensional path in the embedding Euclidean space, where $D \geq d+1$ according to Nash's embedding theorem \cite{nash1954c1}. Therefore, if the invariant distribution were to be found using the Fokker-Planck equation, it must be done so in the $D-$ dimensional space. This is very challenging for the following reasons. The first is, in more cases than otherwise, it is impossible to determine $D$. Even if $D$ can be determined, we must then re-write the  $d-$ dimensional stochastically developed equation in the $D-$ dimensional (Euclidean) embedding space, which is often impossible or difficult whilst defeating  the very purpose of stochastic development.  Finally, we need to redefine the probability space pertaining to the dimension $D$ of the embedding space. 

GALA does not satisfy the Euclidean Fokker-Planck equation, as expected, and still converges for all the problems considered. A tempting possibility would perhaps be to consider the so-called Fokker-Plack equation on Riemannian manifolds \cite{solo2009nonlinear}. GALA does not conform to this either, even though MMALA does (see (\ref{eqn:rmfke:mmala}) in Appendix). But, as numerically evidenced in the consistent divergence of MMALA for most problems considered here, one anticipates that the last cited form of Fokker-Planck equation is perhaps not the right equation to study invariant distributions. Overall, the global properties of diffusions on Riemannian manifolds are far from adequately understood in the literature, though some first steps are taken. For instance, some work on the short time aymptotics of the heat kernel, which is related to the transition probability of a Brownian motion on a Riemannian manifold, has been reported in Chapters 4 and 5 in \cite{HsuElton}. However, a more complete understanding of invariance may require an understanding of the long-term asymptotics which may be not even be well-defined depending on the structure of the curvature tensor.  Accordingly, the question of invariance, though important, remains unresolved as yet.

So far, the discussion was about the theoretical issues arising in the related work. The numerical examples reinforce these observations wherein it is shown that MMALA fails to converge for all problems except the logistic regression case. Unlike in other cases, the LR problem typically has a much smaller connection term (last term in (\ref{eqn:GC})) which becomes smaller still as iterations progress. Even as the MMALA does converge for this problem, it takes longer compared to GALA as reported in Figure \ref{fig:LR} and Table 3. This contrast becomes more pronounced with an increase in the dimension.

Finally, a word about the problems considered in  \cite{GirolamiM2011RmLa}; these are problems in which typically, the connection term is either small or vanishes altogether. Therefore, in such problems, the method either reduces to or asymptotically behaves like pre-conditioned MALA. However, for even a 1D problem when the data size is large or a high-dimensional correlated problem like the Gaussian example with unknown mean and covariance considered in this work, when the connection term becomes important in the proposal step, MMALA diverges as seen in Figures \ref{fig:rayl} and \ref{fig:mvn}.

\begin{comment}
Referring to the recent attempt at a geometrically inspired reformulation of MALA \cite{GirolamiM2011RmLa}, one may compare the relevant Langevin SDE in the last cited article with (\ref{GALA:eqn}) and readily observe that the drift term involving the likelihood $L$ is not properly developed by the authors. A similar inconsistency is again observed in a follow up article \cite{XifaraT2014Ldat}, wherein the authors introduce modifications that would not ensure the solution of the modified Langevin equation to lie on the Riemannian manifold. In what follows, we furnish a few specific illustrations of the GALA based MCMC strategy of general interest to the statistics community and in doing so, we also show relative performances of various competing schemes.  
\end{comment}

\section{Illustrative examples}\label{sec:examples}
In this section we illustrate the workings of GALA on a suite of related methods for two classes of parameter estimation problems. In the first, given a set of realizations from a probability distribution with a known functional form, we estimate the unknown parameters. The second problem concerns logistic regression wherein $N$ number of explanatory variables with the corresponding binary response variables are given and the aim is to reconstruct the regression parameters.
Some of these problems have been considered in \cite{GirolamiM2011RmLa}, though the authors therein work with one-dimensional Gaussian or uncorrelated multivariate Gaussian distributions. Extending the geometric construction from one to multivariate densities is however non-trivial and this is what we accomplish in this section. In addition, we also consider the parameter estimation of Rayleigh, Weibull and Banana-shaped distributions by way of highlighting how an erroneous departure from proper stochastic development could either yield an incorrect solution or failure of the method for the estimation problem involving a non-Gaussian density. We compare the results obtained with GALA with those obtained by MALA, MMALA and HMC methods.

\begin{figure}[h]
\centering
\includegraphics[width=140mm]{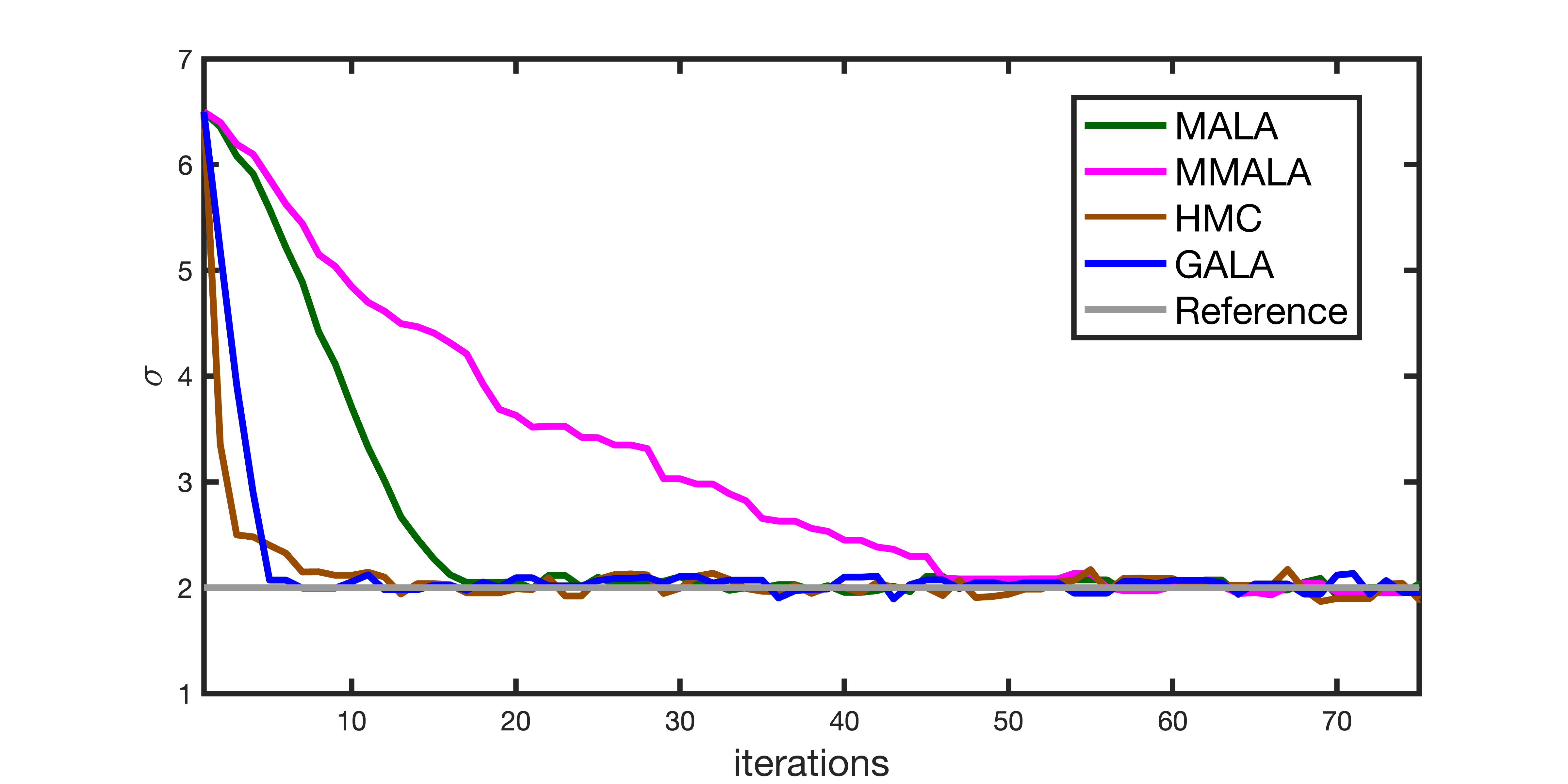}
\caption{Parameter $\sigma$ in the Rayleigh density via GALA ($\Delta t=0.2$), MALA ($\Delta t=0.01$), MMALA ($\Delta t=0.15$) and HMC ($\Delta t=0.04,L=50$) for $N=200$ sample observations}
\label{fig:rayl}
\end{figure}

We consider toy problems ranging from a 1-dimensional Rayleigh to a 65-dimensional multivariate Gaussian and the heavy-tailed Weibull distribution to a highly twisted Banana-shaped distribution. The reason to work with such toy problems is to demonstrate the accuracy of estimation. As will be seen in the results for large dimensional examples, Stan \cite{HoffmanMD2014TNSA,carpenter2017stan} and other methods converge to incorrect parameters. The priors for all examples except for logistic regression is taken as uniform as it slightly increases the problem difficulty and perhaps also leads to a fairer comparison among various methods. The initialization for various methods is kept the same except for Stan in which case its defaults are used. Figures \ref{fig:rayl}-\ref{fig:LR} show the behaviour of various methods in the warmup phase, which helps to visualize the speed of convergence. Tables display a comparison of various performance metrics of all the methods after the warmup phase. Effective sample size (ESS) is often used as a performance metric which is fine for a sampling problem, since indeed samples are desired from a distribution. However, it is perhaps not the right metric for parameter estimation problems considered in this work for the following reason - assuming enough data is available, the posterior distribution would be almost like a Dirac measure at the correct parameter value and an ideal algorithm  should converge to the correct parameter value and stay there. However ESS for such a solution would vanish which is clearly not the right inference. Therefore, we do not consider ESS comparison, but instead use warmup and acceptance rate, since a longer warmup and a high rejection rate lead to wasted computation. The warmup is determined based on the first time the Markov chain enters within a tolerance level of the correct parameter value and stays there. The acceptance percentage represents the number of samples accepted for the entire chain, i.e. including the warmup phase. The estimated mean and sample variance for the parameters are determined based on a certain number (different for different problems, as specified in the caption of each table) of samples after warmup, while the true parameter value for each problem is mentioned in the black bar in the tables.

\subsection{Estimating the parameters of a probability distribution}

% Rayleigh
\textbf{Rayleigh distribution. }
Consider a problem where $N$ samples $\{z\}_{i=1}^N$ are available from a Rayleigh distribution with unknown parameter $\sigma$. 
We first derive the developed equation for the Rayleigh distribution following the steps listed in the pseudo-code in the previous section
(see supplementary material for a detailed derivation)   
\begin{equation}\label{gale:rayl}
d\sigma_t = (-\frac{\sqrt{N}}{2} + \frac{\sum_{i=1}^N{z_i^2}}{4 \sigma^2 \sqrt{N}} + \frac{\sigma_t}{4N})dt + \frac{\sigma_t}{2\sqrt{N}}dB_t
\end{equation}
which may be contrasted with those in MALA and MMALA as \\
\begin{equation}
\text{MALA:} \;\;\; d\sigma_t = (-\frac{N}{\sigma} + \frac{\sum_{i=1}^N}{z_i^2})dt+dB_t
\end{equation}
\begin{equation}
\text{MMALA:} \;\;\;d\sigma_t = (-\frac{\sigma_t}{4} + \frac{\sum_{i=1}^N{z_i^2}}{8 \sigma_t N} + \frac{\sigma_t}{4N})dt + \frac{\sigma_t}{2\sqrt{N}}dB_t
\end{equation}

Results in the warmup phase for parameter reconstruction by various methods are shown in Figure \ref{fig:rayl}. Several performance metrics over 10 repeated simulations are summarized in Table 1. HMC requires 50 steps of Hamiltonian dynamics per sample, which may be loosely considered as a $2\%$  acceptance rate which is not reflected in the figures, this can be contrasted with the acceptance rate for other methods as shown in Table 1. Even though the cost per proposal is low for HMC, it takes longer to obtain the same number of samples overall due to this high acceptance rate; this is reflected in the computation time which is more than twice compared to GALA. The sample variance obtained by GALA is also lower compared to other methods.

\begin{figure*}[h!]
    \centering
    \begin{subfigure}{0.49\textwidth}
        \centering
        \includegraphics[height=1.7in]{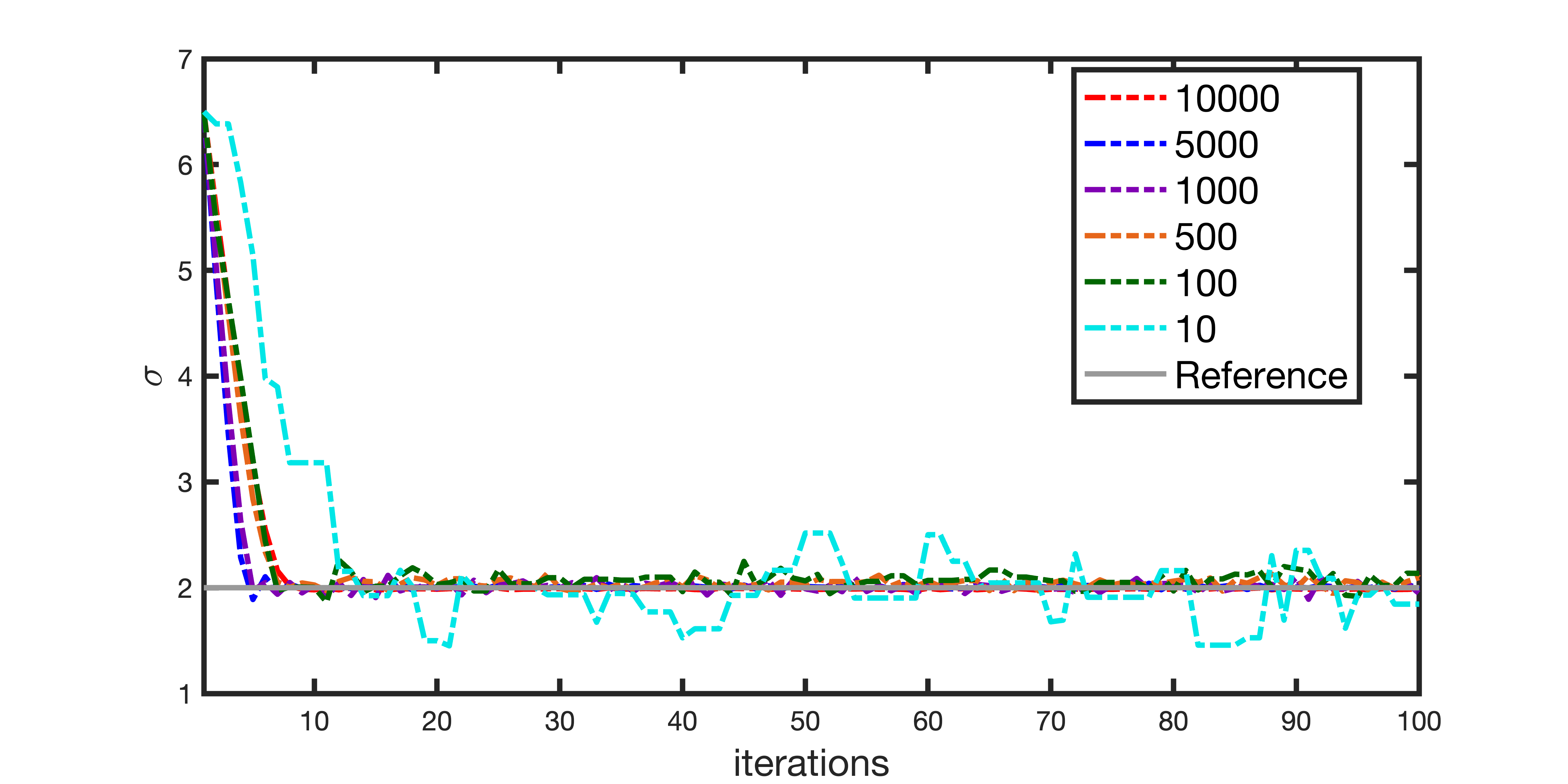}
        \caption{}
    \end{subfigure}%
    ~ 
    \begin{subfigure}{0.5\textwidth}
        \centering
        \includegraphics[height=1.7in]{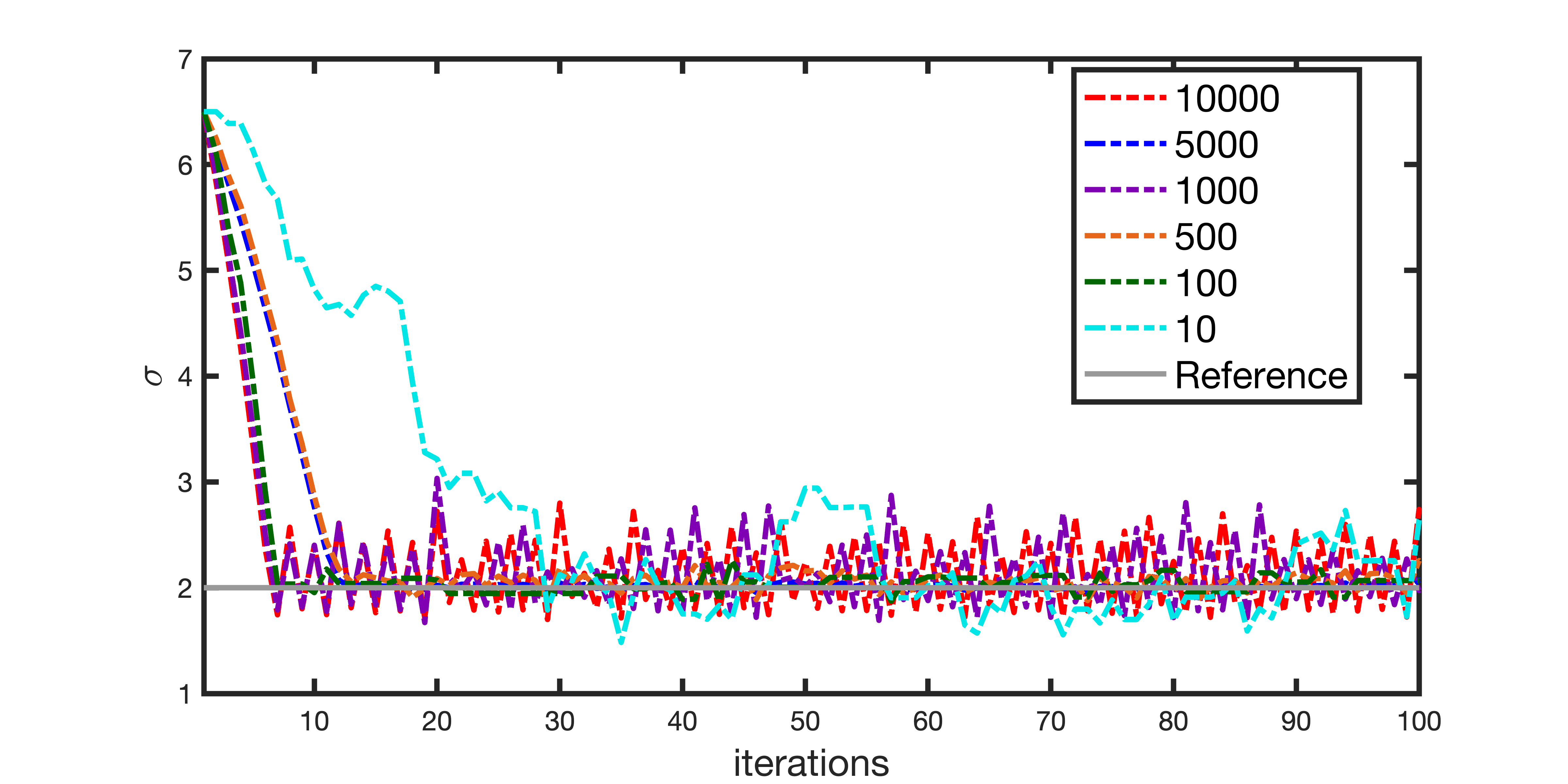}
        \caption{}
    \end{subfigure}
    \begin{subfigure}{0.5\textwidth}
        \centering
        \includegraphics[height=1.7in]{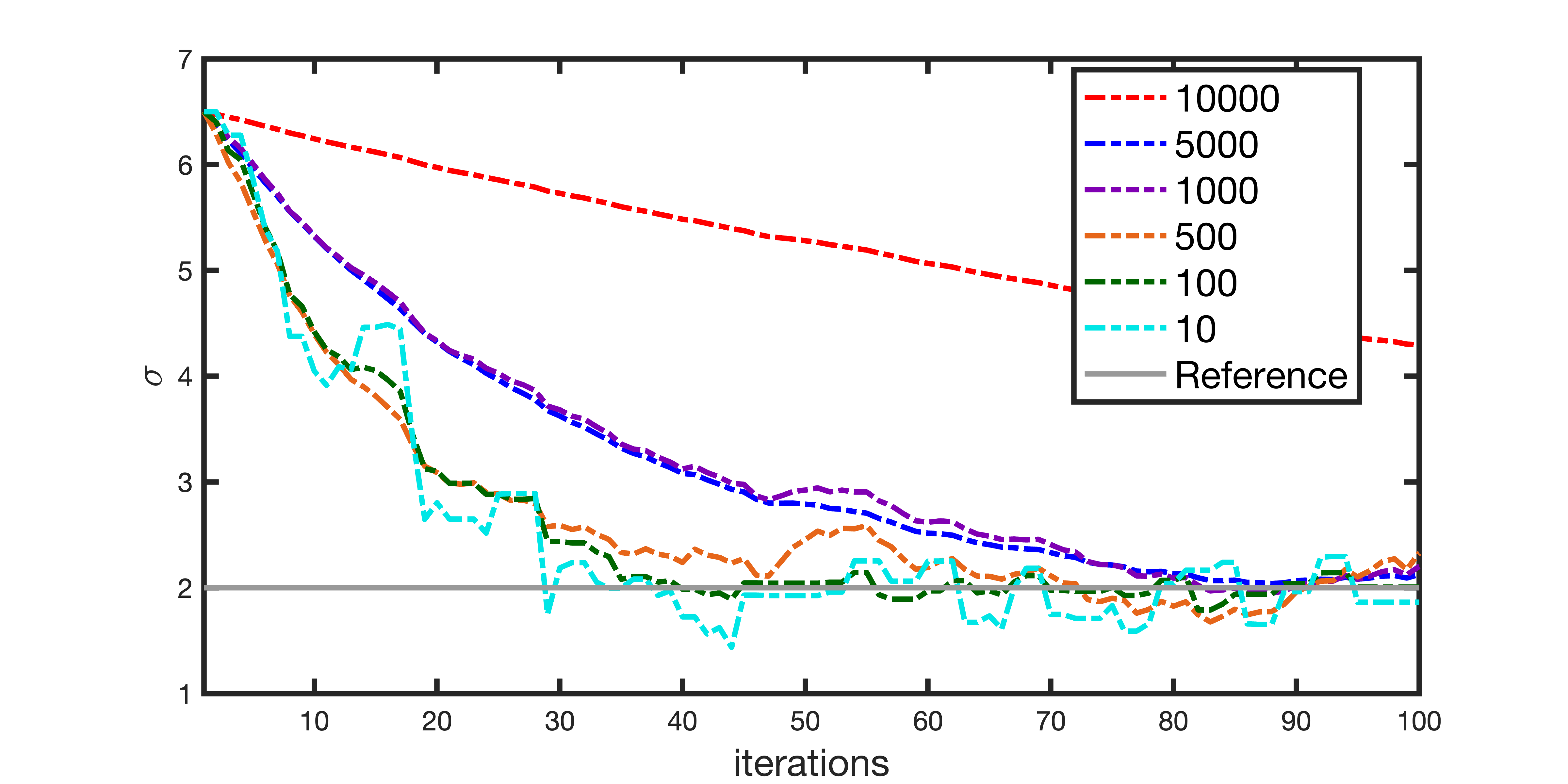}
        \caption{}
    \end{subfigure}%
    ~ 
    \begin{subfigure}{0.5\textwidth}
        \centering
        \includegraphics[height=1.7in]{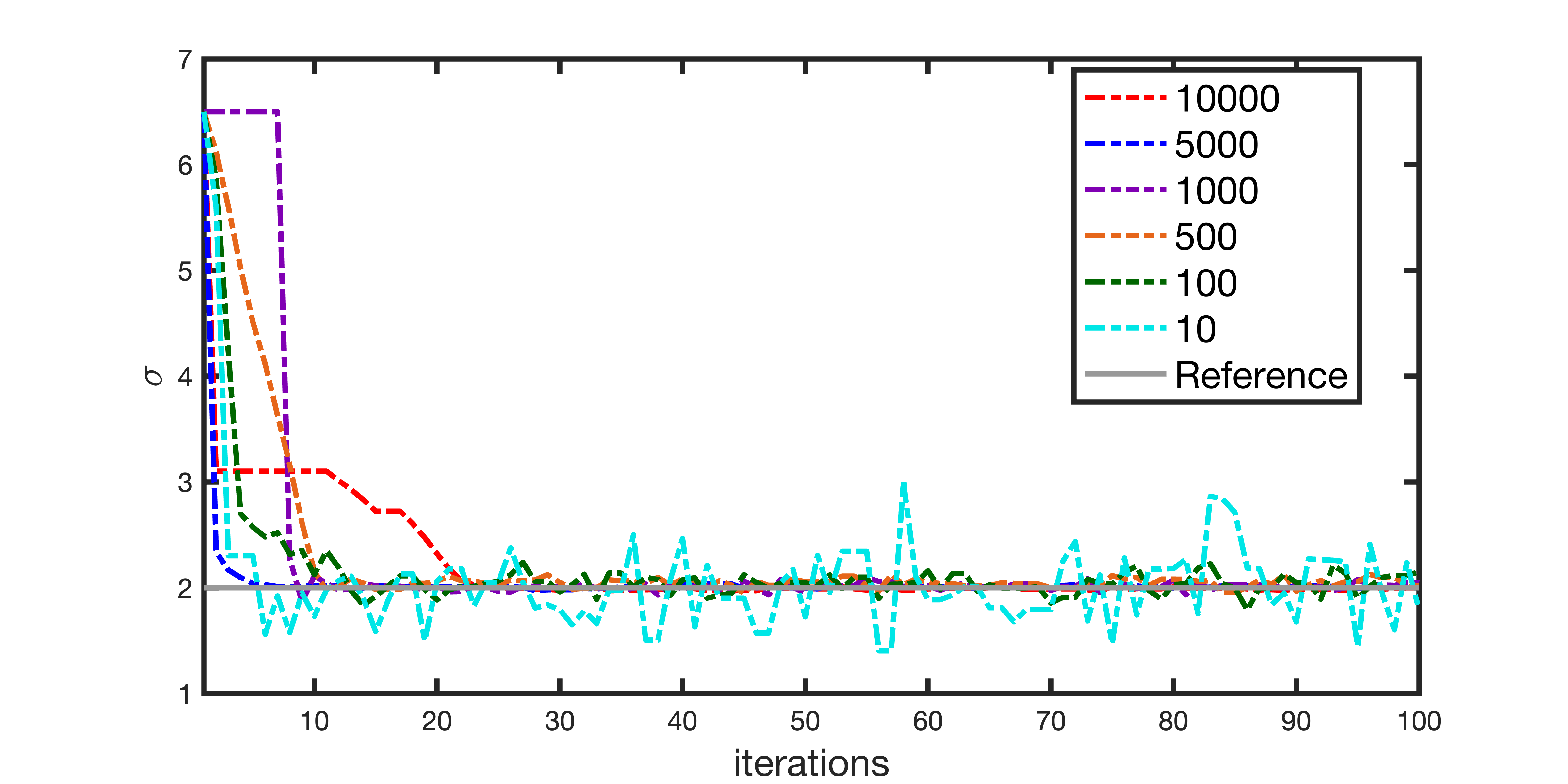}
        \caption{}
    \end{subfigure}
    \caption{Reconstructing the parameter in the Rayleigh density: a comparison of results via different methods with varying number of observations; (a) GALA; (b) MALA; (c) MMALA; (d) HMC}
       \label{fig:compare}
\end{figure*}

As part of our convergence study, we now compare the performances of different methods as the number of observations increases. We present the results for Rayleigh distribution which, though one-dimensional, is unsymmetric enough to be a good test problem. As anticipated and as shown in figure \ref{fig:compare}, the performance of most methods improves, such as in the forms of ESS being higher and burn-in period smaller, with increasing number of observations. The exception is MMALA where it sharply deteriorates; this likely happens as the dynamics is not properly developed on the Riemannian manifold. In other words, MMALA appears to confine the Langevin dynamics somewhat incorrectly, a feature more visible with an increased quantum of observation data. The slight ambiguity in the performance of HMC may be attributed to sub-optimal tuning. 
 
%The same observations hold for all other problems; indeed, the comparisons are more stark as the problem dimension or complexity increases. 
 
% ??The table below compares the Effective Sample Size (ESS) as per \cite{2011HoMC}, for the Markov chains obtained by different schemes??

%  \begin{table}[h]
%  \centering
%  \begin{tabular}{ |c|c|c| } 
% \hline
% Method & ESS (minimum,median,maximum) & computation time in seconds (per run)\\
% \hline
% %\multirow{3}{4em}{Multiple row} & cell2 & cell3 \\ 
% GALA & (529,800,1000) & 0.4397\\ 
%  MALA &  (464,644,743) & 0.3578\\ 
%  MMALA &  (22,45,88) & 0.4548\\ 
%  HMC & (642,770,819) & 9.5851\\
% \hline
% \end{tabular}
% \caption{Comparison of the Effective Sample Size (ESS) values for 1000 posterior samples obtained over 10 independent runs of each method for the Rayleigh problem (using a set of 200 observations)}
% \end{table}

% {|p{0.16\linewidth}|p{0.18\linewidth}|p{0.18\linewidth}|p{0.18\linewidth}|p{0.18\linewidth}|}

 \begin{table}[ht]
\centering
\begin{tabular}{|>{\footnotesize}l|>{\tiny}l|T|T|T|}\hline
&\textbf{\small GALA} & \textbf{\small MALA} & \textbf{\small MMALA} & \textbf{\small HMC}\\
\hline
\rowcolor{black} \multicolumn{5}{|c|}{\textbf{\color{white} \small Rayleigh (True $\sigma=2$)}}  \\
\hline
%\multirow{3}{4em}{Multiple row} & cell2 & cell3 \\ 
{Warmup} & 48,54,68 &27,29,48   &338,397,500 & 11,13,23 \\
{Acceptance (\%)} & 90.05,90.75,92.4& 76.5, 77.85,79.95& 86.85,88.95,90.25 &NA  \\
{Estimated mean} & 2.0298,2.0353,2.0396 & 2.031,20.361,2.0399 &2.0206,2.0523,2.0384 & 2.0345,2.0390,2.0440  \\
{Sample variance} &  $[1.8,2.1,2.7]\times 10^{-3}$ &$[2.9,3.1,3.3]\times 10^{-3}$ &  $[2.8,4.5,8]\times 10^{-3}$ & $[4.6,5.1,6.1] \times 10^{-3}$ \\
{Runtime (seconds)} &1.0605 & 0.9821 &   1.0904& 2.1387\\
\hline
\rowcolor{black} \multicolumn{5}{|c|}{\textbf{\color{white} \small Banana (True $B=0.1$) }} \\
\hline
{Warmup} &9,11,12 &12,13,14 & - &54,59,63 \\
{Acceptance (\%)} &100,100,100 & 83.7,84.7,86.7 &- & NA \\
{Estimated mean} & 0.1005,0.1005,0.1005 & 0.1006,0.1007,0.1008 & - & 0.1005,0.1005,0.1006 \\
{Sample variance} & $[0.82,3]\times 10^{-8}$ & $[2.4, 3.1, 4.03] \times 10^{-6}$&- &  $[2,2.2,2.5] \times 10^{-5}$ \\
{Runtime (seconds)} & 0.2&  0.25&  0.42 & 0.48\\
\hline
\end{tabular}
\caption{Comparison of various performance metrics (minimum, median and maximum) for 2000 posterior samples obtained over 10 independent runs of each method. 200  and 10  observations are used for the Rayleigh and Banana distribution respectively. The mean and sampling variance are calculated based on 1000 samples after discarding the warmup samples for each method.}
\end{table}
 
%   \begin{table}[h]
%  \centering
%  \begin{tabular}{ |c|c|c|c|c|c| } 
% \hline
% Method & Warmup & Accepted & Estimated mean & Sample variance & computation time in seconds\\
% \hline
% %\multirow{3}{4em}{Multiple row} & cell2 & cell3 \\ 
% GALA & 9,11,12 & 2000,2000,2000 & 0.1005,0.1005,0.1005 & $8 10^{-9}, 2 10^{-8},310^{-8}$ & 0.2\\
%  MALA & 12,13,14 & 1674,1694,1734 & 0.1006,0.1007,0.1008 & $2.4 10^{-6}, 3.1 10^{-6}, 4.03 10^{-6}$&0.25\\
%  MMALA &  - &2000,2000,2000 & 1.001,1.298,1.3907 & $1.1 10^{-12},1.5 10^{-12},2.47 10^{-12}$ & 0.42\\
%  HMC & 54,59,63 & -& 0.1005,0.1005,0.1006 & $210^{-5},2.210^{-5},2.510^{-5}$ & 0.48\\
% \hline
% \end{tabular}
% \caption{Comparison of various performance metrics (minimum, median and maximum) for 2000 posterior samples obtained over 10 independent runs of each method for the Banana-shaped distribution (using a set of 20 observations)}
% \end{table}
 
% Banana
\textbf{Banana-shaped distribution.} Next, consider the 2-dimensional banana-shaped distribution, the joint probability density of which is given by
\begin{equation}
p_{xy}(x,y;B) = \exp\{-\frac{x^2}{200} - \frac{1}{2}(y+Bx^2 - 100B)^2\}
\end{equation}
The banana-shaped distribution is basically a twisted Gaussian distribution with a twist parameter B and forms a good test ditribution in the context of problem geometry. Detailed derivation of the stochastically developed equation for this distribution is given in the supplementary material. Figure \ref{fig:banana} compares results in the warmup phase for the twist parameter reconstructed by GALA, MALA and HMC given a set of only 10 sample points. The MMALA fails for this problem. Similar to the Rayleigh distribution, in this problem too the acceptance rate in HMC is about $2\%$ while for GALA it is $100\%$ which does not get reflected in the Figure. Again, Table 1 gives a summary of the various performance metrics for the methods considered. GALA performs better compared to other methods for all the metrics considered, particularly in the sampling variance which is 4 orders of magnitude lower than HMC and 3 orders lower than MALA whilst taking the least computation time. 
\begin{figure}[h!]
\centering
\includegraphics[width=140mm]{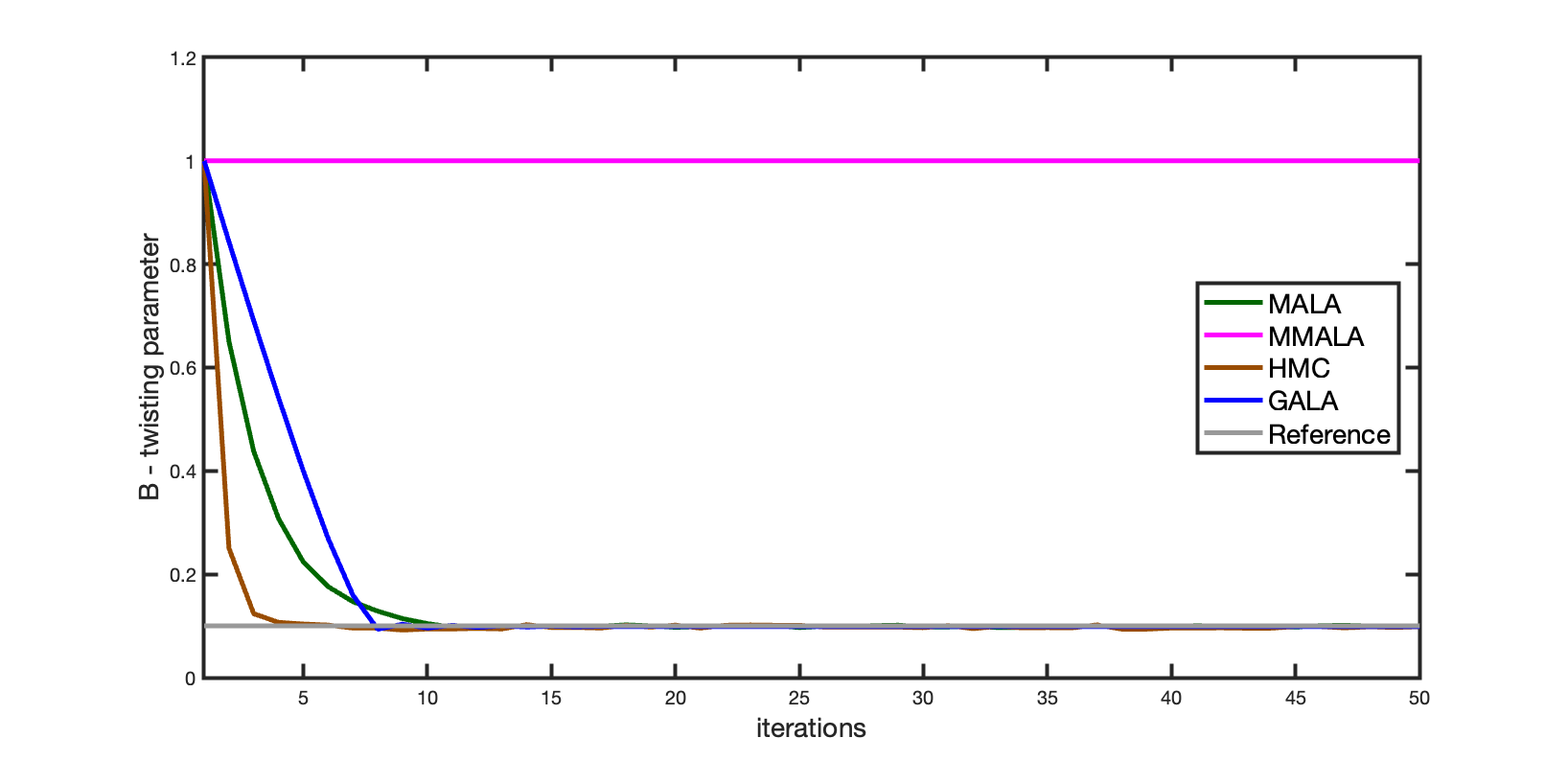}
\caption{Parameter $B$ in the Banana-shaped distribution via GALA ($\Delta t=0.1$), MALA ($\Delta t=0.000005$), RMMALA ($\Delta t=0.1$) and HMC ($\Delta t=0.0001,L=50$) for $N=10$ sample observations}
\label{fig:banana}
\end{figure}
Indeed, given the fully connected nature of curves in one dimension, the full potential of a Riemannian geometric method such as GALA is not realized for 1D cases, as in the Rayleigh and banana-shaped distribution problems. In what follows, we consider a few higher dimensional illustrations to showcase the potential benefits of GALA.

%Weibull
\textbf{Weibull distribution.} 
The Weibull distribution is 1-dimensional and characterised by two parameters. This heavy-tailed distribution is important as it can be used to represent many different shapes by appropriately choosing the two parameters (viz. the shape parameter, $k$ and the scale parameter, $\lambda$). The shape of the probability density is very sensitive to changes in the parameter $k$. A detailed derivation of the developed equation for this distribution is given in the supplementary material. Figure \ref{fig:wbl} gives a comparison of results through GALA, MALA, MMALA and HMC in the warmup phase. The estimation by GALA, which is manifestly of a superior quality vis-{\'a}-vis MALA and MMALA, is only matched by the HMC. Table 2 gives a comparison of the various performance metrics. GALA performs better than other methods overall, particularly the variance which is at least 1 order of magnitude lower than the other methods. However, owing to a complex nature of the gradients with respect to the desired parameters, the expectations appearing in the Fisher information metric have been numerically evaluated for GALA and MMALA, see supplementary material for the expressions. Note that this issue could either possibly be solved analytically, or accelerated numerically (since numerical expectations can be parallelised), and only appears for very specific distributions. This aspect may be borne in mind whilst assessing the reported comparisons of these two methods with MALA and HMC, particularly the computation time. 

%%%%%%%%%%%%%%%%%%%

\begin{figure*}[h!]
    \centering
    \begin{subfigure}{0.9\textwidth}
        \centering
        \includegraphics[height=2in]{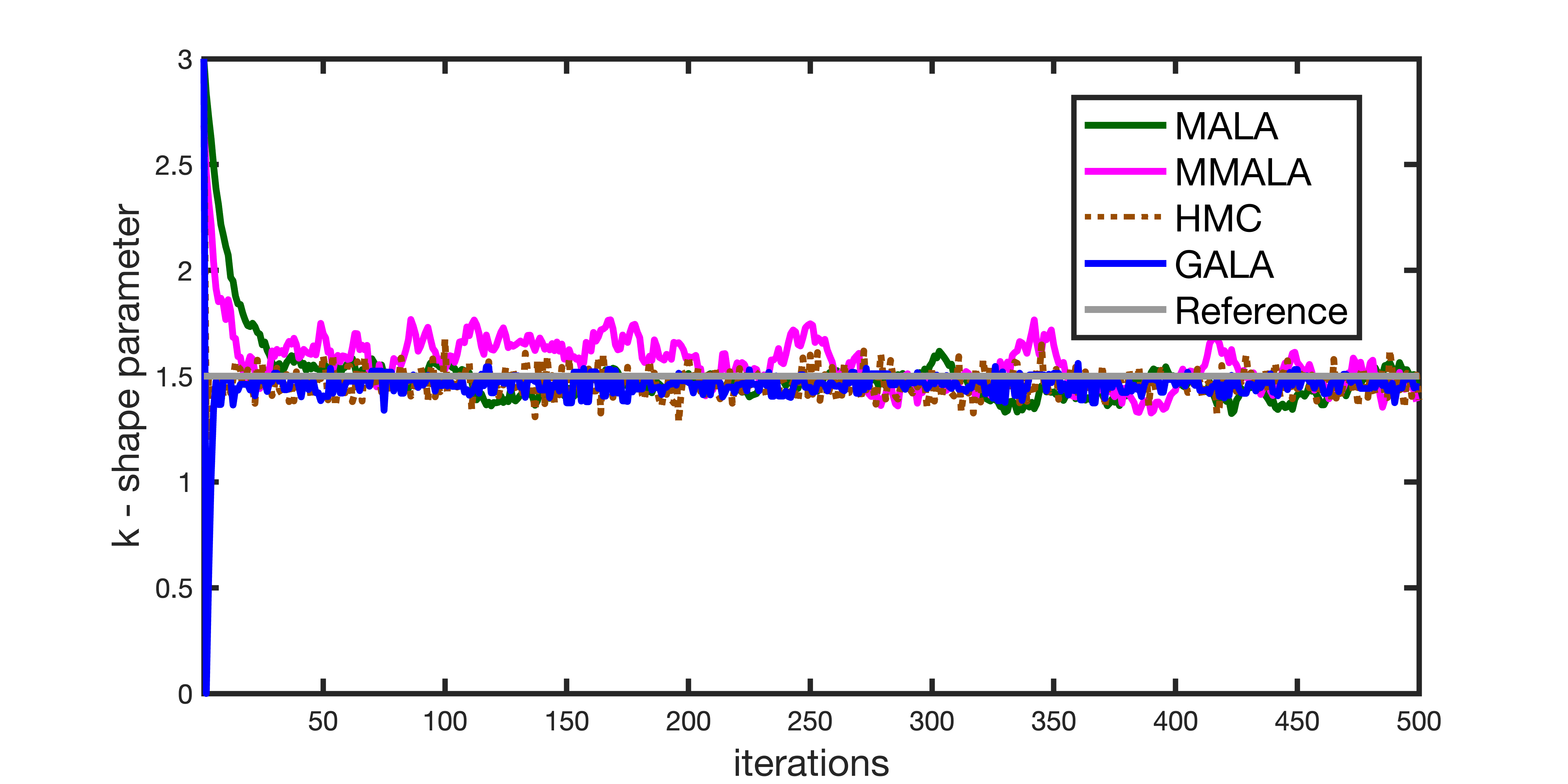}
        \caption{$k$}
    \end{subfigure}
    \begin{subfigure}{0.9\textwidth}
        \centering
        \includegraphics[height=2in]{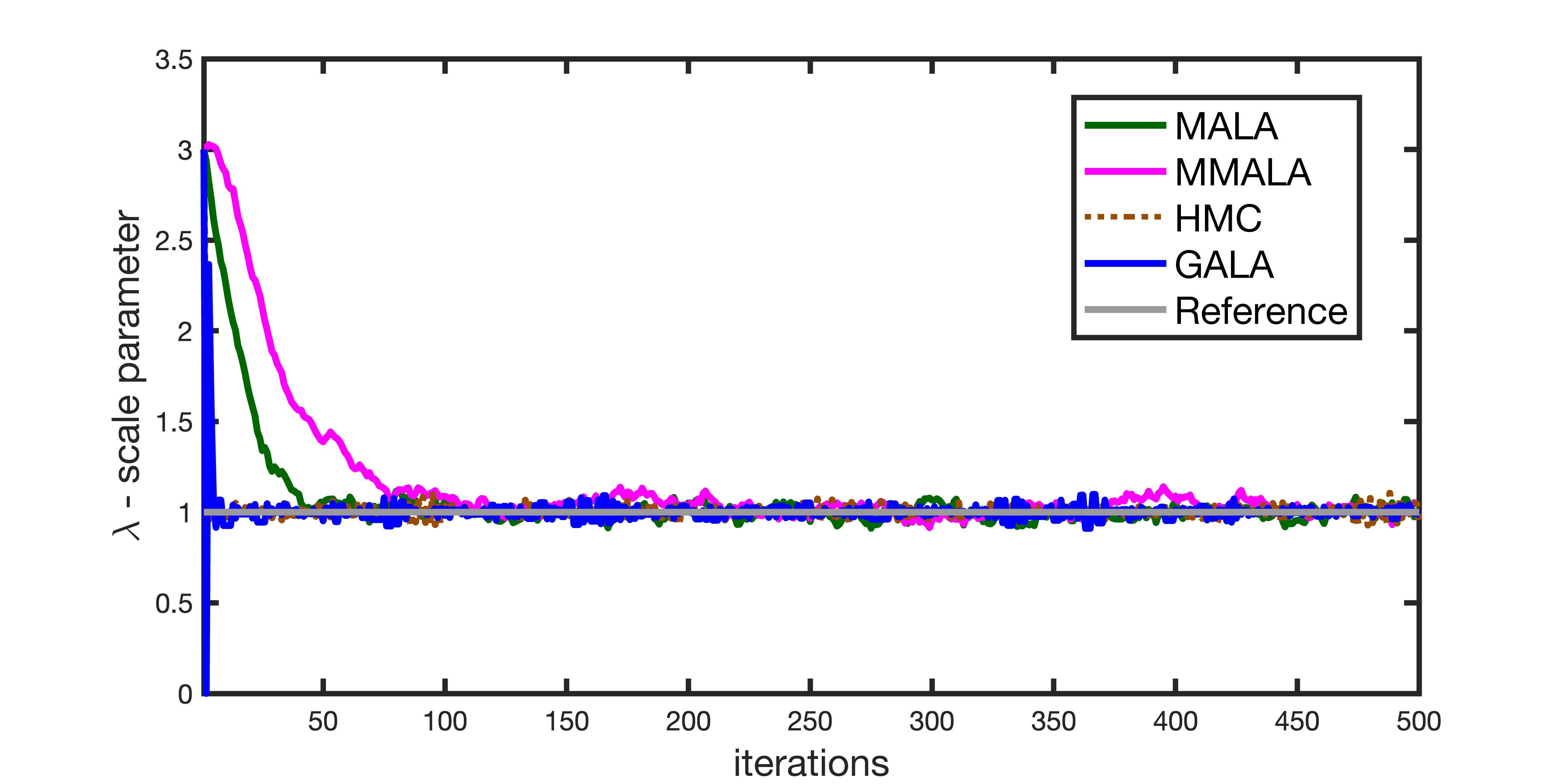}
        \caption{$\lambda$}
    \end{subfigure}
    \caption{Parameters of the Weibull distribution via GALA ($\Delta t=0.1$), MALA($\Delta t=0.0005$), MMALA ($\Delta t=0.1$) and HMC ($\Delta t=0.01,L=10$) for $N=400$ sample observations; (a) shape parameter - $k$; (b) scale parameter - $\lambda$}
\label{fig:wbl}
\end{figure*}

\begin{table}[ht]
\centering
\begin{tabular}{|>{\footnotesize}l|>{\tiny}l|T|T|T|}
\hline
&\textbf{ \small GALA} & \textbf{\small MALA} & \textbf{\small MMALA} & \textbf{\small HMC}\\
\hline
\rowcolor{black}
\multicolumn{5}{|c|}{\textbf{\color{white}  \small Weibull (True $\lambda=1$, True $k=1.5$)}}  \\
\hline
 {Warmup} & 8,8,9	 &24,33,56	&137,180,254	 &4,16,39  \\
 {Acceptance (\%)}  &95.15,95.8,96.45 & 100,100,100 & 100,100,100 & NA \\
 {Estimated mean $\lambda$} &1.03,1.031,1.031 & 1.029,1.031,1.036 &0.999,1.043,1.075 & 1.03,1.032,1.033  \\
 {Sample variance $\lambda$} &  $[5.9,6.6,7.4] \times 10^{-5}$&  $[0.98,1.1,1.2] \times 10^{-3}$ & $[1.3,2.5,4.3] \times 10^{-3}$& $[0.89,1.1,1.3] \times 10^{-3}$  \\ 
 {Estimated mean $k$} & 1.532,1.533,1.535&1.52,1.53,1.54 &1.458,1.537,1.565 & 1.53,1.534,1.538\\
 {Sample variance $k$} &  $[1.8,2,2.2] \times 10^{-4}$ & $[2.2,2.8,4]. \times 10^{-3}$ & $[3.9,6,9.8]\times 10^{-3}$ & $[2.6,3,3.1]\times 10^{-3}$ \\
 {Runtime (seconds)}  & $29.13^{*}$ &8.23 & $27.45^{*}$ & 5.32 \\
\hline
\end{tabular}
\caption{Comparison of various performance metrics (minimum, median and maximum) for 2000 posterior samples obtained over 10 independent runs of each method. 400  observations are used for estimation. The mean and sampling variance are calculated based on 1000 samples after discarding the warmup samples for each method. The runtime for GALA and MMALA is unreasonably high due to the numerical expectations used for the Riemannian metric and its derivatives based on 2000 Weibull samples generated every iteration.}
\end{table}

\begin{figure*}[h!]
    \centering
    \begin{subfigure}{0.5\textwidth}
        \centering
        \includegraphics[height=1.5in]{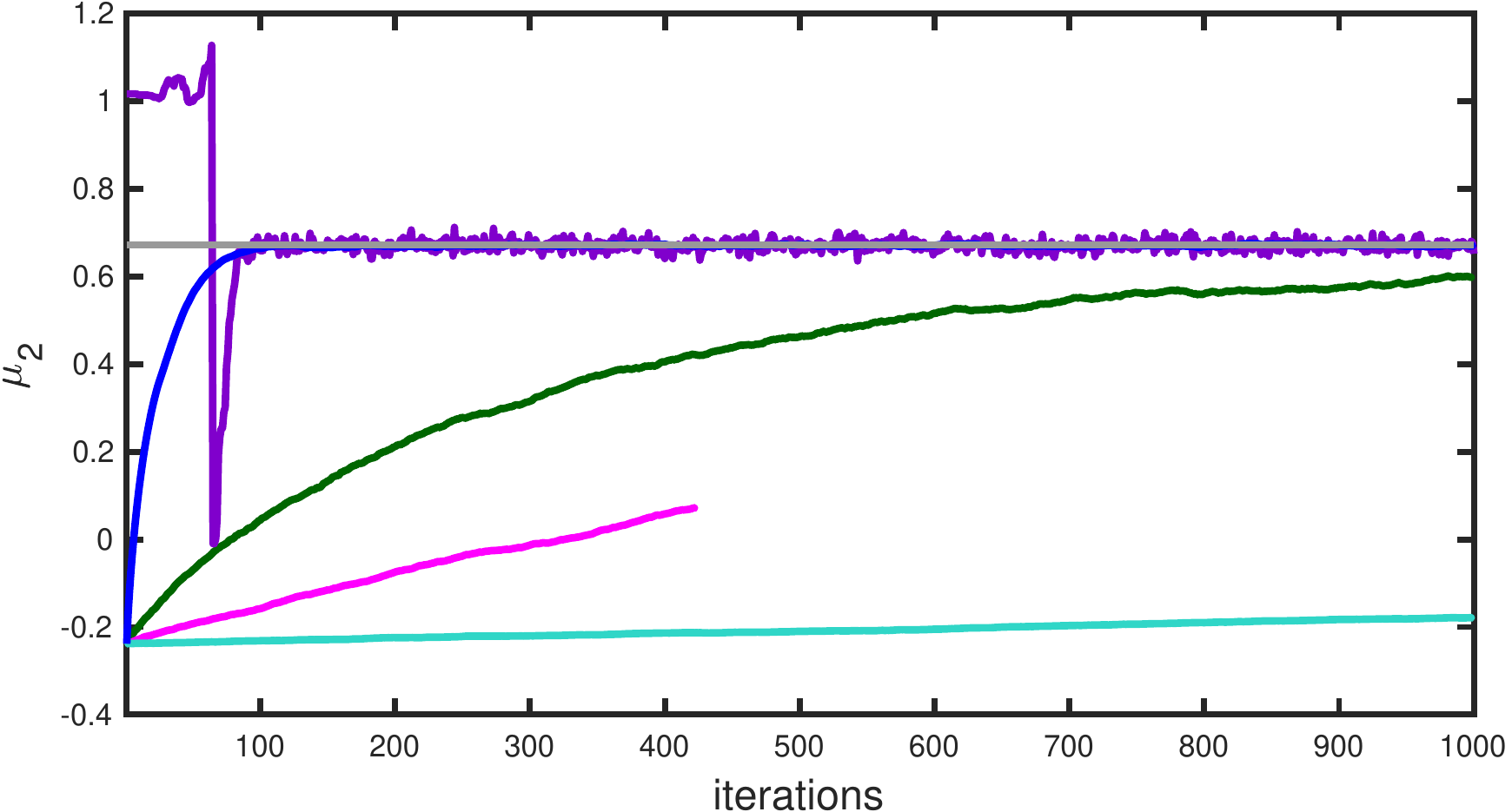}
        % \caption{$\mu_1$}
    \end{subfigure}%
    ~
    \begin{subfigure}{0.5\textwidth}
        \centering
        \includegraphics[height=1.5in]{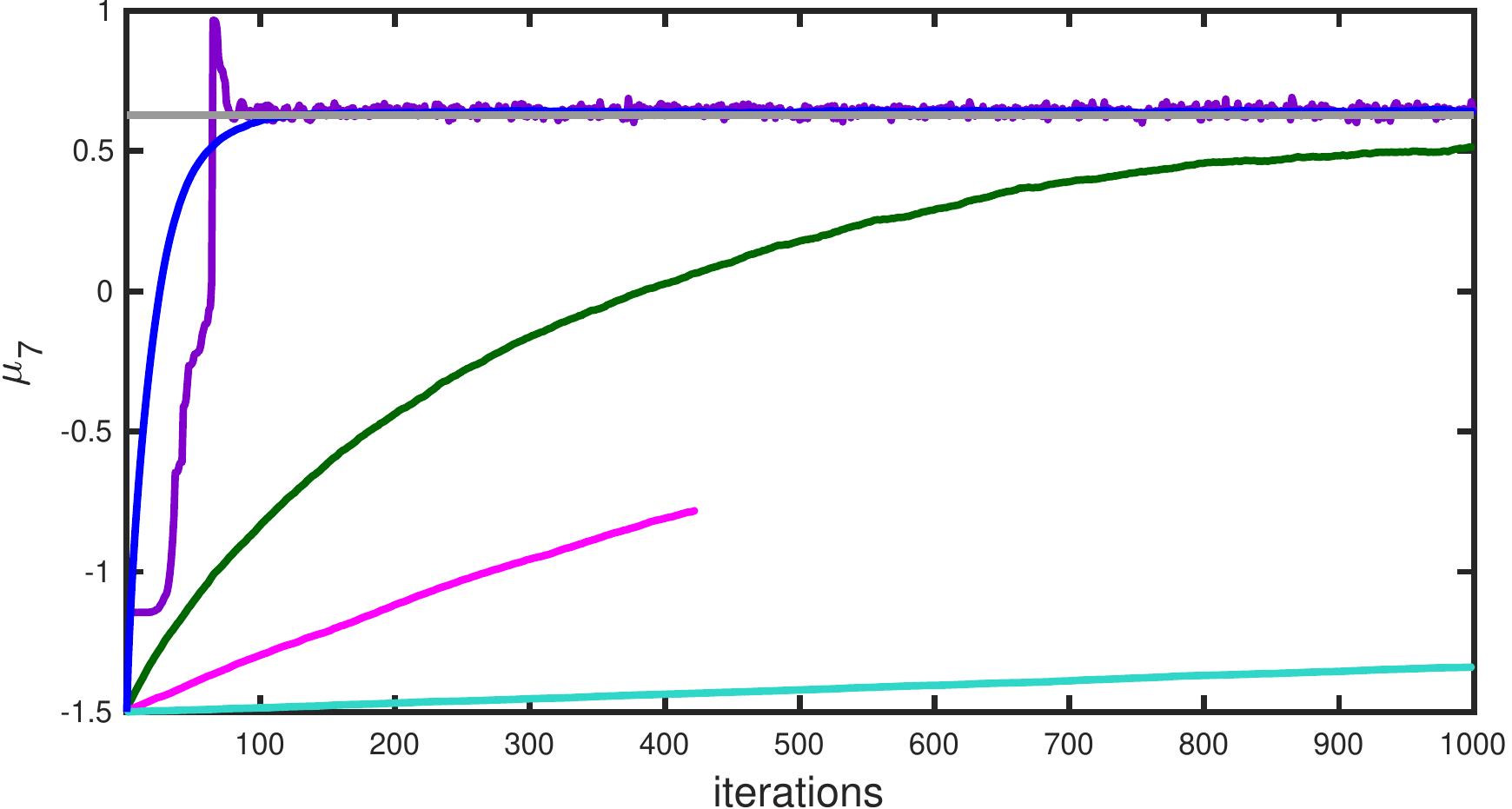}
        % \caption{$\mu_2$}
    \end{subfigure}
    \begin{subfigure}{0.5\textwidth}
        \centering
        \includegraphics[height=1.5in]{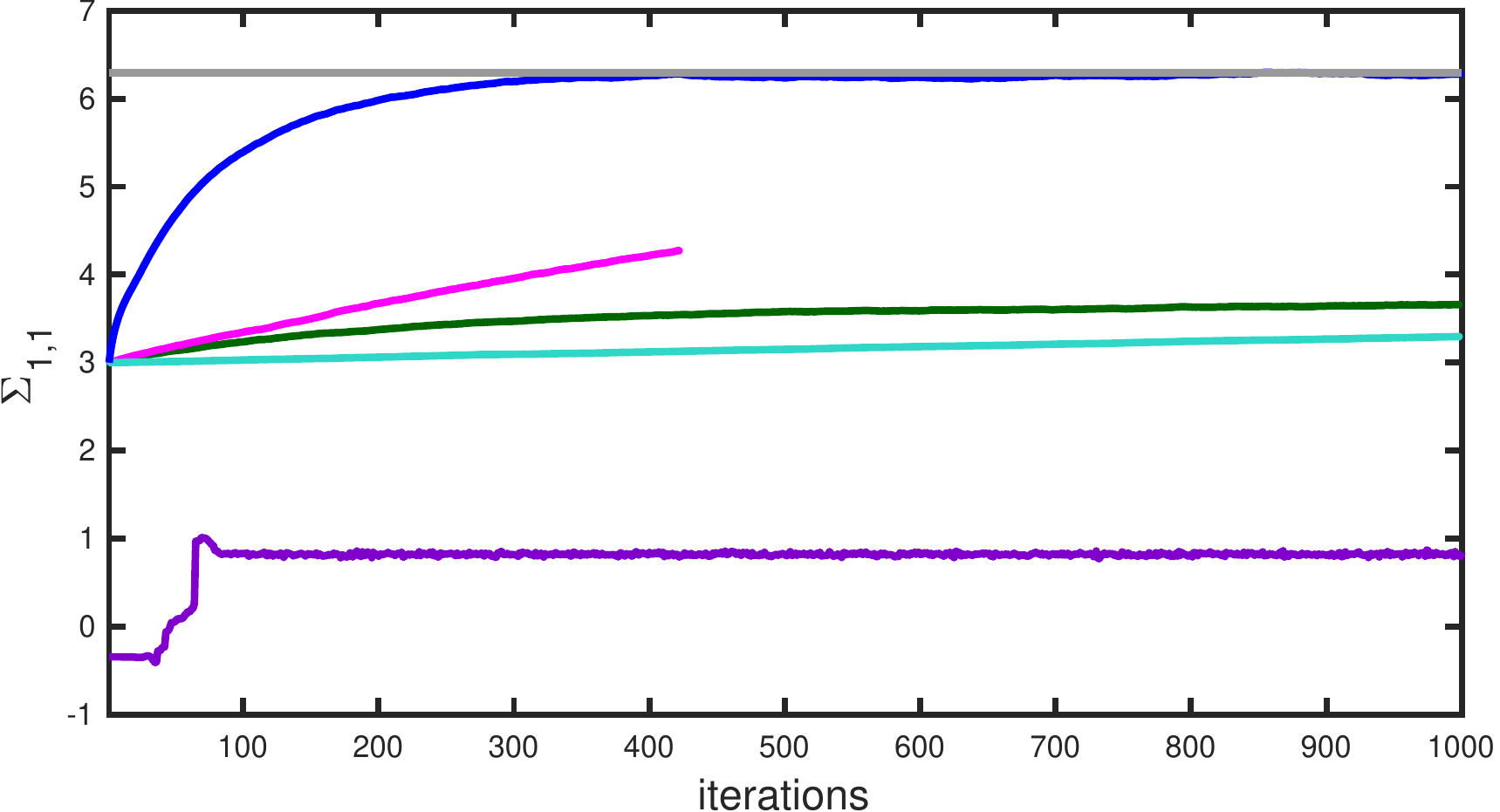}
        % \caption{$\Sigma_{22}$}
    \end{subfigure}%
    ~
    \begin{subfigure}{0.5\textwidth}
        \centering
        \includegraphics[height=1.5in]{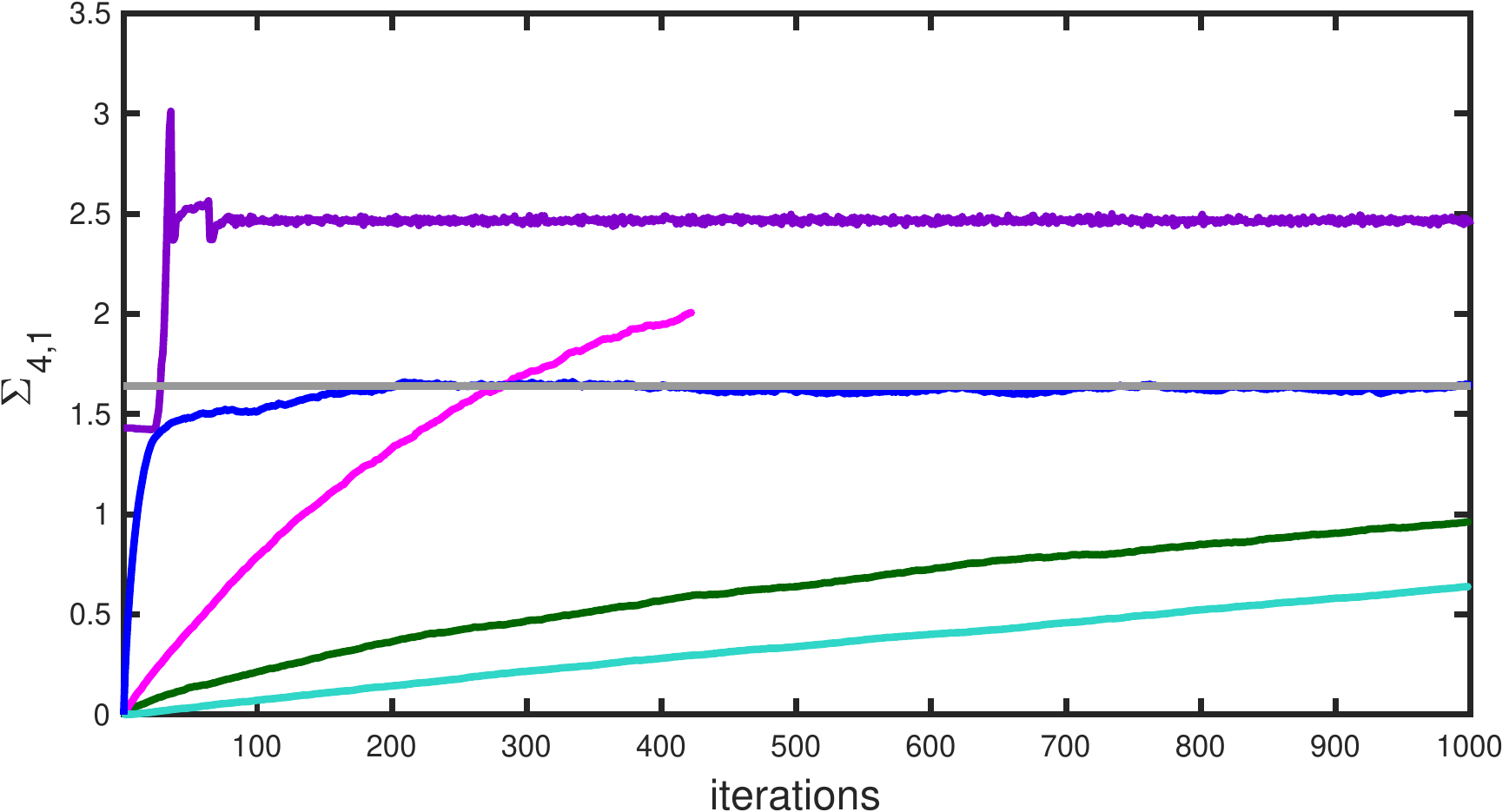}
        % \caption{$\Sigma_{32}$}
    \end{subfigure}
     \begin{subfigure}{0.5\textwidth}
        \centering
        \includegraphics[height=1.5in]{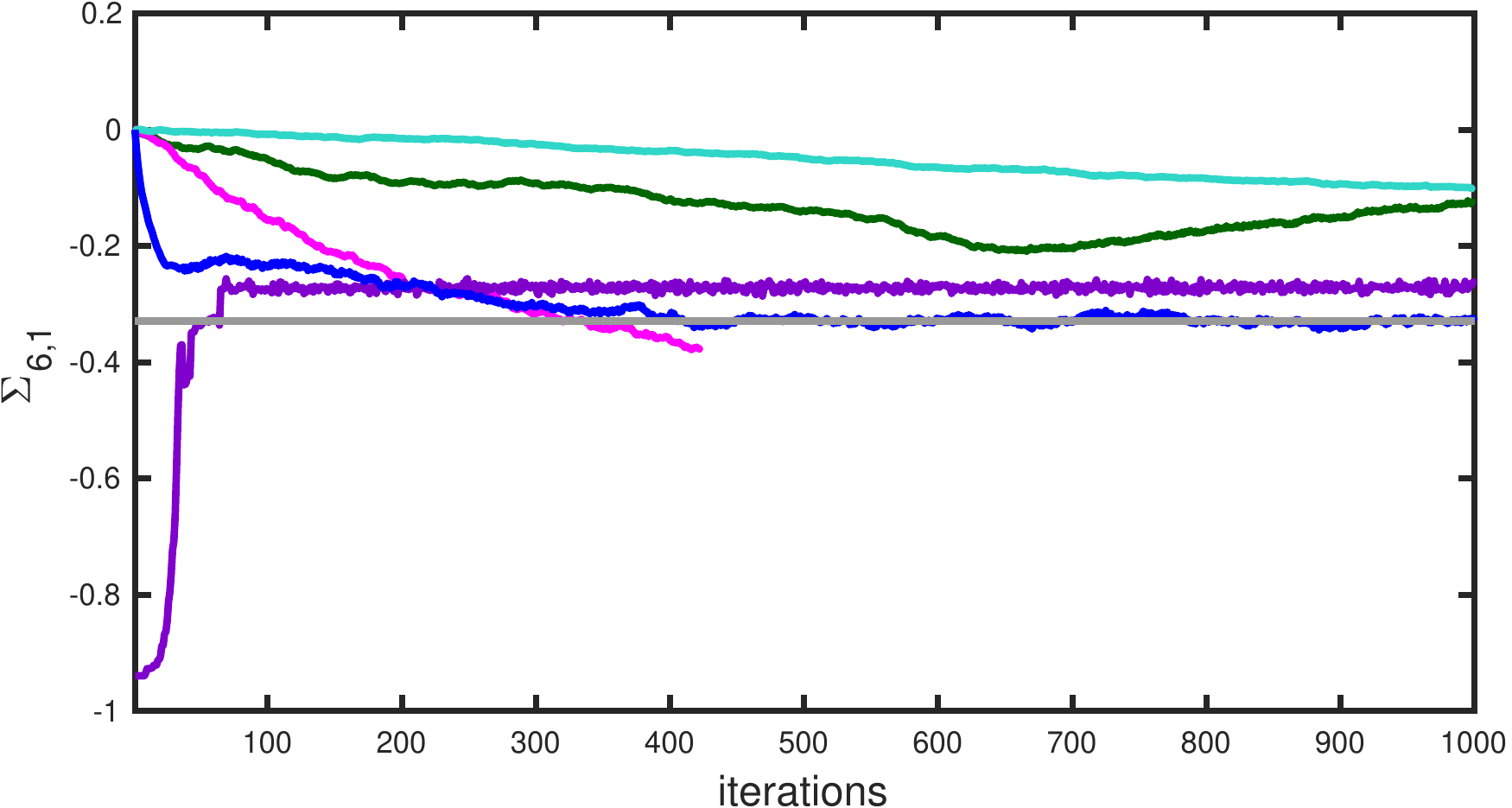}
        % \caption{$\Sigma_{54}$}
    \end{subfigure}%
    ~
    \begin{subfigure}{0.5\textwidth}
        \centering
        \includegraphics[height=1.5in]{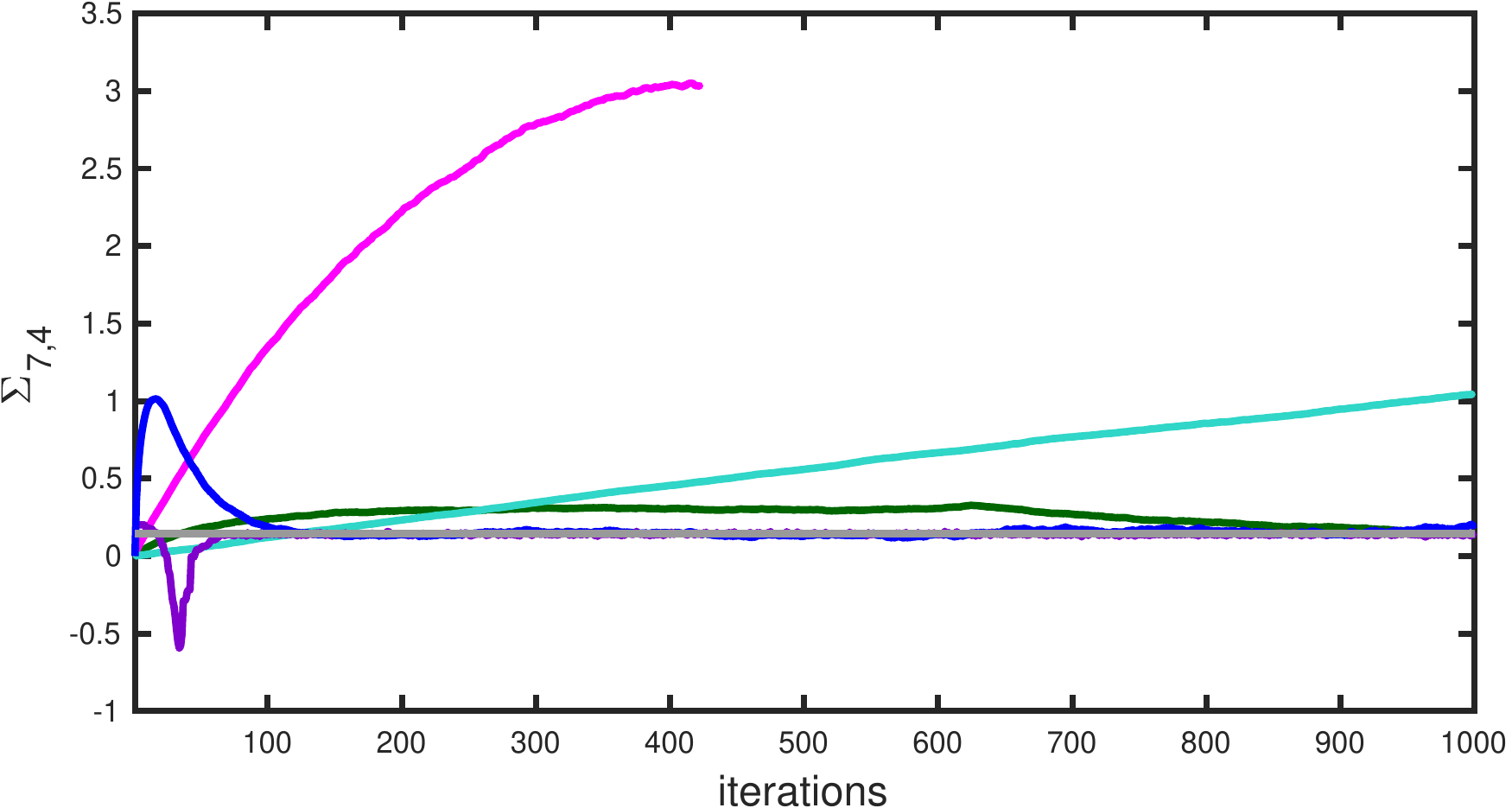}
        % \caption{$\Sigma_{54}$}
    \end{subfigure}
    \begin{subfigure}{0.5\textwidth}
        \centering
        \includegraphics[height=1.5in]{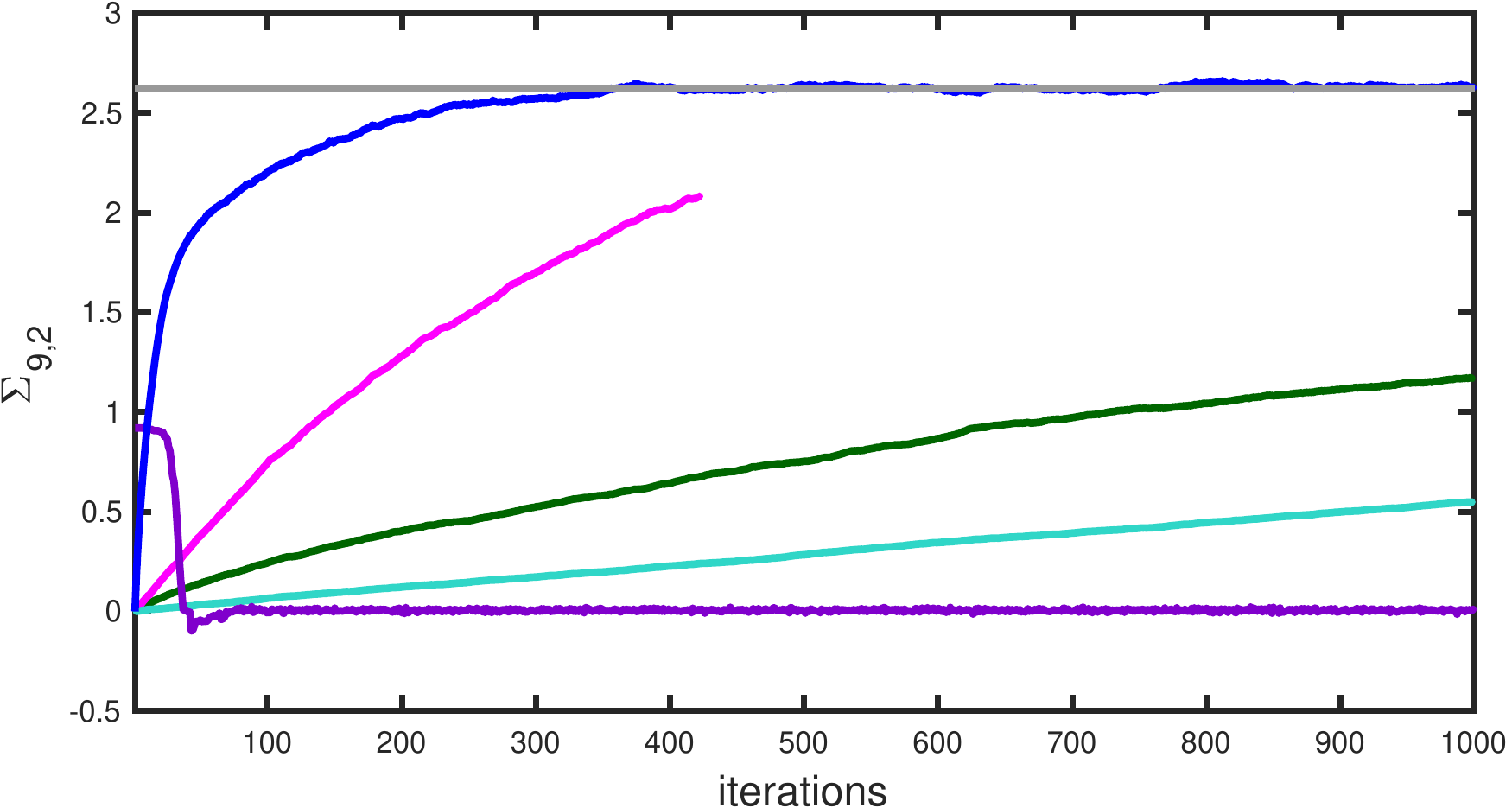}
        % \caption{$\Sigma_{54}$}
    \end{subfigure}%
    ~
    \begin{subfigure}{0.5\textwidth}
        \centering
        \includegraphics[height=1.5in]{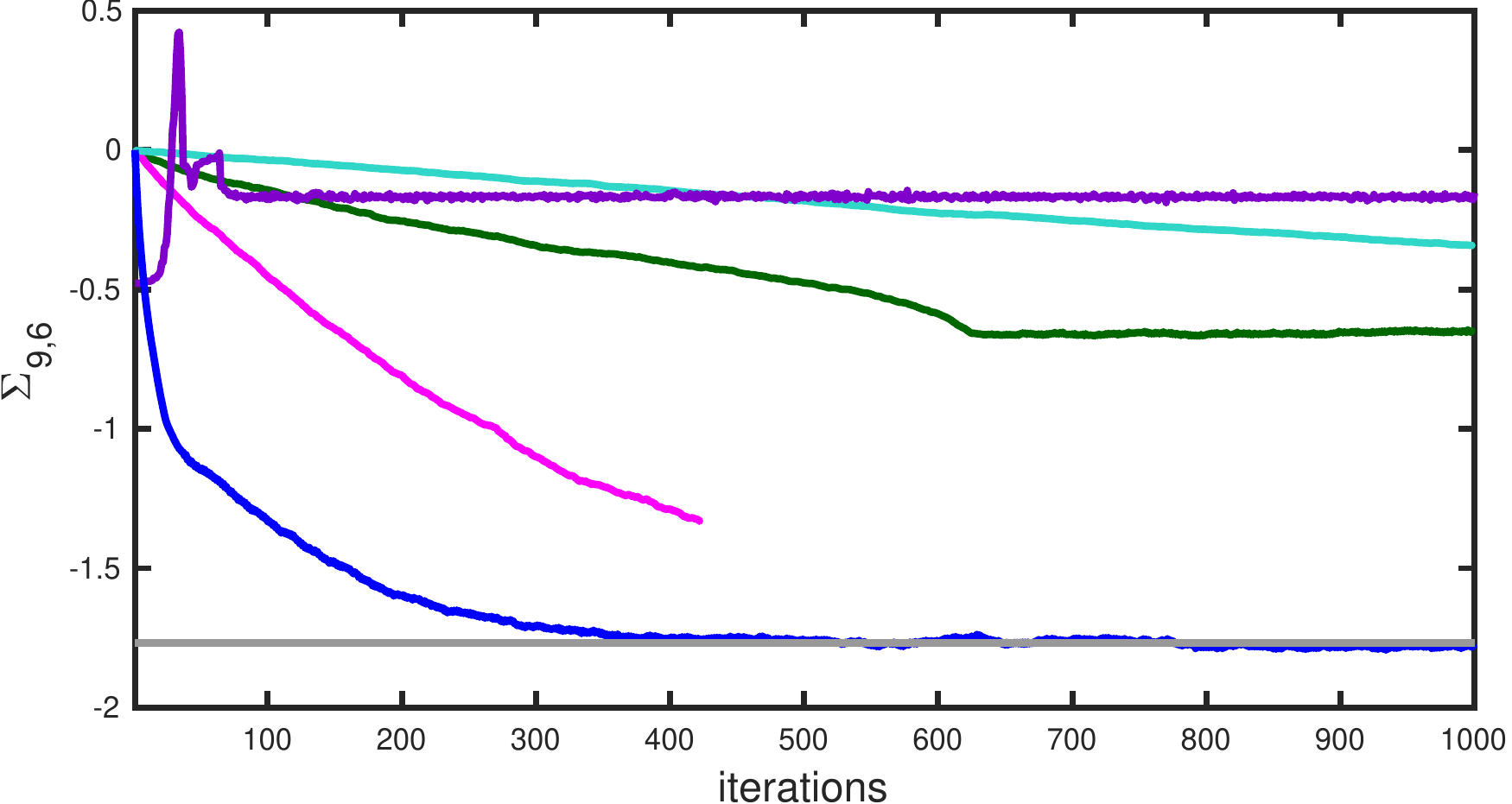}
        % \caption{$\Sigma_{54}$}
    \end{subfigure}
    \begin{subfigure}{0.5\textwidth}
        \centering
        \includegraphics[height=0.85in]{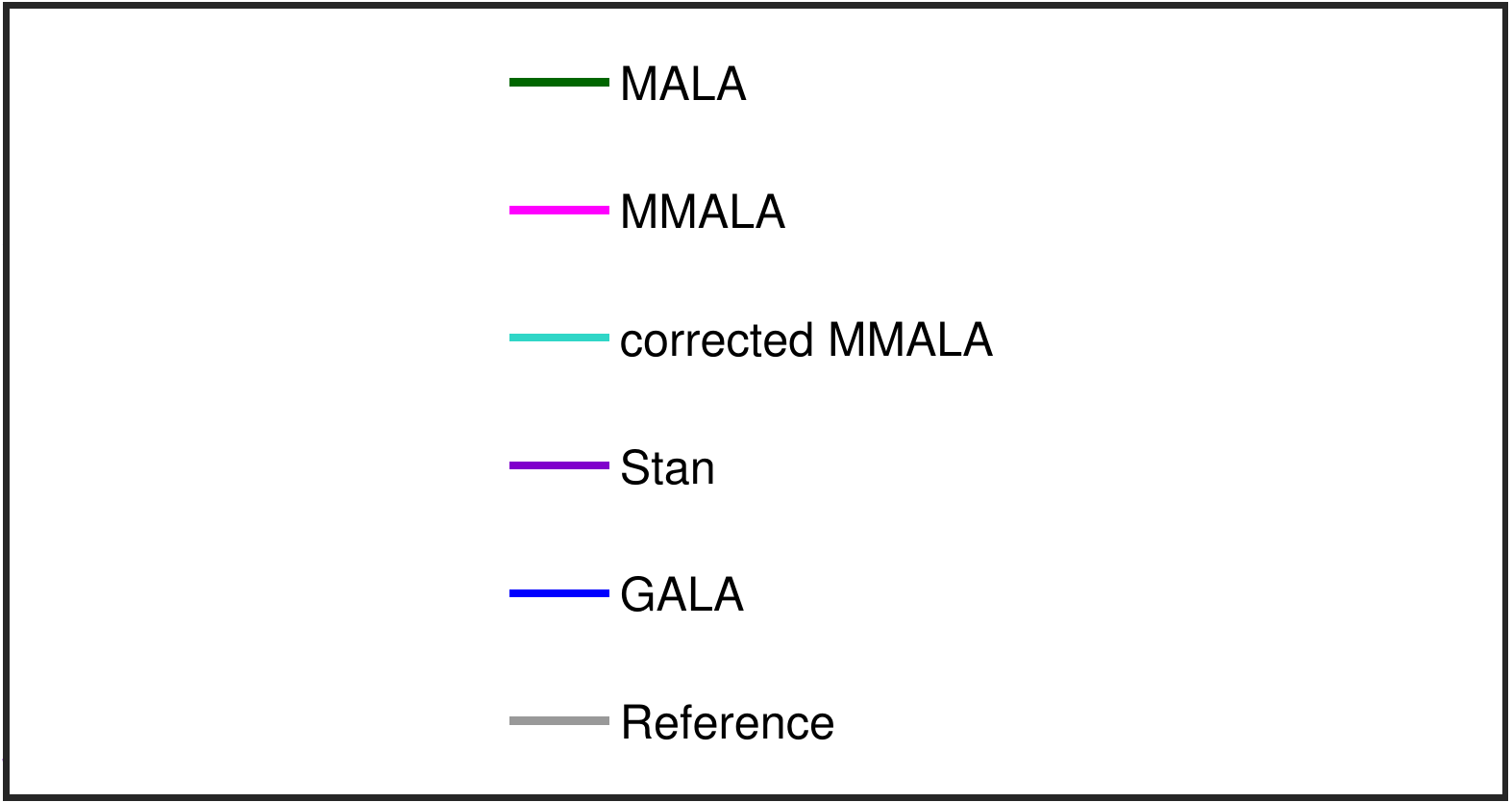}
    \end{subfigure}%
   \caption{A few components of the mean vector and covariance matrix for the 65 parameter multivariate Gaussian distribution via several competing methods; the legend indicating different methods used is shown separately.}
    % \caption{A few components of the covariance matrix $\Sigma$ in the multivariate Gaussian distribution via GALA ($\Delta t=0.02$), MALA ($\Delta t=0.005$) and HMC ($\Delta t=0.01,L=20$) for $N=1000$ sample observations; (a) $\Sigma_{22}$; (b) $\Sigma_{32}$; (c) $\Sigma_{54}$}
\label{fig:mvn}
\end{figure*}

% Gaussian
\textbf{Multivariate Gaussian distribution.}
Now consider a multivariate Gaussian distribution. Again, the detailed derivation for the developed equation is included in the supplementary material. In order to better understand the performance variation of different methods with increasing dimensionality, we consider a sequence of problems with number of parameters varying from 5 to 65. Figure \ref{fig:mvn} shows the chain plots for the 65-dimensional parameter problem with 10 unknown mean and 55 covariance matrix components (i.e. for a 10-dimensional Gaussian distributed dataset) obtained by GALA, MALA, MMALA, corrected MMALA \cite{XifaraT2014Ldat} and Stan, for a few components of the mean vector and the covariance matrix. All methods except GALA fail for the 65-dimensional problem. For most of the components, MALA does not converge in the 1000 steps considered. MMALA diverges, and all samples after about 400 iterations are rejected. Stan (brms package in R) converges for all the mean components; however out of the 55 components of the covariance matrix, it only converges for one or two. The computation time with Stan is also (at least) five times more than GALA for the Gaussian problem.

\begin{figure*}
    \centering
    \begin{subfigure}{0.95\textwidth}
        \centering
        \includegraphics[width=1\textwidth]{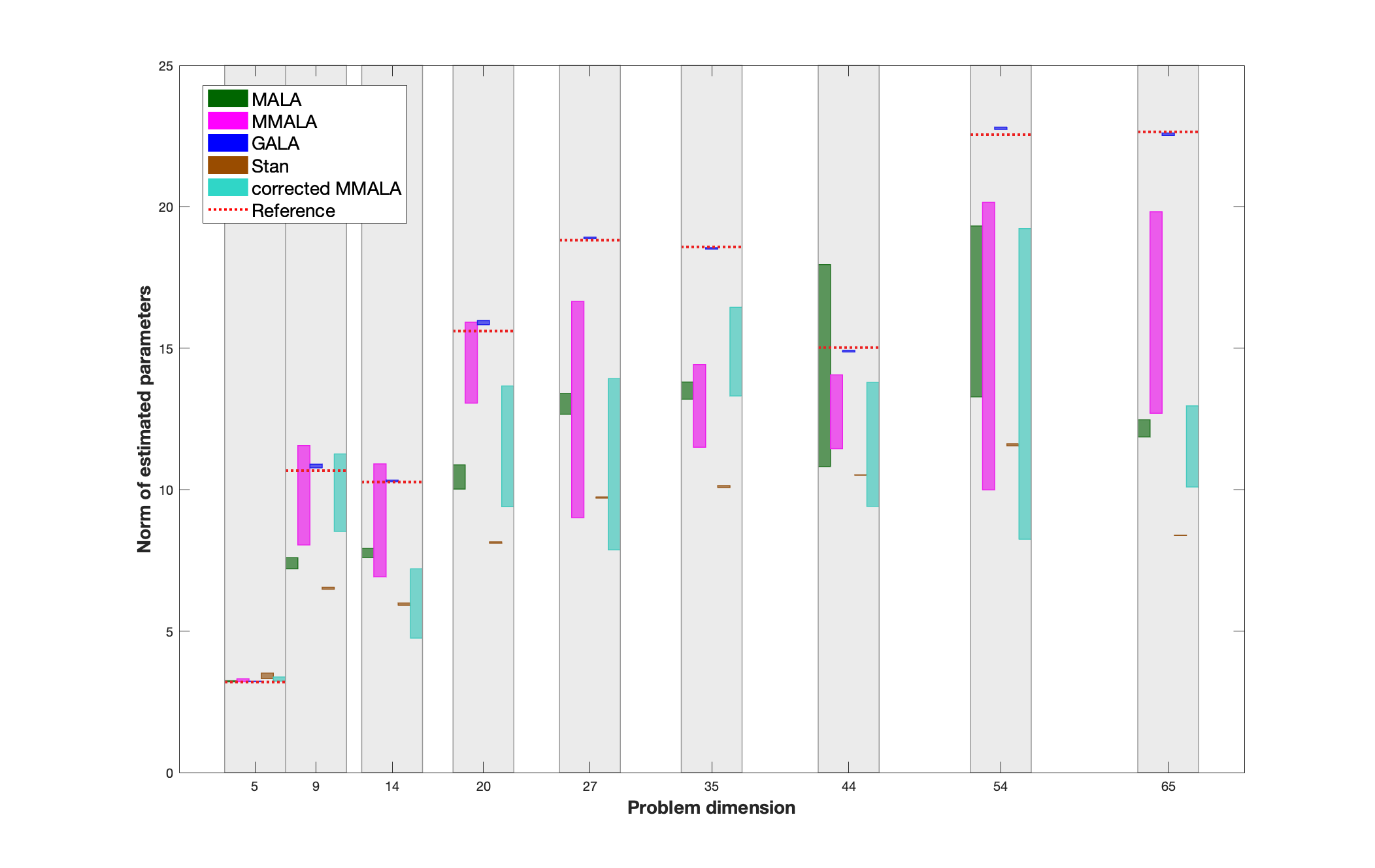}
        % \caption{$\mu_1$}
    \end{subfigure}%
    
    \begin{subfigure}{0.95\textwidth}
        \centering
        \includegraphics[width=1\textwidth]{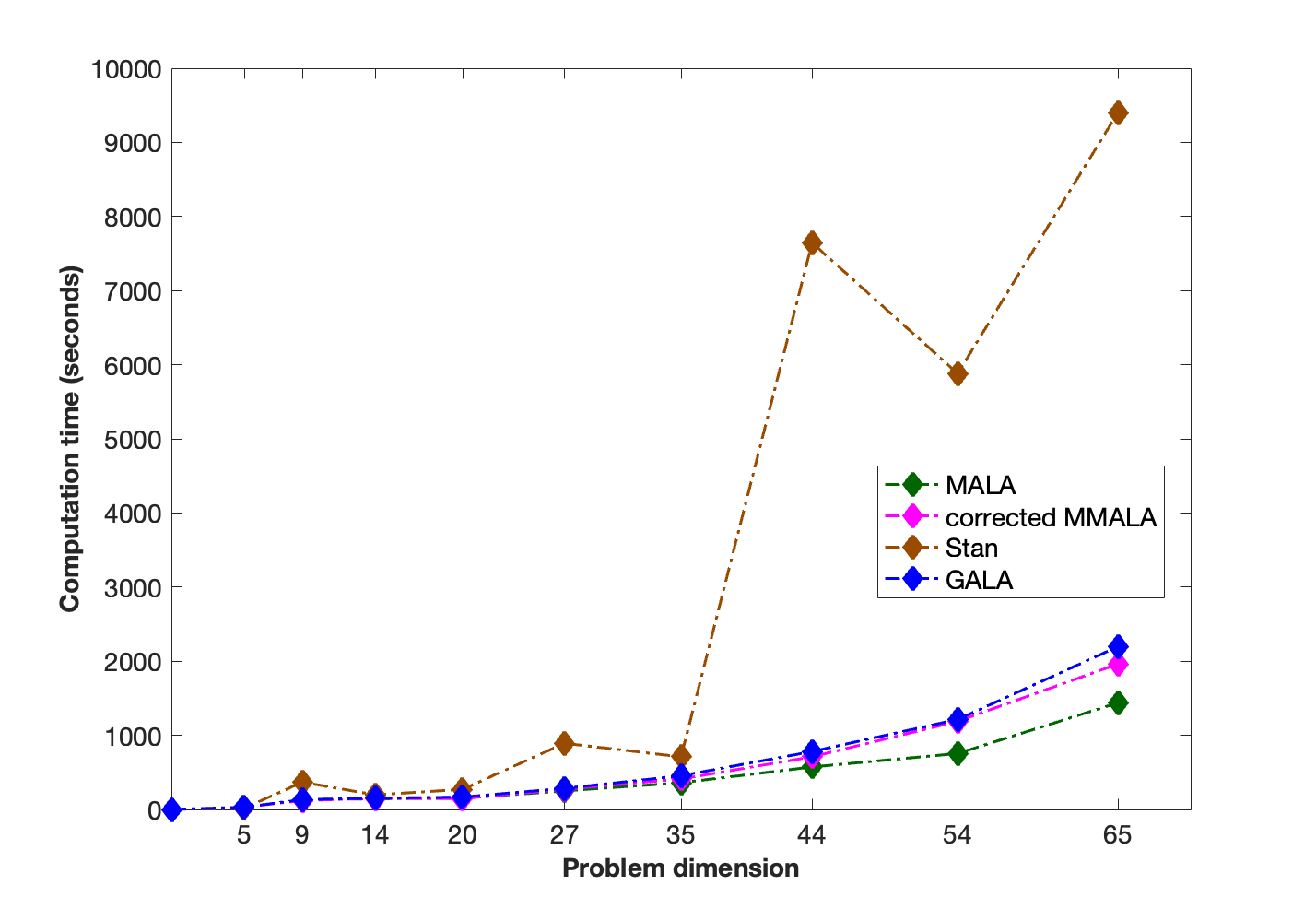}
        % \caption{$\mu_2$}
    \end{subfigure}
    \caption{Top: Minimum and maximum of estimated parameter norms across 4 independent chains of length 1000 each for Gaussian problems with varying dimensions. The means are based on the samples after warmup if convergence occurs, otherwise it is determined using the last 100 samples. Bottom: A comparison of computation time. MMALA is not included since after a few steps, all samples are typically rejected which renders a comparison inappropriate.}
    \label{fig:mvn:alldim}
    \end{figure*}

Figure \ref{fig:mvn:alldim} gives the variation in performance as well as computation time of several methods with increasing dimension. Specifically, it gives the minimum and maximum norms of the estimated mean parameter vector across 4 independent chains of length 1000 each (with the last 100 samples if there is no convergence, otherwise with all the samples following warmup). The gradual performance deterioration of most methods with GALA being the sole exception is a highlight of this figure. For instance, all methods but Stan (which fails to converge for the cross-covariance term $\sigma_{12}$) converge to the correct solution for the 5-parameter (2D Gaussian) problem. For the 9-parameter problem (3D Gaussian) case, all methods but corrected MMALA and GALA fail even as we observe a markedly slower rate of convergence with corrected MMALA. For still higher dimensional cases, all methods except GALA fail (at least for the 1000 steps over which the simulations are presently performed). Figure \ref{fig:mvn:alldim} also displays the computational time of all methods (except MMALA due to rejection of all samples after a few steps) according to dimension. Stan stands out as the method whose computational time increases the fastest with dimension whereas MALA, corrected MMALA, and GALA have similar computational times across 5-65 dimensions for this problem.

\begin{figure*}
    \centering
    \begin{subfigure}{0.5\textwidth}
        \centering
        \includegraphics[height=2.1in]{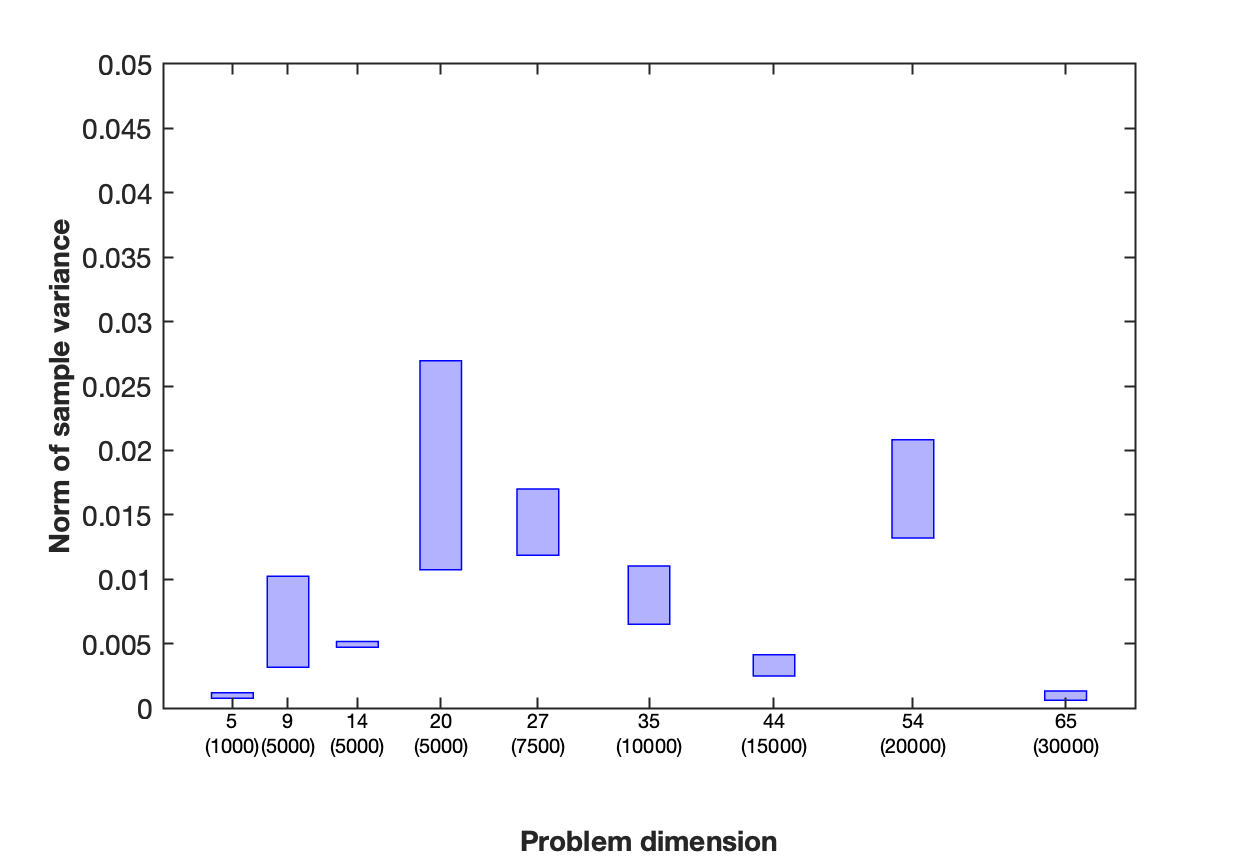}
        % \caption{$\mu_1$}
    \end{subfigure}%
    ~
    \begin{subfigure}{0.5\textwidth}
        \centering
        \includegraphics[height=2.1in]{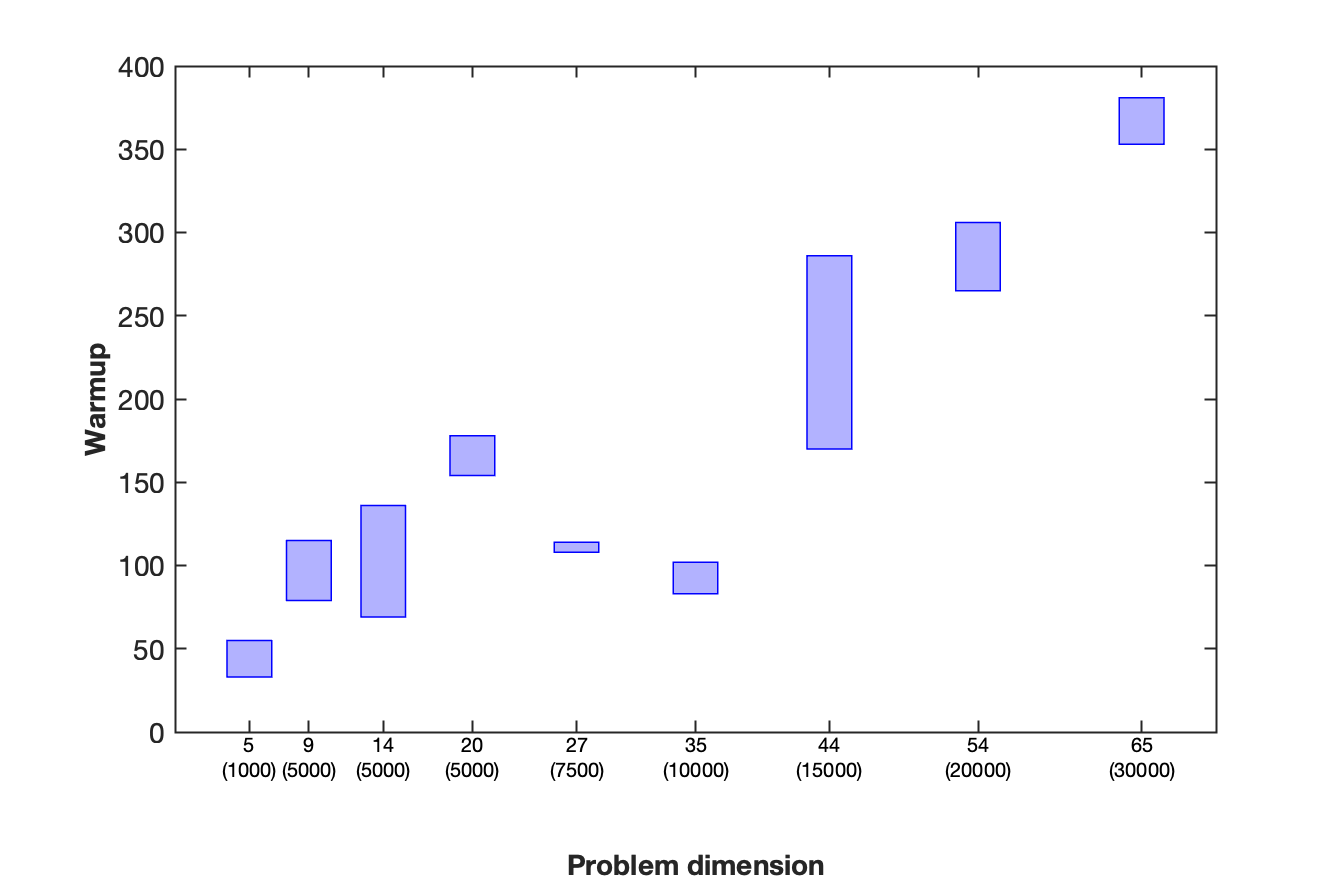}
        % \caption{$\mu_2$}
    \end{subfigure}
    \caption{Left: ranges of sample variance after warmup for GALA across the 4 independent chains for Gaussian problems with varying dimensions. Right: ranges of warmup for GALA for across 4 independent chains. The bracketed labels on $X-$ axis indicate the size of dataset used.}
    \label{fig:mvn:gala:var:warm}
    \end{figure*}

Figure \ref{fig:mvn:gala:var:warm} shows the ranges of sample variance and warmup length for GALA across dimensions 5-65. Again it may be noted that this is only a representative trend with increasing dimension for only 4 independent chains and may vary with initial conditions. The sample variances do not vary much across dimension, which is an impressive robustness across dimensions. The warmup lengths increases with the problem dimension but seemingly linearly. A 5-D problem with 1000 sample size requires a warmup of 50 iterations, whereas a 65-D problem with 30000 sample size requires a warmup of only 400 iterations. These two metrics of performance are demonstrating the unique strength of GALA, as it is the only one to converge towards the solution, and its exceptional efficiency and scalability.

 \subsection{Application to Logistic Regression}
In this subsection, we take up a logistic regression problem, one that arises frequently in diverse fields like machine learning, social and medical sciences. Let $X_{D \times N} = \{ X^{(1)}, X^{(2)}, X^{(3)} ...X^{(N)}\}$ represent $N$ samples of the $D$-dimensional explanatory variables that are available along with the binary response variable $t_{1 \times N}$. Here each $t_i$ is a Bernoulli random variable with the probability of success depending on $X^{(i)}$. Assuming that the true regression coefficients  are represented by $\beta_{(D+1)\times 1}$, the  probability of success for each $X_{D \times 1}^{(i)}$  is given by 
$$p \left( X^{(i)}| \beta \right)=p \left(t_i=1 \right) = \frac{1}{1+\exp \left(-\beta_0 + \sum_{j=1}^D \beta_j X_j^{(i)}\right)}$$
The likelihood of the data is then the product of likelihoods over the $N$ data points. Assuming a prior density on $\beta$ as $\mathcal{N}(0,\alpha I)$, where $\alpha$ is chosen appropriately, the Fisher information matrix and its derivative for the posterior are given by
\begin{equation}
G_{pq} = \sum_{i=1}^N \frac{\exp \left( -\beta_0 + \sum_{j=1}^D \beta_j X_j^{(i)} \right) X_p^{(i)}X_q^{(i)}}{1+\exp \left(-\beta_0 + \sum_{j=1}^D \beta_j X_j^{(i)}\right)^2}  + \alpha^{-1}\delta_{pq}
\end{equation}
\begin{equation}
\frac{\partial G_{pq}}{\partial \beta_r} = - \sum_{i=1}^N   \frac{ \exp \left(-\beta^T \bar{X}^{(i)} \right)  \bar{X}^{(i)}_p \bar{X}^{(i)}_q X^{(i)}_r}{\left( 1+\exp\left(-\beta^T \bar{X}^{i} \right) \right)^2}   + 2 \sum_{i=1}^N   \frac{\left( \exp \left( -\beta^T \bar{X}^{(i)} \right) \right)^2  \bar{X}^{(i)}_p \bar{X}^{(i)}_q \bar{X}^{(i)}_r}{\left (1+\exp \left(-\beta^T \bar{X}^{(i)} \right) \right)^3 } 
\end{equation}

\begin{figure*}
    \centering
    \begin{subfigure}{0.5\textwidth}
        \centering
        \includegraphics[height=1.5in]{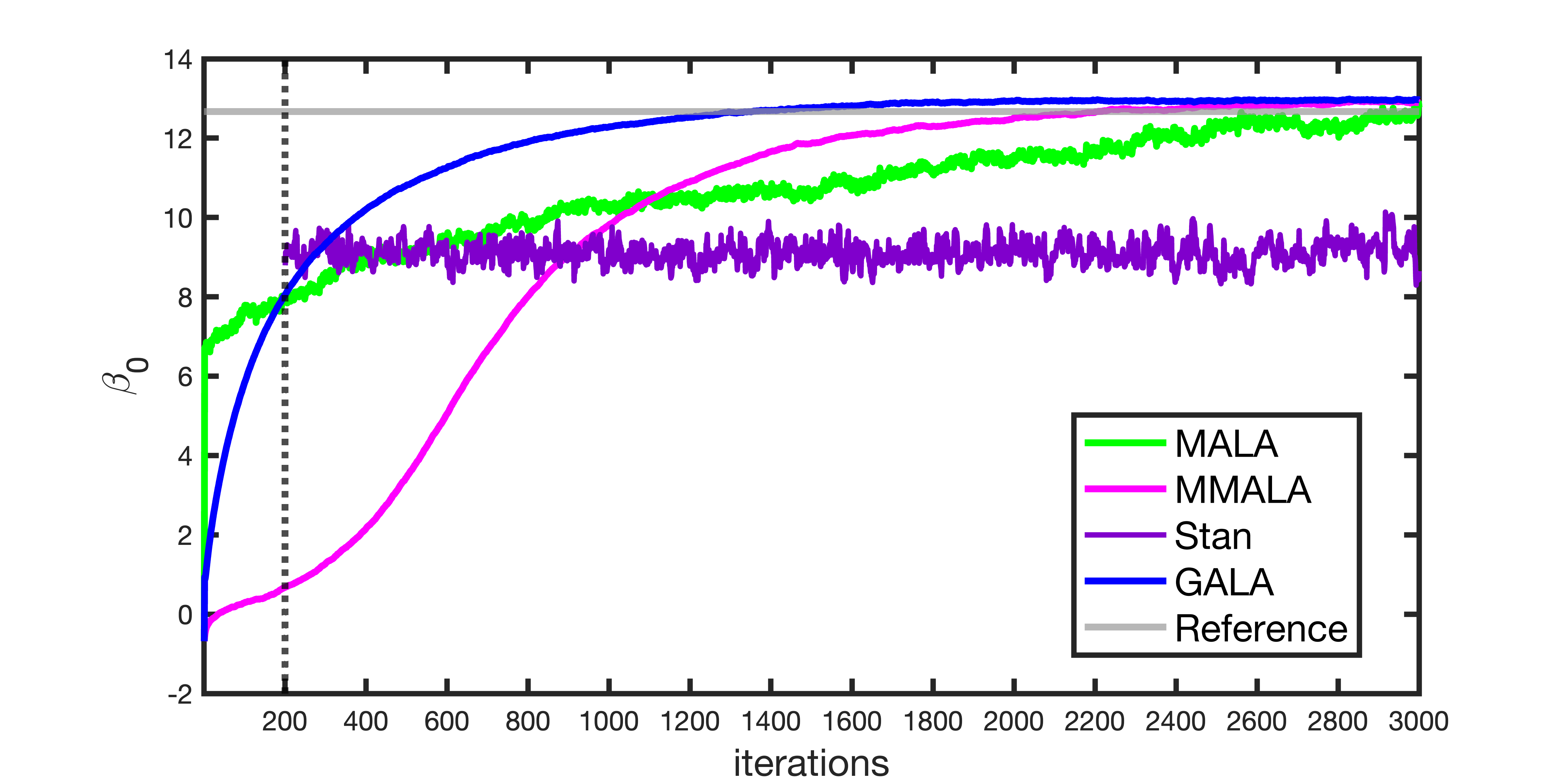}
        % \caption{$\beta_0$}
    \end{subfigure}%
    ~ 
    \begin{subfigure}{0.5\textwidth}
        \centering
        \includegraphics[height=1.5in]{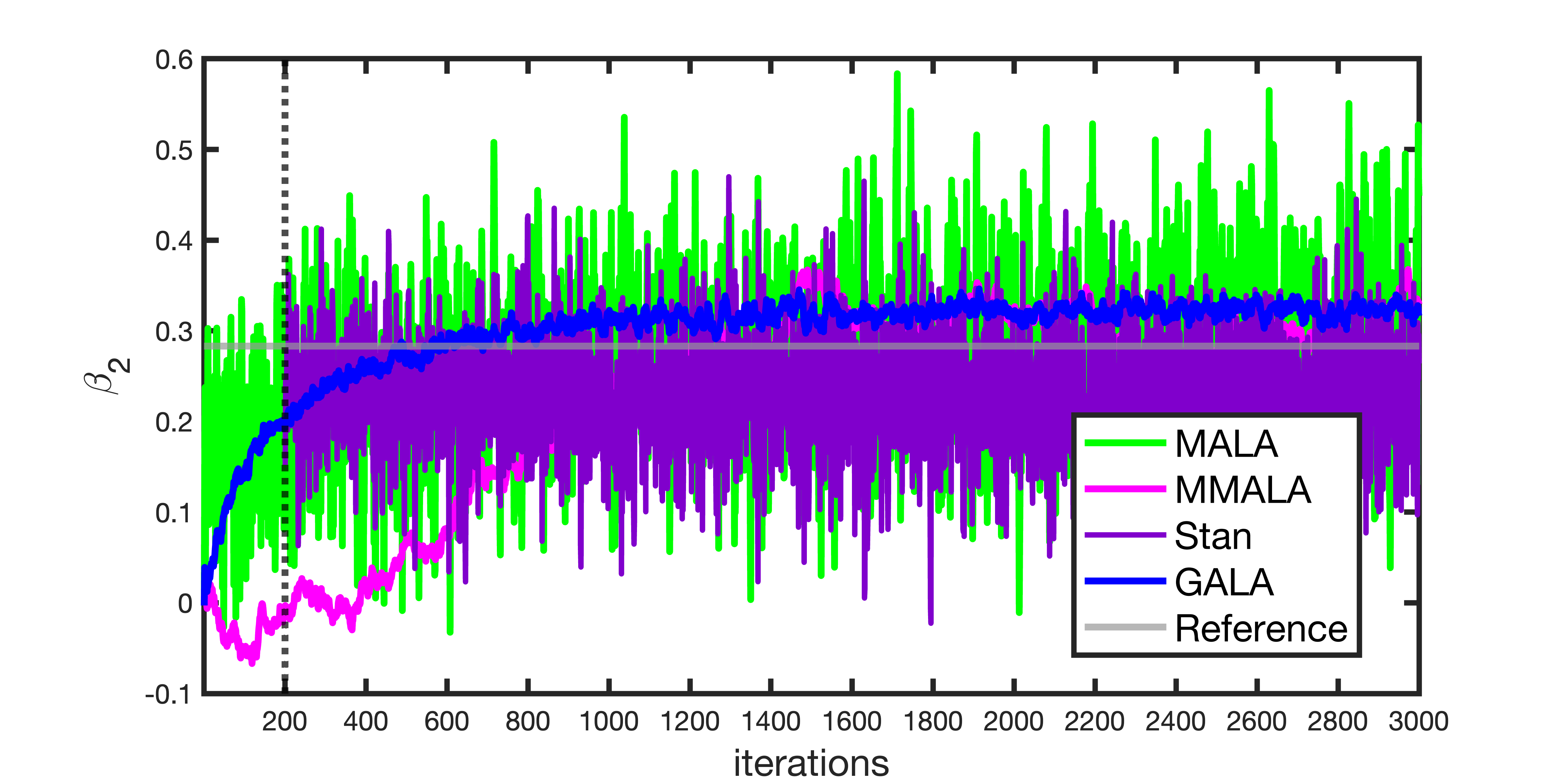}
        % \caption{$\beta_2$}
    \end{subfigure}
    \begin{subfigure}{0.5\textwidth}
        \centering
        \includegraphics[height=1.5in]{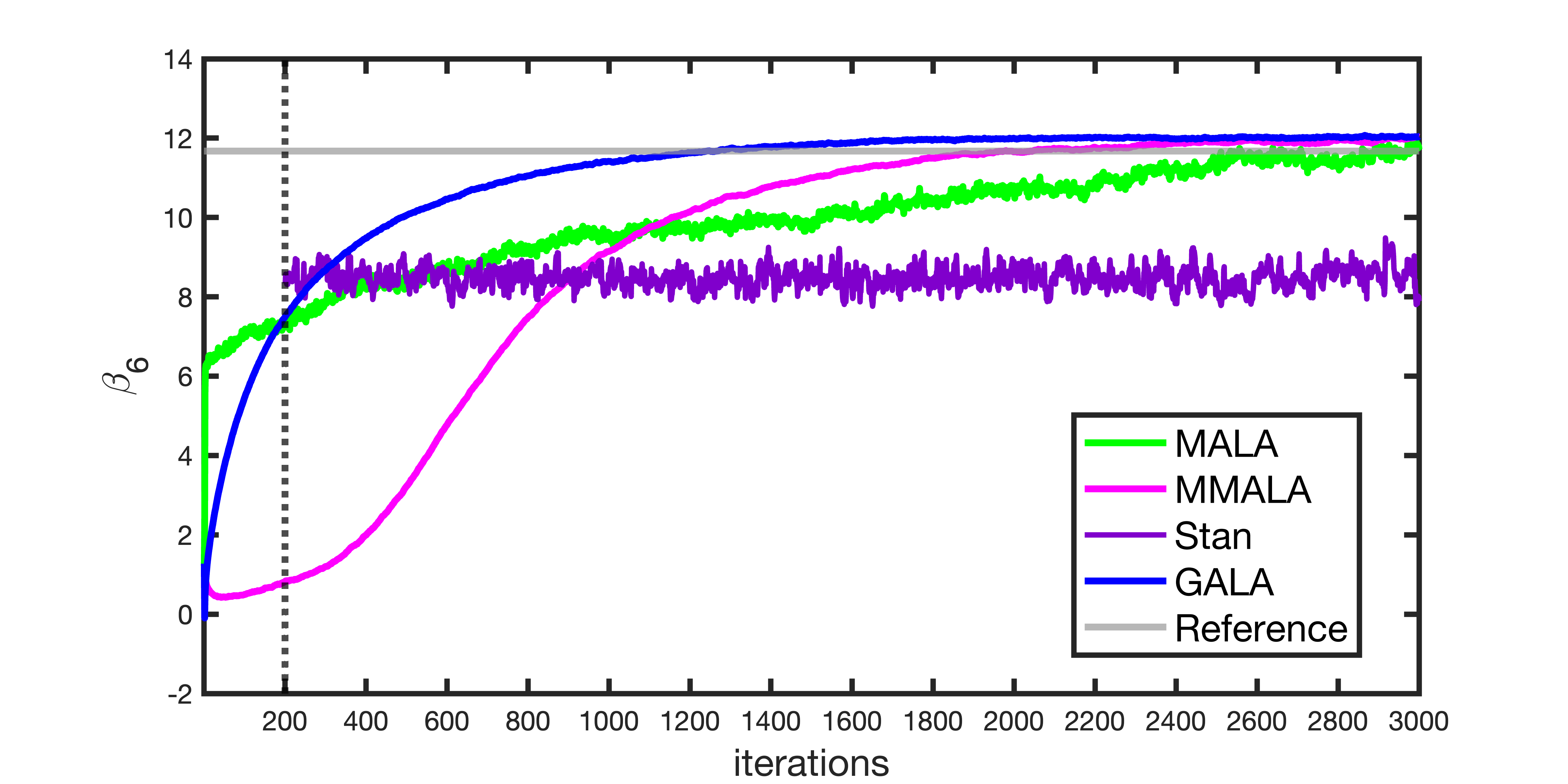}
        % \caption{$\beta_6$}
    \end{subfigure}%
    ~ 
    \begin{subfigure}{0.5\textwidth}
        \centering
        \includegraphics[height=1.5in]{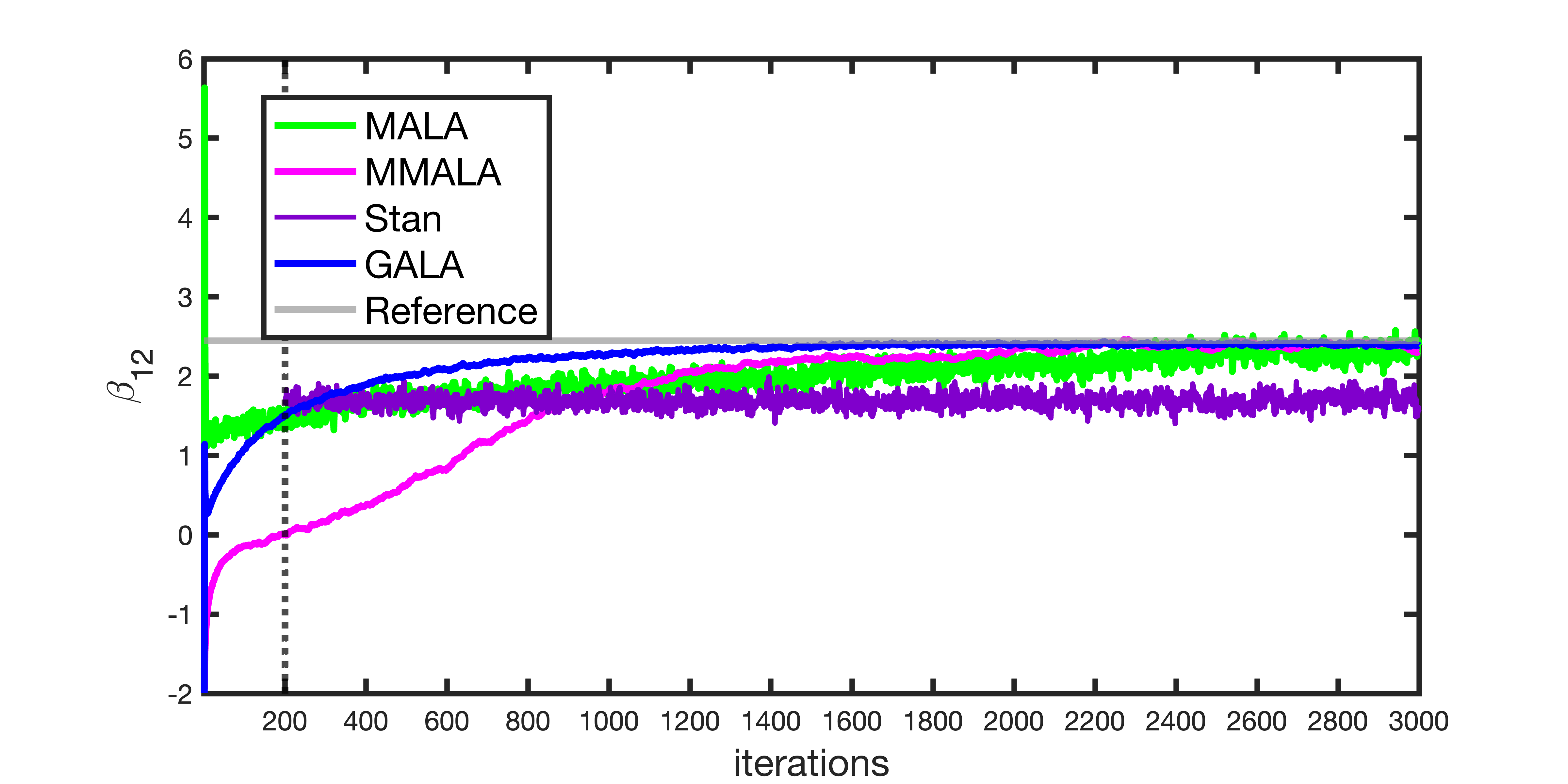}
        % \caption{$\beta_{12}$}
    \end{subfigure}
    \begin{subfigure}{0.5\textwidth}
        \centering
        \includegraphics[height=1.5in]{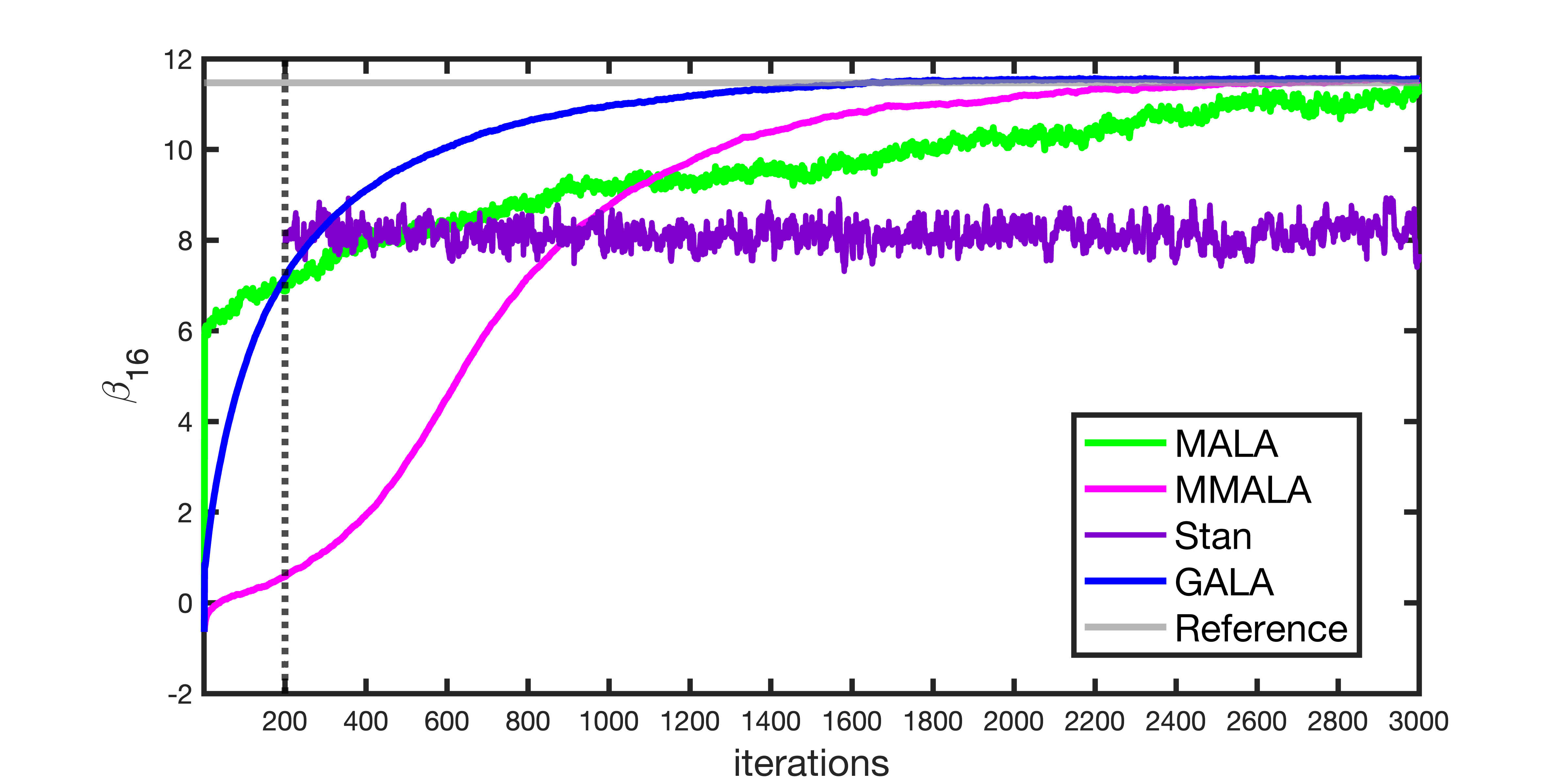}
        % \caption{$\beta_{16}$}
    \end{subfigure}%
    ~ 
    \begin{subfigure}{0.5\textwidth}
        \centering
        \includegraphics[height=1.5in]{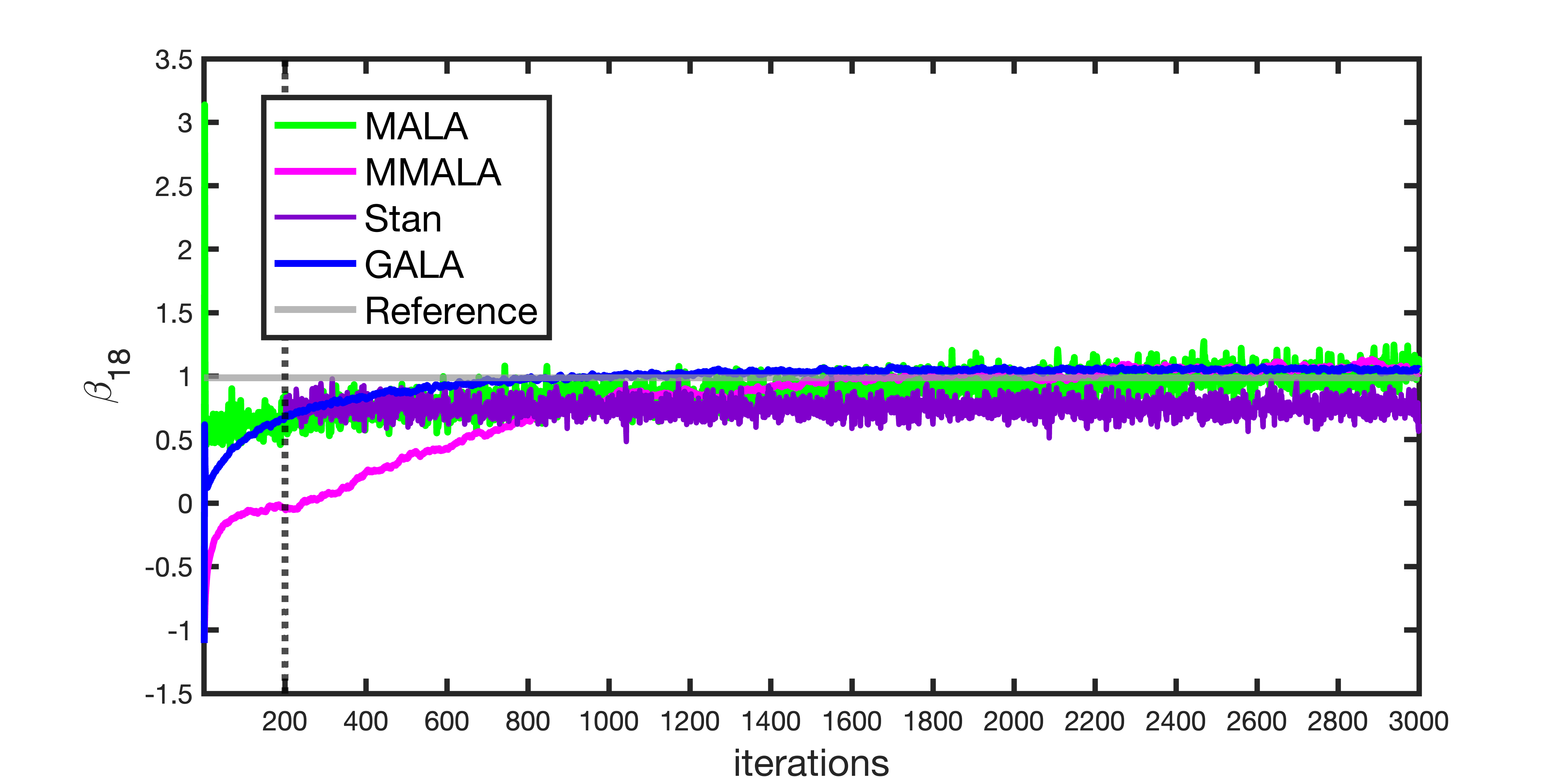}
        % \caption{$\beta_{18}$}
    \end{subfigure}
    
    \begin{subfigure}{0.5\textwidth}
        \centering
        \includegraphics[height=1.5in]{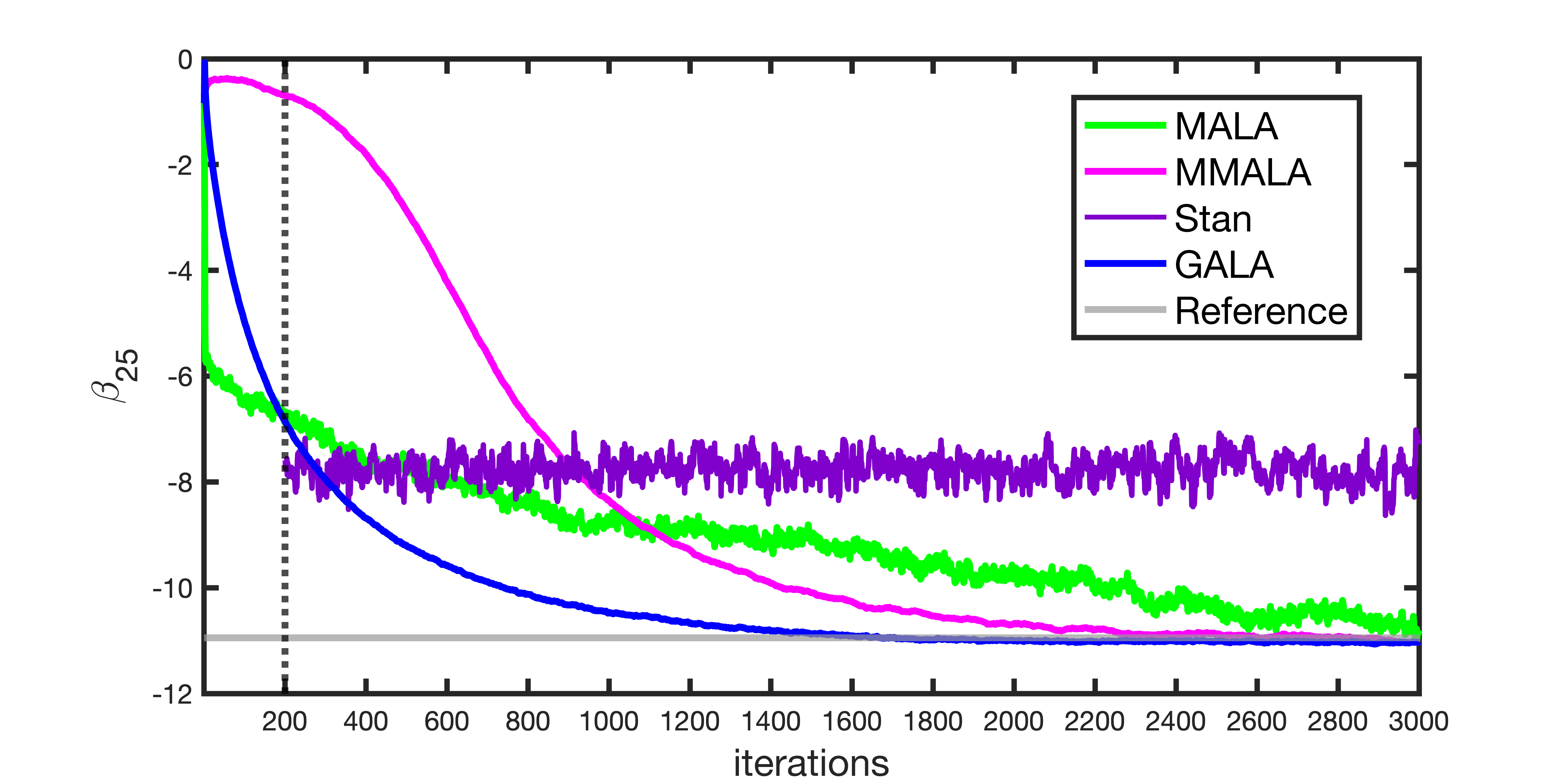}
        % \caption{$\beta_{25}$}
    \end{subfigure}%
    ~ 
    \begin{subfigure}{0.5\textwidth}
        \centering
        \includegraphics[height=1.5in]{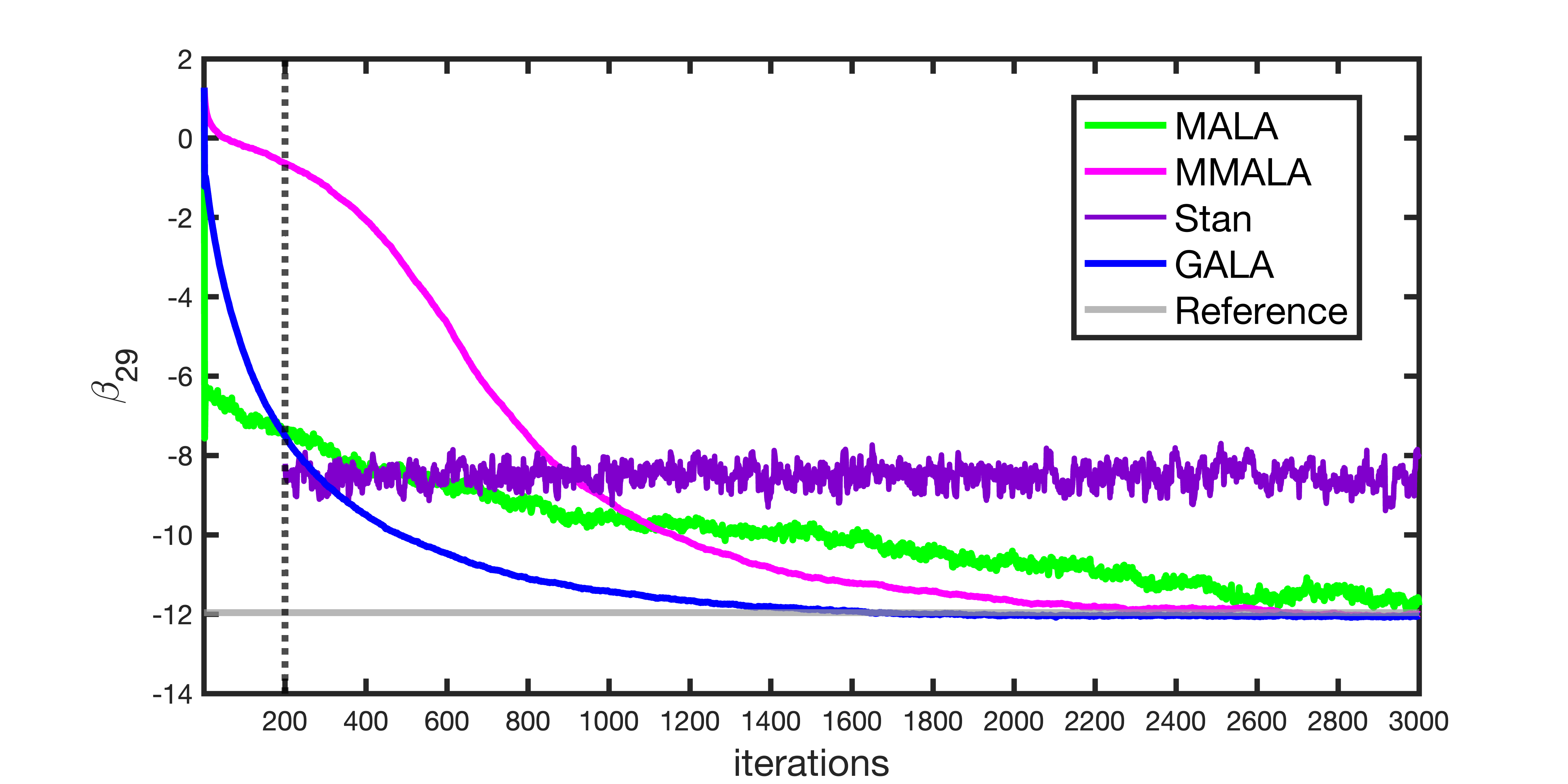}
        % \caption{$\beta_{29}$}
    \end{subfigure}
    \caption{A few of the reconstructed regression parameters for a 30 dimensional logistic regression problem via several methods.}
    % \caption{A 30-D logistic regression problem: comparisons of a few reconstructed regression parameters via GALA($\Delta t=0.02$), MALA($\Delta t=0.01$), R.M.MALA($\Delta t=0.02$) and HMC ($\Delta t=0.01,L=10$), with $N=20000$ observations ; (a) $\beta_0$; (b) $\beta_1$; (c) $\beta_2$; (d) $\beta_3$; (e) $\beta_4$; (f) $\beta_5$}
\label{fig:LR}
\end{figure*}

See the supplementary material for a detailed derivation. Thus, all the input quantities needed for GALA are now  determined. We show in  Figure \ref{fig:LR} a comparison of results via GALA, MALA, MMALA and Stan (rstanarm package in R) for a few regression parameters. For the 30-dimensional problem considered, the parameters are chosen so that 25 of them are uniformly distributed between $[0,15]$ while the remaining 5 are uniformly distributed in $[-15,-10]$. This is done to make the problem slightly more challenging. MALA just about converges in the 3000 steps for this problem. MMALA is faster than MALA though much slower than GALA; whereas Stan, even though it is the fastest, fails for this problem. The warmup with Stan are automatically discarded, which is reflected in the figures. Again, Table 3 summarizes the various performance metrics. Similar to the Gaussian example, in this problem too, the norm of the 30-dimensional mean and variance is given for convenience. 

%Note that as the number of observations increases, the performance of all the methods improves with the exception of MMALA, which exhibits a stark deterioration in the rate of convergence. The results shown in figure \ref{fig:LR} are for a high value (4000) of observations as the contrast in accuracies of various methods is more clear as compared to the case where observations are of the order of a few hundreds.
%%%%%%%%%%%%%%%%%%%%%%%

%%%%%%%%%%%%%%%%%%%%%%%%
 
 %% table for LR and mvn
 \begin{table}
\centering
\begin{tabular}{|>{\footnotesize}l|>{\tiny}l|T|T|T|}\hline
&\textbf{\small GALA} & \textbf{\small MALA} & \textbf{\small MMALA} & \textbf{\small Stan}\\
\hline
% \rowcolor{black}
% \multicolumn{5}{|c|}{\textbf{\color{white}  Multivariate Gaussian (65 dimensional), norm of true mean $=22.65$}}  \\
% \hline
% %\multirow{3}{4em}{Multiple row} & cell2 & cell3 \\ 
% % \multirow{2}{*}{GALA}& 380,381, \\ & 353,358\\

% \textbf{Warmup} & 380,381,353,358 & -&-&- \\
% \textbf{Acceptance (\%)} &100,100,100,100 &-&-&- \\
% \textbf{Norm of estimated mean} &22.5,22.51,22.45,22.51 & -&-&- \\
% \textbf{Norm of sample variance} &$5.5e^{-3},5.3e^{-3},3.8e^{-3},3.8e^{-3}$ &-&-&- \\
% \textbf{Runtime (minutes)}& 36.58 & 23.96 & 36.72 & $\approx$185 \\
% \hline
\rowcolor{black} \multicolumn{5}{|c|}{\textbf{ \color{white} \small Logistic Regression (30 dimensional),  norm of true mean $=48.62$}} \\
\hline
% %\multirow{3}{4em}{Multiple row} & cell2 & cell3 \\ 
{Warmup} & 1348,1363,1294,1336 & -& 2222,2203,2112,2185 & -\\
{Acceptance (\%)} & 100,100,100,100 &-&100,100,100,100 &-\\
{Norm of estimated mean} & 48.81,48.8,48.82,48.92 &-&48.9,48.85,48.82,48.76 &-\\
{Norm of sample variance} & 0.01,0.0091,0.01,0.02 &-&0.016,0.01,0.02,0.01 &-\\
{Runtime (seconds)}& 966 & 819 & 852 &$\approx$ 160 \\

%   MMALA & 2222,2203, & 3000,3000, &  48.9,48.85,&0.016,0.01, \\
%   & 2112,2185 & 3000,3000 &  & 48.82,48.76 &0.02,0.01 & \\
% Stan  & - & - &  - & - &  \\ 
\hline
\end{tabular}
\caption{Comparison of various performance metrics for 3000 samples obtained for the logistic regression problem for each of the 4 independent runs of each method. Whereas for the logistic regression problem, the estimates of the norm of mean and sampling variance are based on 500 samples after discarding warmup. The logistic regression codes for GALA and MMALA are parallelised to acheieve a 40 \% reduction in computation time while that for MALA is not.}
\end{table}

\section{Concluding remarks}\label{sec:conclusions}
Exploiting the Fisher-Information matrix as a Riemannian metric and the associated Riemannian connection, we have stochastically developed a given SDE in the standard Euclidean setting. Unlike the known equation for a Brownian motion developed on the Riemannian manifold, the SDE that we geometrically adapt has a non-trivial drift term as well. We have specifically used this novel construction to modify the Langevin SDE and hence MALA. Our anticipation had been that a restriction of solutions to the Riemannian hypersurface should yield significantly higher accuracy and faster convergence even though Brownian noise processes have unbounded variations. That this feature can indeed be realized is demonstrated through a couple of applications, e.g. estimating the parameters in a probability density given a set of observations and solving the logistic regression problem. For both problems, the GALA based approach far outperforms the standard MALA and HMC, both in faster burn-in and estimation accuracy alike (e.g. sampling variance smaller by an orders of magnitude). This relative superiority of performance is generally more pronounced as the problem dimension increases, and so does the superiority in computational cost compared to Stan. This is particularly noticeable in the estimation of covariance in multivariate normal distributions where it is the only successful method.

% Beyond performance, and scalability to high dimensions, we also want to highlight the accessibility of GALA compared to HMC/STAN/NUTS. Indeed, Stan NUTS requires solving an optimization of the hyperparameters $L$ and $dt$ driving the algorithm. In some situations, just to set up the problem requires... and .... and fails whenever... GALA does not necessitate such optimisation, only a reasonable choice of $dt$ that allows the algorithm to function. We found this ease of use to be. We provide a 

Beyond performance, and scalability to high dimensions, we also want to highlight the accessibility of GALA compared to HMC and NUTS (Stan). Indeed, HMC requires tuning of two parameters for it to work efficiently, which becomes a struggle with an increase in dimension. The NUTS sampler was developed with the objective to get around this very difficulty, and claimed to perform at least as well as HMC. However, we did not find Stan to be accessible. The installation on a Linux or Windows machine for the MATLAB or R implementation of Stan (software to implement the NUTS sampler) failed despite multiple attempts and several hours of professional help from research software engineers. We finally moved to R on Mac OS to successfully implement the 'rstan' package. Even so, except for a few standard distributions,  just to set up the problem requires a fair bit of knowledge to write a program in Stan. To overcome this, we worked with 'rstanarm' for the logistic regression problem and 'brms' package in R for the multivariate Gaussian problem.  A few packages including the 'rstanarm' and 'brms'  \cite{burkner2017brms,muth2018user} were developed to bypass the need for a user to program in Stan, which is useful but one package may be more straightforward than another for a given problem. Leaving aside the difficulties in installation and the posing of the problem in Stan, we observe from the results, that it converges to incorrect values for both large dimensional examples considered in this work, viz. the 30-dimensional logistic regression problem and the 65-dimensional Gaussian parameter estimation problem. Indeed, this issue was explored in \cite{sanz2020nuts}, and it was found that NUTS does not converge to the correct invariant distribution, although it could be achieved with some modification. In contrast, GALA is easy to implement, only a reasonable choice of $dt$ allows the algorithm to function efficiently, the value of $dt$ we used is typically one or two orders of magnitude higher than that used for MALA. Unlike all first gradient based methods, GALA requires the derivative of the Fisher-Information metric, which we provide in the supplementary material for all the examples considered in this work, for convenience.

A word about a possible future direction before we conclude this article. The continuous but non-differentiable structure of the B.M. requires that we write the SDEs in terms of differentials and not the usual derivatives. The derivation of the developed equations on a Riemannian manifold, as in Section \ref{sec:derivation}, therefore required the language of exterior calculus and Cartan's structure equations. Although the curvature tensor, or more precisely the curvature 2-form which is a fundamental tensor field of incompatibilty on a Riemannian manifold, has appeared in our developed equation, we have presently neglected it as a higher order term.  An understanding and exploitation of this term in the context of Monte Carlo algorithms is, to our understanding, an important element of future study. A related curiosity also lies in a possible extension of the geometric framework to a Riemann-Cartan manifold, which would enable the developed dynamics to be further enriched by the torsion 2-form. Overall, the mathematical machinery of Cartan's moving frame appears to be a powerful tool in an insightful understanding of the role of geometry for stochastic development, possibly opening up routes to more efficient Monte Carlo algorithms.

% COMMENT

\bibliographystyle{abbrv}
%\bibliography{convexBodyVol}
\bibliography{GALEref}

\section*{Appendices}
\appendix
\section{Invariant distribution of simplified MMALA}
The proposal SDE for simplified MMALA is
\begin{equation}
dX_t = \frac{1}{2}G^{-1}(X_t) \nabla \log(\pi(X_t)) dt + \sqrt{G^{-1}(X_t) }dB_t
\end{equation}
 The Fokker-Planck equation \cite{RoyDebasish} for this SDE is 
 \begin{equation} \label{eqn:fkp}
\frac{\partial p(x,t)}{\partial t} = -\frac{\partial }{\partial x} (\frac{1}{2}G^{-1}(x) \nabla \log(\pi(x))p(x,t)) + \frac{\partial^2 }{\partial x^2}(\frac{G^{-1}(x)}{2} p(x,t))
\end{equation}

Let $\pi(x)$ be its stationary solution so that the LHS of \ref{eqn:fkp} vanishes. Accordingly, subsituting $p(x,t) = \pi(x)$ on the RHS, we have
\begin{eqnarray}
&& -\frac{\partial }{\partial x} (\frac{1}{2}G^{-1}(x) \nabla \log(\pi(x))\pi(x)) + \frac{\partial^2 }{\partial x^2}(\frac{G^{-1}(x)}{2} \pi(x)) \nonumber \\
=&& -\frac{1}{2}\frac{\partial }{\partial x} (G^{-1}(x) \frac{\partial \pi(x)}{\partial x} ) +\frac{\pi(x)}{2} \frac{\partial^2 G^{-1}(x)}{\partial x^2}  + \frac{G^{-1}(x)}{2} \frac{\partial^2 \pi(x)}{\partial x^2}   \nonumber \\
=&& - \frac{1}{2}\frac{\partial G^{-1}(x)}{\partial x} \frac{\partial \pi(x)}{\partial x} - \frac{1}{2} G^{-1}(x)\frac{\partial^2 \pi(x)}{\partial x^2}+\frac{\pi(x)}{2} \frac{\partial^2 G^{-1}(x)}{\partial x^2}  + \frac{G^{-1}(x)}{2} \frac{\partial^2 \pi(x)}{\partial x^2}  \nonumber  \\
=&&  - \frac{1}{2}\frac{\partial G^{-1}(x)}{\partial x} \frac{\partial \pi(x)}{\partial x} + \frac{\pi(x)}{2} \frac{\partial^2 G^{-1}(x)}{\partial x^2}  \label{eqn:smm}
\end{eqnarray}

The RHS does not vanish, meaning the assumption of $\pi(x)$ being the invariant distribution of the simplified MMALA SDE is incorrect.

\section{Invariant distribution of MMALA}
For the SDE in equation \ref{eqn:solo:1}, the Fokker-Planck equation on Riemannian manifold is given by equation \ref{eqn:solo:3} \cite{solo2009nonlinear}
\begin{equation}\label{eqn:solo:1}
dx^i = a^i dt + \sigma^{ij}dB_j 
\end{equation}
\begin{equation}\label{eqn:solo:3}
\frac{\partial p}{\partial t} =\frac{1}{2 \sqrt{g}} \frac{\partial }{\partial x^i}(\sqrt{g} g^{ij} \frac{\partial p}{\partial x^j}) - \frac{1}{\sqrt{g}}\frac{\partial }{\partial x^i}(p \sqrt{g}[a^i + \frac{1}{2} g^{jk} \gamma^i_{jk}])  
\end{equation}
The equation above assumes $\sigma^{im} (\sigma^T)^{mj}= g^{ij}$ .The drift term in MMALA is $a^i =\frac{ g^{ij}}{p} \frac{\partial p}{\partial x^j} - \frac{1}{2}g^{jk} \Gamma^i_{jk} $, substituting in \ref{eqn:solo:3}, we have
\begin{eqnarray}\label{eqn:rmfke:mmala}
\frac{\partial p}{\partial t} &=& \frac{1}{\sqrt{g}} \frac{\partial}{\partial x^i} \{ \frac{1}{2}\sqrt{g} g^{ij} \frac{\partial p}{\partial x^j} -  p \sqrt{g}[\frac{ g^{ij}}{p} \frac{\partial p}{\partial x^j} - \frac{1}{2}g^{jk} \Gamma^i_{jk} + \frac{1}{2} g^{jk} \gamma^i_{jk}] \}  \\
&=& 0 \nonumber
\end{eqnarray}
Thus, the MMALA method converges to the stationary distribution of the Fokker-Planck equation on Riemannian manifolds.\\
\section{Stochastic development}
 $H$ in equation (\ref{eqn:omega2})  is the horizontal vector field given by
\begin{equation} \label{horz-vector}
H_i = Q^j_iX_j - Q^j_i Q^l_m \gamma^k_{jl} X_{km} \quad i \in [1,d]
\end{equation}
where $X_i = \frac{\partial }{\partial x^i}$ and $X_{km} = \frac{\partial }{\partial Q^k_m}$ are the basis of the tangent space of the frame bundle $F(M)$. Therefore, each vector in equation (\ref{eqn:omega2}) of the form $Hv$ where $v$ lies in $\mathbb{R}^d$ may be written as 
\begin{equation}
H v = v^iQ^j_iX_j -  v^iQ^j_i Q^l_m \gamma^k_{jl} X_{km} 
\end{equation}
For convenience, let $\overset{.}{\bar{y}^{\bar{x}(t)}_{\alpha}}(0) = \eta$, then the integrand in equation (\ref{eqn:omega2}) may be simplified in terms of the curvature $(3,1)$-tensor $R^i_{jkl}$ as
\begin{align}
(\mathcal{K}_{\alpha})^c_d = {}&\frac{1}{2}R^c_{dab}  dx^a \wedge dx^b (H\eta, H\alpha) \\
={}& \frac{1}{2}R^c_{dab}  dx^a \wedge dx^b  (  \eta^iQ^j_iX_j -  \eta^iQ^j_i Q^l_m \gamma^k_{jl} X_{km} \; ,  \alpha^pQ^q_p X_q -  \alpha^pQ^q_p Q^s_t \gamma^k_{qs}(x) X_{rt}) \nonumber  \\
={}& \frac{1}{2}R^c_{dab} (  \eta^iQ^j_i\delta^a_j  \alpha^p Q^q_p \delta^b_q -    \eta^iQ^j_i\delta^b_j \alpha^p Q^q_p \delta^a_q ) \nonumber  \\
={}& \frac{1}{2}R^c_{dab} ( \eta^iQ^a_i \alpha^p Q^b_p - \eta^iQ^b_i \alpha^pQ^a_p) \nonumber  \\
={}& \frac{1}{2}R^c_{dab} ( \eta^iQ^a_i \alpha^p Q^b_p - \eta^iQ^b_i \alpha^pQ^a_p) \nonumber 
\end{align}

\vspace{10pt}

\textbf{\Large Supplementary material}\\
\section{Rayleigh distribution} \label{app:rayl}
%\textbf{Rayleigh distribution}

%\begin{equation}
%\text{Probability density function:} \;\;\; p(x;\sigma) = \frac{x}{\sigma^2}\exp(-\frac{x^2}{2\sigma^2})
%\end{equation}

Probability density function: $p(x;\sigma) = \frac{x}{\sigma^2}\exp(-\frac{x^2}{2\sigma^2})$\\
Mean: $\sigma \sqrt{\frac{\pi}{2}}$ \\
Variance: $\frac{(4-\pi)\sigma^2}{2}$ \\
Log-likelihood: $L=\log(p(x;\sigma)) = \log(x) - 2\log(\sigma) - \frac{x^2}{2\sigma^2}$\\
 Gradient of log-likelihood: $\frac{\partial L}{\partial \sigma}= -\frac{2}{\sigma} + \frac{x^2}{\sigma^3}$\\
Fisher-Information matrix:
\begin{eqnarray}
G &=& E(\frac{\partial L}{\partial \sigma} \frac{\partial L}{\partial \sigma}) \\
&=& E((-\frac{2}{\sigma} + \frac{x^2}{\sigma^3})^2) \nonumber \\
&=& E(\frac{4}{\sigma^2} + \frac{x^4}{\sigma^6} - \frac{4x^2}{\sigma^4}) \nonumber \\
\text{where:} \;\;\; E(x^2) &=& \int_0^\infty (x^2) (\frac{x}{\sigma^2})\exp(-\frac{x^2}{2 \sigma^2}) = 2\sigma^2 \\
E(x^4) &=& \int_0^\infty (x^4) (\frac{x}{\sigma^2})\exp(-\frac{x^2}{2 \sigma^2}) = 8 \sigma^4
\end{eqnarray}
Therefore, $G = E(\frac{4}{\sigma^2} + \frac{8\sigma^4}{\sigma^6} - \frac{4 \times 2\sigma^2}{\sigma^4}) = \frac{4}{\sigma^2}$\\
Derivative of Fisher-Information matrix:$\frac{\partial G}{\partial \sigma} = -\frac{8}{\sigma^3}$\\
Connection: $\gamma = G^{-1}\frac{\partial G}{\partial \sigma} = -\frac{\sigma^2}{4}\frac{8}{\sigma^3} = -\frac{2}{\sigma}
$
% Banana
\section{Banana-shaped distribution} \label{app:banana}
Probability density function:$p(z_1,z_2;B) \propto \exp(-\frac{z_1^2}{200} - \frac{1}{2}(z_2 + Bz_1^2 - 100B)^2)$\\
In the above equation $z_1$ and $z_2$ are distributed as $  \sim \mathcal{N}(0,\Sigma)$, where 
\begin{equation}
\Sigma = \begin{bmatrix} 100 & 0 \\ 0 & 1 \end{bmatrix} 
\end{equation}
%With the following transformation of variables, 
%\begin{eqnarray}
%x_1 &= & z_1 \label{x1x2} \\
%x_2 & = & - (z_2 + Bz_1^2 - 100B) \nonumber 
%\end{eqnarray}
%$x_1$ and $x_2$ are banana-distributed with twisting parameter B
%Inverting equation \ref{x1x2}, we have
%\begin{eqnarray}
%z_1 &=& x_1 \\
%z_2 & = & 100B - Bx_1^2 - x_2 \nonumber
%\end{eqnarray}
Log-likelihood: $ L=\log(p(z_1,z_2;B)) = -\frac{z_1^2}{200} - \frac{1}{2}(z_2 + Bz_1^2 - 100B)^2$\\
Gradient of log-likelihood:$\frac{\partial L}{\partial B}= - (z_2+ Bz_1^2 - 100B)(z_1^2 - 100)$
Fisher-Information matrix:
\begin{eqnarray}
G &=& E[\frac{\partial L}{\partial B} \frac{\partial L}{\partial B}] \label{Gbanana} \\
&=& E[(z_2+ Bz_1^2 - 100B)^2(z_1^2 - 100)^2] \nonumber \\
&=& E[(z_2^2 + B^2z_1^4 + 10000B^2 + 2Bz_1^2z_2 -  200 B z_2 - 200B^2 z_1^2)(z_1^4 + 10000 - 200z_1^2)] \nonumber \\
&=& E[(z_1^4z_2^2 + B^2z_1^8 + 10000B^2z_1^4 + 2Bz_1^6z_2 -  200 B z_2 z_1^4 - 200B^2 z_1^6) \nonumber \\
&& + 10000(z_2^2 + B^2z_1^4 + 10000B^2 + 2Bz_1^2z_2 -  200 B z_2 - 200B^2 z_1^2) \nonumber \\
&& -200(z_1^2z_2^2 + B^2z_1^6 + 10000B^2 z_1^2 + 2Bz_1^4z_2 -  200 B z_1^2z_2 - 200B^2 z_1^4) \nonumber \\
&=& E[(z_1^4z_2^2 + B^2z_1^8 + 10000B^2z_1^4   - 200B^2 z_1^6)  + 10000(z_2^2 + B^2z_1^4 + 10000B^2  - 200B^2 z_1^2) \nonumber \\
&& -200(z_1^2z_2^2 + B^2z_1^6 + 10000B^2 z_1^2  - 200B^2 z_1^4)] \nonumber \\
&=& 3\sigma_1^4 \sigma_2^2 + 7B^2\sigma_1^8 + 3 \times 10000 B^2 \sigma_1^4 - 200 \times 5 B^2 \sigma_1^6  + 10000(\sigma_2^2 + 3B^2 \sigma_1^4 + 10000 B^2 - 200 B^2 \sigma_1^2)  \nonumber \\
&& - 200 (\sigma_1^2 \sigma_2^2 + 5 B^2 \sigma_1^6 + 10000 B^2 \sigma_1^2 - 200 \times 3 B^2 \sigma_1^4) \nonumber \\
&=& 3 \times 10^4 + 7 \times 10^8 B^2 + 3 \times 10^8 B^2  - 10^9 B^2 + 10^4 + 3 \times 10^8 B^2 + 10^8 B^2 - 2 \times 10^8 B^2 \nonumber \\
&& -200(100 + 5 \times 10^6 B^2 +10^6 B^2 - 6 \times 10^6 B^2) \nonumber \\
&=& 3 \times 10^4 + 10^4 + 2 \times 10^8 B^2 - 2 \times 10^4 \nonumber \\
&=& 2 \times 10^4  + 2 \times 10^8 B^2 
\end{eqnarray}
$G$ for the product of likelihoods over N observations is $N\times G$
Therefore,
Derivative of Fisher-Information matrix: $N \times 4 \times 10^8 B$\\
Connection:
\begin{eqnarray}
\gamma &=& G^{-1} \frac{\partial G}{\partial B} \nonumber \\
&=& [N \times (2 \times 10^4  + 2 \times 10^8 B^2 )]^{-1} (N \times 4 \times 10^8 B) \nonumber \\
&=& \frac{4 \times 10^8 B}{(2 \times 10^4  + 2 \times 10^8 B^2 )}  \nonumber \\
&=& \frac{4 \times 10^4 B}{(2  + 2 \times 10^4 B^2 )}  \nonumber \\
\end{eqnarray}
% Weibull
\section{Weibull distribution} \label{app:wbl}
Probability density function:
\begin{eqnarray}
p(x;\lambda,k) &=& \frac{k}{\lambda} (\frac{x}{\lambda})^{k-1}\exp^{-(\frac{x}{\lambda})^k} ; x \geq 0 \nonumber \\
&=& 0 \;\;\;\;\;\;\;\;\;\;\;\;\ ;  x<0 \nonumber
\end{eqnarray}
Log- likelihood: 
\begin{eqnarray}
 L=\log(p(x;\lambda,k)) &=& \log(k) - \log(\lambda) + (k-1)\log(x) - (k-1)\log(\lambda) -(\frac{x}{\lambda})^k \nonumber \\
 &=& \log(k) -k \log(\lambda) +(k-1)\log(x) -(\frac{x}{\lambda})^k \nonumber
 \end{eqnarray}
Gradient of log-likelihood:
\begin{eqnarray}
\frac{\partial L}{\partial \lambda}&=& -\frac{k}{\lambda} - k (\frac{x}{\lambda})^{k-1}\frac{-x}{\lambda^2} \nonumber \\
&=& -\frac{k}{\lambda} + \frac{k}{\lambda} (\frac{x}{\lambda})^{k} \nonumber \\
&=& [ (\frac{x}{\lambda})^{k} - 1]\frac{k}{\lambda} \nonumber \\
\frac{\partial L}{\partial k} &=& \frac{1}{k} - \log(\lambda) + \log(x) - (\frac{x}{\lambda})^k\log(\frac{x}{\lambda})
\end{eqnarray}
Therefore
\begin{equation}
\nabla L =  \begin{bmatrix} 
[ (\frac{x}{\lambda})^{k} - 1]\frac{k}{\lambda} \\
\frac{1}{k} - \log(\lambda) + \log(x) - (\frac{x}{\lambda})^k\log(\frac{x}{\lambda})
\end{bmatrix} 
\end{equation}
Fisher-Information matrix:
\begin{eqnarray}
G &=& E\left(\begin{bmatrix}
\frac{\partial L}{\partial \lambda}\frac{\partial L}{\partial \lambda} & \frac{\partial L}{\partial \lambda}\frac{\partial L}{\partial k} \\[12pt]
\frac{\partial L}{\partial k}\frac{\partial L}{\partial \lambda} & \frac{\partial L}{\partial k}\frac{\partial L}{\partial k}
\end{bmatrix}\right)  \\
&=& \begin{bmatrix}
E(\frac{\partial L}{\partial \lambda}\frac{\partial L}{\partial \lambda}) & E(\frac{\partial L}{\partial \lambda}\frac{\partial L}{\partial k}) \\[12pt]
E(\frac{\partial L}{\partial k}\frac{\partial L}{\partial \lambda}) & E(\frac{\partial L}{\partial k}\frac{\partial L}{\partial k})
\end{bmatrix} \nonumber\\
&=&  \begin{bmatrix}
G_{11} & G_{12} \\ G_{21} & G_{22} 
\end{bmatrix} \nonumber
\end{eqnarray}
Where
\begin{eqnarray}
G_{11} &=& E( [ (\frac{x}{\lambda})^{k} - 1]^2(\frac{k}{\lambda})^2)  \\
&=& (\frac{k}{\lambda})^2 E( [ (\frac{x}{\lambda})^{k} - 1]^2) \nonumber \\
G_{12}=G_{21}&=&E( [ (\frac{x}{\lambda})^{k} - 1]\frac{k}{\lambda} [\frac{1}{k} - \log(\lambda) + \log(x) - (\frac{x}{\lambda})^k\log(\frac{x}{\lambda})])  \\
&=& E( [ (\frac{x}{\lambda})^{k} - 1]\frac{k}{\lambda} (\frac{1}{k} - \log(\lambda) )) + E( [ (\frac{x}{\lambda})^{k} - 1]\frac{k}{\lambda} ( \log(x) - (\frac{x}{\lambda})^k\log(\frac{x}{\lambda}))) \nonumber \\
&=& \frac{k}{\lambda} (\frac{1}{k} - \log(\lambda) )E( [ (\frac{x}{\lambda})^{k} - 1]) + \frac{k}{\lambda} E( [ (\frac{x}{\lambda})^{k} - 1][ \log(x) - (\frac{x}{\lambda})^k\log(\frac{x}{\lambda})]) \nonumber \\
G_{22}&=& E(\{\frac{1}{k} - \log(\lambda) + \log(x) - (\frac{x}{\lambda})^k\log(\frac{x}{\lambda})\}^2)
\end{eqnarray}
Derivative of the Fisher-Information matrix\\
Derivative w.r.t $\lambda$:
\begin{equation}
\frac{\partial G}{\partial \lambda} = \begin{bmatrix}
\frac{\partial G_{11}}{\partial \lambda} & \frac{\partial G_{12}}{\partial \lambda} \\[12pt]
\frac{\partial G_{21}}{\partial \lambda} & \frac{\partial G_{22}}{\partial \lambda}
\end{bmatrix}
\end{equation}
\begin{eqnarray}
\frac{\partial G_{11}}{\partial \lambda}&=& \frac{\partial \{ (\frac{k}{\lambda})^2 E( [ (\frac{x}{\lambda})^{k} - 1]^2)\}}{\partial \lambda}  \\
&=& \frac{\partial  (\frac{k}{\lambda})^2 }{\partial \lambda}E( [ (\frac{x}{\lambda})^{k} - 1]^2) +   (\frac{k}{\lambda})^2 \frac{\partial E( [ (\frac{x}{\lambda})^{k} - 1]^2)}{\partial \lambda} \nonumber \\
&=& \frac{\partial  (\frac{k}{\lambda})^2 }{\partial \lambda}E( [ (\frac{x}{\lambda})^{k} - 1]^2) +   (\frac{k}{\lambda})^2 E( \frac{\partial[ (\frac{x}{\lambda})^{k} - 1]^2)}{\partial \lambda} \nonumber \\
&=& -\frac{2k^2}{\lambda^3} E( [ (\frac{x}{\lambda})^{k} - 1]^2)  - (\frac{k}{\lambda})^2E(2k ((\frac{x}{\lambda})^k-1)\frac{x^k}{\lambda^{k+1}})  \nonumber \\
&=& -\frac{2k^2}{\lambda^3} E( [ (\frac{x}{\lambda})^{k} - 1]^2)  
- \frac{2k^3}{\lambda^{k+3}}E[(\frac{x}{\lambda})^k-1)] \nonumber \\
\frac{\partial G_{12}}{\partial \lambda}&=& \frac{\partial \{ \frac{k}{\lambda} (\frac{1}{k} - \log(\lambda) )E( [ (\frac{x}{\lambda})^{k} - 1]) + \frac{k}{\lambda} E( [ (\frac{x}{\lambda})^{k} - 1][ \log(x) - (\frac{x}{\lambda})^k\log(\frac{x}{\lambda})])\}}{\partial \lambda}  \\
&=& \frac{\partial \{ \frac{k}{\lambda} (\frac{1}{k} - \log(\lambda) )E( [ (\frac{x}{\lambda})^{k} - 1]\}}{\partial \lambda} + \frac{\partial \{\frac{k}{\lambda} E( [ (\frac{x}{\lambda})^{k} - 1][ \log(x) - (\frac{x}{\lambda})^k\log(\frac{x}{\lambda})])\}}{\partial \lambda} \nonumber \\
&=& -\frac{k}{\lambda^2}(\frac{1}{k} - \log(\lambda) )E( [ (\frac{x}{\lambda})^{k} - 1] - \frac{k}{\lambda}\frac{1}{\lambda}E( [ (\frac{x}{\lambda})^{k} - 1]  + \frac{k}{\lambda}(\frac{1}{k}-\log(\lambda)) E(\frac{\partial [(\frac{x}{\lambda})^k -1]}{\partial \lambda}) \nonumber \\
&& - \frac{k}{\lambda^2} E( [ (\frac{x}{\lambda})^{k} - 1][ \log(x) - (\frac{x}{\lambda})^k\log(\frac{x}{\lambda})])  + \frac{k}{\lambda} E( \frac {\partial [ (\frac{x}{\lambda})^{k} - 1]}{\partial \lambda}[ \log(x) - (\frac{x}{\lambda})^k\log(\frac{x}{\lambda})]) \nonumber \\
&& + \frac{k}{\lambda}  E( [ (\frac{x}{\lambda})^{k} - 1] \frac{ \partial [ \log(x) - (\frac{x}{\lambda})^k\log(\frac{x}{\lambda})]}{\partial \lambda}) \nonumber \\
&=& -\frac{k}{\lambda^2}(\frac{1}{k} - \log(\lambda) )E( [ (\frac{x}{\lambda})^{k} - 1] - \frac{k}{\lambda}\frac{1}{\lambda}E( [ (\frac{x}{\lambda})^{k} - 1]  + \frac{k}{\lambda}(\frac{1}{k}-\log(\lambda)) E(\frac{\partial [(\frac{x}{\lambda})^k]}{\partial \lambda}) \nonumber \\
&& - \frac{k}{\lambda^2} E( [ (\frac{x}{\lambda})^{k} - 1][ \log(x) - (\frac{x}{\lambda})^k\log(\frac{x}{\lambda})])  + \frac{k}{\lambda} E( \frac {\partial [ (\frac{x}{\lambda})^{k} ]}{\partial \lambda}[ \log(x) - (\frac{x}{\lambda})^k\log(\frac{x}{\lambda})]) \nonumber \\
&& - \frac{k}{\lambda}  E( [ (\frac{x}{\lambda})^{k} - 1] \frac{ \partial [ (\frac{x}{\lambda})^k\log(\frac{x}{\lambda})]}{\partial \lambda}) \nonumber \\
&=& -\frac{k}{\lambda^2}(\frac{1}{k} - \log(\lambda) )E( [ (\frac{x}{\lambda})^{k} - 1] - \frac{k}{\lambda}\frac{1}{\lambda}E( [ (\frac{x}{\lambda})^{k} - 1]  - \frac{k}{\lambda}(\frac{1}{k}-\log(\lambda)) E(\frac{kx^k}{\lambda^{k+1}}) \nonumber \\
&& - \frac{k}{\lambda^2} E( [ (\frac{x}{\lambda})^{k} - 1][ \log(x) - (\frac{x}{\lambda})^k\log(\frac{x}{\lambda})])  - \frac{k}{\lambda} E( \frac{kx^k}{\lambda^{k+1}}[ \log(x) - (\frac{x}{\lambda})^k\log(\frac{x}{\lambda})]) \nonumber \\
&& - \frac{k}{\lambda}  E( [ (\frac{x}{\lambda})^{k} - 1] [-\frac{k}{\lambda}(\frac{x}{\lambda})^k \log(\frac{x}{\lambda}) - (\frac{x}{\lambda})^k \frac{1}{\lambda}]) \nonumber \\
\frac{\partial G_{22}}{\partial \lambda}&=& \frac{\partial [E(\{\frac{1}{k} - \log(\lambda) + \log(x) - (\frac{x}{\lambda})^k\log(\frac{x}{\lambda})\} ^2)] }{\partial \lambda} \\
&=& E[\frac{\partial (\{\frac{1}{k} - \log(\lambda) + \log(x) - (\frac{x}{\lambda})^k\log(\frac{x}{\lambda})\} ^2)}{\partial \lambda}] \nonumber \\
&=& 2E[\{\frac{1}{k} - \log(\lambda) + \log(x) - (\frac{x}{\lambda})^k \log(\frac{x}{\lambda})\} \{ \frac{-1}{\lambda} + [\frac{k}{\lambda}(\frac{x}{\lambda})^k \log(\frac{x}{\lambda}) + \frac{x^k}{\lambda^{k+1}}] \}] \nonumber 
\end{eqnarray}
Derivative w.r.t $k$:
\begin{equation}
\frac{\partial G}{\partial k} = \begin{bmatrix}
\frac{\partial G_{11}}{\partial k} & \frac{\partial G_{12}}{\partial k} \\[12pt]
\frac{\partial G_{21}}{\partial k} & \frac{\partial G_{22}}{\partial k}
\end{bmatrix}
\end{equation}
\begin{eqnarray}
\frac{\partial G_{11}}{\partial k} &=& \frac{\partial \{ (\frac{k}{\lambda})^2 E( [ (\frac{x}{\lambda})^{k} - 1]^2)\} }{\partial k} \\
&=&  \frac{\partial (\frac{k}{\lambda})^2 }{\partial k} E( [ (\frac{x}{\lambda})^{k} - 1]^2)  + (\frac{k}{\lambda})^2 \frac{\partial \{E( [ (\frac{x}{\lambda})^{k} - 1]^2) \}}{\partial k}  \nonumber \\
&=& \frac{2k}{\lambda^2}E( [ (\frac{x}{\lambda})^{k} - 1]^2)  + (\frac{k}{\lambda})^2 E \frac{  \partial( [ (\frac{x}{\lambda})^{k} - 1]^2)}{\partial k}  \nonumber \\
&=&\frac{2k}{\lambda^2}E( [ (\frac{x}{\lambda})^{k} - 1]^2)  + (\frac{k}{\lambda})^2 E(2[(\frac{x}{\lambda})^k-1]\log(\frac{x}{\lambda})(\frac{x}{\lambda})^k) \nonumber \\
% G12
\frac{\partial G_{12}}{\partial k}  &=&  \frac{\partial \{ \frac{k}{\lambda} (\frac{1}{k} - \log(\lambda) )E( [ (\frac{x}{\lambda})^{k} - 1])\} }{\partial k}  \\
&& + \frac{\partial \{ \frac{k}{\lambda} E( [ (\frac{x}{\lambda})^{k} - 1][ \log(x) - (\frac{x}{\lambda})^k\log(\frac{x}{\lambda})]) \} }{\partial k} \nonumber \\
&=&\frac{\partial   (\frac{1}{\lambda} - \frac{k \log(\lambda)}{\lambda} )}{\partial k}E( [ (\frac{x}{\lambda})^{k} - 1]) + (\frac{1}{\lambda} - \frac{k \log(\lambda)}{\lambda} )\frac{\partial \{ E( [ (\frac{x}{\lambda})^{k} - 1]) \}}{\partial k}   \nonumber \\
&&+ \frac{1}{\lambda}E( [ (\frac{x}{\lambda})^{k} - 1][ \log(x) - (\frac{x}{\lambda})^k\log(\frac{x}{\lambda})]) + \frac{k}{\lambda} E(\frac{\partial ( [ (\frac{x}{\lambda})^{k} - 1][ \log(x) - (\frac{x}{\lambda})^k\log(\frac{x}{\lambda})])}{\partial k}) \nonumber \\
&=& -\frac{\log(\lambda)}{\lambda} E( [ (\frac{x}{\lambda})^{k} - 1]) + (\frac{1}{\lambda} - \frac{k \log(\lambda)}{\lambda} ) E[(\frac{x}{\lambda})^k \log(\frac{x}{\lambda})]  + \frac{1}{\lambda}E( [ (\frac{x}{\lambda})^{k} - 1][ \log(x) - (\frac{x}{\lambda})^k\log(\frac{x}{\lambda})])  \nonumber \\
&&+ \frac{k}{\lambda}E((\frac{x}{\lambda})^k \log(\frac{x}{\lambda})[\log(x) -  (\frac{x}{\lambda})^k\log(\frac{x}{\lambda}) ] - [(\frac{x}{\lambda})^k-1][(\frac{x}{\lambda})^k \log(\frac{x}{\lambda}) \log(\frac{x}{\lambda})]) \nonumber \\
&=& -\frac{\log(\lambda)}{\lambda} E( [ (\frac{x}{\lambda})^{k} - 1]) + (\frac{1}{\lambda} - \frac{k \log(\lambda)}{\lambda} ) E[(\frac{x}{\lambda})^k \log(\frac{x}{\lambda})]  + \frac{1}{\lambda}E( [ (\frac{x}{\lambda})^{k} - 1][ \log(x) - (\frac{x}{\lambda})^k\log(\frac{x}{\lambda})])  \nonumber \\
&& + \frac{k}{\lambda} E((\frac{x}{\lambda})^k \log(\frac{x}{\lambda})[\log(x) - 2(\frac{x}{\lambda})^k \log(\frac{x}{\lambda}) + \log(\frac{x}{\lambda}) ]) \nonumber \\
\frac{\partial G_{22}}{\partial k} &=& \frac{\partial [E(\{\frac{1}{k} - \log(\lambda) + \log(x) - (\frac{x}{\lambda})^k\log(k)\}^2)]}{\partial k}  \\
&=& E\frac{ \partial(\{\frac{1}{k} - \log(\lambda) + \log(x) - (\frac{x}{\lambda})^k  \log(\frac{x}{\lambda})  \}^2)}{\partial k} \nonumber \\
&=& E[2 \{\frac{1}{k} - \log(\lambda) + \log(x) - (\frac{x}{\lambda})^k  \log(\frac{x}{\lambda})  \} \{ -\frac{1}{k^2} - (\frac{x}{\lambda})^k \log(\frac{x}{\lambda})  \log(\frac{x}{\lambda}) \}]  \nonumber 
\end{eqnarray}
The expectations for this distribution appearing in the expressions for $G$ and derivatives of $G$ are evaluated numerically at every step of the Markov chain (for GALA and MMALA) by sampling from the Weibull distribution with parameters equal to $k$ and $\lambda$ at each step. 
Connection: $ \gamma^k_{ij} = G^{kl} (\partial_i g_{jl}+ \partial_j g_{il} - \partial_l g_{ij}) $
% %%%%%%%%%%%%%%MVN%%%%%%%%%%%%%%%%%%

\section{Multivariate Gaussian distribution} \label{app:mvn}

Probability density function: $p(y;\mu,\Sigma) = \frac{1}{2 \pi}  |\Sigma|^{\frac{-1}{2}} \exp(-\frac{1}{2}(y-\mu)^T \Sigma^{-1} (y-\mu)) $\\
Two major derivations are required. Namely, the derivation of \textbf{gradient} and the \textbf{Hessian} of log-likelihood with respect to the \textbf{square root} of covariance matrix.\\
Log - Likelihood: $\log(p(y;\mu,\Sigma)) = L= -\log(2\pi) - \frac{1}{2}\log(|\Sigma|) - \frac{1}{2}(y-\mu)^T \Sigma^{-1} (y-\mu)  $\\
%As shown in the equation \ref{log-likelihood} , the second term is independent of $\mu$
For a d-dimensional distribution, let 
\begin{eqnarray}
\theta & = & (\theta_1,\theta_2,...,\theta_{D}) \label{theta}  \;,\; \text{where} \;\;D= \frac{d^2+3d}{2} \\
 & = & (\mu_1,\mu_2,... \mu_d, \Sigma_{11},\Sigma_{21},\Sigma_{22},\Sigma_{31},\Sigma_{32},...,\Sigma_{d(d-1)},\Sigma_{dd}) \nonumber
\end{eqnarray}
Gradient of Log-likelihood:\\
\textbf{Gradient of Log-likelihood with respect to $\mu$}
\begin{eqnarray}
\frac{\partial L}{\partial \mu}= \frac{1}{2}[\Sigma^{-1} + (\Sigma^{-1})^T ](y-\mu)  \label{grad:mu} \\
\text{i.e.} \;\;\frac{\partial L}{\partial \theta_i}= \frac{1}{2}\{[\Sigma^{-1} + (\Sigma^{-1})^T ] (y-\mu)\}_i  \;,\; i \in [1,d]\nonumber 
\end{eqnarray}\\
\textbf{Gradient of log-likelihood with respect to $\Sigma$}
\begin{equation}
\frac{\partial L}{\partial \Sigma} = - \frac{1}{2}\frac{\partial (\log(|\Sigma|))}{\partial \Sigma} - \frac{1}{2}\frac{\partial[(y-\mu)^T \Sigma^{-1} (y-\mu)]}{\partial \Sigma}  \end{equation}
\begin{equation}  \text{where} \frac{\partial (\log(|\Sigma|))}{\partial \Sigma} = (\Sigma^{-1})^T \label{del log det} \;\; \text{ \cite{petersen2008matrix} }\end{equation}   %from eqn (57), matrixcookbook.
\begin{equation} \label{P}
\text{Let}\;\;P( \Sigma)=[(y-\mu)^T  \Sigma^{-1} (y-\mu)]
\end{equation}
Re-writing in indical notation as follows
\begin{eqnarray}
P(\Sigma)&=&(y-\mu)_p \{ \Sigma^{-1} (y-\mu)\}_p \nonumber \\
&=&(y-\mu)_p \Sigma^{-1} _{pq}(y-\mu)_q \nonumber \\
\text{Therefore}\;\; \frac{\partial L}{\partial \Sigma}  =&=&-\frac{1}{2}(\Sigma^{-1})^T - \frac{1}{2}\frac{\partial P(\Sigma)}{\partial \Sigma}  \label{LP}
\end{eqnarray}
%Evaluating $\frac{\partial P(\Sigma)}{\partial \Sigma} $ in indical notations as follows:
\begin{eqnarray}
\text{Now,}\;\; \frac{\partial P(\Sigma)}{\partial \Sigma_{ij}}  &=& \frac{\partial [(y-\mu)_p \Sigma^{-1} _{pq}(y-\mu)_q] }{\partial \Sigma_{ij}} \\
&=& (y-\mu)_p\frac{\partial \Sigma^{-1} _{pq} }{\partial \Sigma_{ij}}(y-\mu)_q \nonumber \\
&=& - (y-\mu)_p\Sigma^{-1} _{pi}\Sigma^{-1} _{jq}(y-\mu)_q \nonumber \\
&=& - (\Sigma^{-1})^T_{ip}(y-\mu)_p(y-\mu)_q(\Sigma^{-1})^T _{qj} \nonumber \\
&=& - (\Sigma^{-1})^T_{ip}[(y-\mu)(y-\mu)^T]_{pq}(\Sigma^{-1})^T _{qj} \nonumber \\
&=& - [(\Sigma^{-1})^T[(y-\mu)(y-\mu)^T](\Sigma^{-1})^T]_{ij} \nonumber 
\end{eqnarray}
Thus \ref{LP} can be written in matrix form again as follows
\begin{equation} \label{grad:sig}
\frac{\partial L}{\partial \Sigma}  = -\frac{1}{2}(\Sigma^{-1})^T + \frac{1}{2}[(\Sigma^{-1})^T(y-\mu)(y-\mu)^T(\Sigma^{-1})^T] 
\end{equation}
Thus, we can evaluate the partial derivatives with respect to $\theta_{d+1}$ through $\theta_D$ using \ref{grad:sig} together with \ref{theta}.\\
%\begin{eqnarray}
%\frac{\partial L}{\partial \theta_3}=  [\frac{\partial L}{\partial \Sigma}]_{11} \\
%\frac{\partial L}{\partial \theta_4}=  [\frac{\partial L}{\partial \Sigma}]_{21} \nonumber \\
%\frac{\partial L}{\partial \theta_5}=  [\frac{\partial L}{\partial \Sigma}]_{22} \nonumber 
%\end{eqnarray}
\textbf{Fisher Information Matrix:} $G = E[\: (\partial_i \log\,p(x;\theta))\:(\partial_j \log\,p(x;\theta))^T\: ]$\\
%Now, as per subsection 3.5 of Malagò \& Giovanni (2015), we have
%\begin{equation}\label{G}
%I_{\xi}(\xi) = E_{\xi}[\: (\partial_i \log\,p(x;\xi))\:(\partial_j \log\,p(x;\xi))^T\: ]
%\end{equation}
%where $\theta$ is the parameter vector by which the statistical model is parameterized. 
From equations \ref{grad:mu} and\ref{grad:sig}, we have 
\begin{equation}
\frac{\partial \log\,p(y;\theta)}{\partial \theta} = [\frac{\partial L}{\partial \mu_1},\frac{\partial L}{\partial \mu_2},...\frac{\partial L}{\partial \mu_d},\frac{\partial L}{\partial \Sigma_{p_1q_1}},\frac{\partial L}{\partial \Sigma_{p_2q_2}}.... \frac{\partial L}{\partial \Sigma_{p_{D-d}q_{D-d}}}]^T
\end{equation}
\vspace{15pt}
where $ (p_1q_1), \: (p_2q_2), \: ..(p_{D-d}q_{D-d})$ correspond to the pairs of indices of the lower triangular part of the covariance matrix (including diagonals).
Thus $\frac{\partial \log\,p(y;\theta)}{\partial \theta} $ is of the form $[\vec{a_1}\;\vec{a_2}]$ where $\vec{a_1} = \{\frac{\partial L}{\partial \mu}\}_{d \times 1}$ and $ \vec{a_2} = vec\{\frac{\partial L}{\partial \Sigma}\}_{(D-d) \times 1}$
Hence, 
\begin{equation}
G_{ij} = E[\: (\partial_i \log\,p(x;\theta))\:(\partial_j \log\,p(x;\theta)) \: ]
\end{equation}
Consider the following cases \\
1. $i,j \;<\; d$ \\
2. $i>d ,\; j<d$ \\
3. $i<d, \; j>d$ \\
4. $ i,j \;> d $ \\
$G_{ij}$ has to be evaluated separately for each of the above cases.\\
Case  1. $i,j \;<\; d$ 
\begin{eqnarray}
G_{ij} &=& E[\: \frac{\partial L}{\partial \mu_i} \times \frac{\partial L}{\partial \mu_j} \: ] \label{G:upper} \\
&=& \frac{1}{4}E[\{[\Sigma^{-1} + (\Sigma^{-1})^T ](y-\mu)\}_i \times \{[\Sigma^{-1} + (\Sigma^{-1})^T ](y-\mu)\}_j] \nonumber \\
&=& E[\{\Sigma^{-1}(y-\mu)\}_i \times \{ \Sigma^{-1}(y-\mu)\}_j] \nonumber \\
&=& E[[\{\Sigma^{-1}(y-\mu)\}\{ \Sigma^{-1}(y-\mu)\}^T ]_{ij}] \nonumber \\
&=& E[\Sigma^{-1}(y-\mu)\}(y-\mu)^T\Sigma^{-1}]_{ij} \nonumber \\
&=& [\Sigma^{-1}E[(y-\mu)\}(y-\mu)^T]\Sigma^{-1}]_{ij} \nonumber \\
&=& [\Sigma^{-1}\Sigma \Sigma^{-1}]_{ij} \nonumber \\
&=& \Sigma^{-1}_{ij} \nonumber
\end{eqnarray}
Case 2. $i>d ,\; j<d$ 
%\begin{eqnarray}
%G_{ij} &=& E[\: \frac{\partial L}{\partial \Sigma_{p_{(i-d)}q_{(i-d)}}} \times \frac{\partial L}{\partial \mu_j} \: ] \\
%&=&\frac{1}{2}E[ \{ -\Sigma^{-1} + [\Sigma^{-1}(y-\mu)(y-\mu)^T\Sigma^{-1}]\}_{p_{(i-d)}q_{(i-d)}} \{\Sigma^{-1}(y-\mu)\}_j] \nonumber \\
%&=& \frac{1}{2}E[ \{ -\Sigma^{-1} ]\}_{p_{(i-d)}q_{(i-d)}} \{\Sigma^{-1}(y-\mu)\}_j]  + \frac{1}{2}E[ \{ [\Sigma^{-1}(y-\mu)(y-\mu)^T\Sigma^{-1}]\}_{p_{(i-d)}q_{(i-d)}} \{\Sigma^{-1}(y-\mu)\}_j] \nonumber
%\end{eqnarray}
For brevity of notation, let $p_{(i-d)}=\tilde{p}_i, \;q_{(i-d)} = \tilde{q}_i $. Thus
\begin{eqnarray}
G_{ij} &=& E[\: \frac{\partial L}{\partial \Sigma_{\tilde{p}_i, \tilde{q}_i}   } \times \frac{\partial L}{\partial \mu_j} \: ] \\
&=&\frac{1}{2}E[ \{ -\Sigma^{-1} + [\Sigma^{-1}(y-\mu)(y-\mu)^T\Sigma^{-1}]\}_{\tilde{p}_i, \tilde{q}_i}  \{\Sigma^{-1}(y-\mu)\}_j] \nonumber \\
%&=& \frac{1}{2}E[ \{ -\Sigma^{-1} ]\}_{p_{(i-d)}q_{(i-d)}} \{\Sigma^{-1}(y-\mu)\}_j]  + \frac{1}{2}E[ \{ [\Sigma^{-1}(y-\mu)(y-\mu)^T\Sigma^{-1}]\}_{p_{(i-d)}q_{(i-d)}} \{\Sigma^{-1}(y-\mu)\}_j] \nonumber \\
&=& \frac{1}{2}E[ \{ -\Sigma^{-1} ]\}_{\tilde{p}_i,\tilde{q}_i} \{\Sigma^{-1}(y-\mu)\}_j]  + \frac{1}{2}E[ \{ [\Sigma^{-1}(y-\mu)(y-\mu)^T\Sigma^{-1}]\}_{\tilde{p}_i\tilde{q}_i} \{\Sigma^{-1}(y-\mu)\}_j] \\
&=& 0 + \frac{1}{2}E[ \{ [\Sigma^{-1}(y-\mu)(y-\mu)^T\Sigma^{-1}]\}_{\tilde{p}_i\tilde{q}_i} \{\Sigma^{-1}(y-\mu)\}_j] \nonumber\\
&=& \frac{1}{2}E[  \Sigma^{-1}_{\tilde{p}_ir}(y-\mu)_r(y-\mu)_s\Sigma^{-1}_{s\tilde{q}_i} \Sigma^{-1}_{jk}(y-\mu)_k] \nonumber\\
&=& \frac{1}{2}\Sigma^{-1}_{\tilde{p}_ir}\Sigma^{-1}_{s\tilde{q}_i} \Sigma^{-1}_{jk}E[(y-\mu)_r(y-\mu)_s(y-\mu)_k] \nonumber 
\end{eqnarray}
Case 3. $i<d, \; j>d$ \\
Again, let $p_{j-d} = \tilde{p}_j \; , \; q_{j-d} = \tilde{q}_j $
\begin{eqnarray}
G_{ij} &=& E[\: \frac{\partial L}{\partial \mu_i} \times  \frac{\partial L}{\partial \Sigma_{\tilde{p}_j \tilde{q}_j }} \: ] \\
&=&\frac{1}{2}E[  \{\Sigma^{-1}(y-\mu)\}_i \{ -\Sigma^{-1} + [\Sigma^{-1}(y-\mu)(y-\mu)^T\Sigma^{-1}]\}_{\tilde{p}_j \tilde{q}_j }] \nonumber \\
&=& \frac{1}{2}E[ \{ -\Sigma^{-1} ]\}_{\tilde{p}_j \tilde{q}_j } \{\Sigma^{-1}(y-\mu)\}_i]  + \frac{1}{2}E[ \{ [\Sigma^{-1}(y-\mu)(y-\mu)^T\Sigma^{-1}]\}_{\tilde{p}_j \tilde{q}_j } \{\Sigma^{-1}(y-\mu)\}_i] \nonumber\\
&=& 0 + \frac{1}{2}E[ \{ [\Sigma^{-1}(y-\mu)(y-\mu)^T\Sigma^{-1}]\}_{ \tilde{p}_j \tilde{q}_j } \{\Sigma^{-1}(y-\mu)\}_i] \nonumber\\
&=& 0 \nonumber
\end{eqnarray}
Case 4. $ i,j \;> d $ \\
\begin{eqnarray}
G_{ij} &=& E[\: \frac{\partial L}{\partial \Sigma_{ \tilde{p}_i \tilde{q}_i }} \times  \frac{\partial L}{\partial \Sigma_{ \tilde{p}_j \tilde{q}_j }}\: ] \label{I:lower} \\
&=& \frac{1}{4} E[\:\{ -\Sigma^{-1} + [\Sigma^{-1}(y-\mu)(y-\mu)^T\Sigma^{-1}]\}_{ \tilde{p}_i \tilde{q}_i } \{ -\Sigma^{-1} + [\Sigma^{-1}(y-\mu)(y-\mu)^T\Sigma^{-1}]\}_{ \tilde{p}_j \tilde{q}_j }\: ]  \nonumber \\
&=&\frac{1}{4} E[\Sigma^{-1}_{ \tilde{p}_i \tilde{q}_i } \Sigma^{-1}_{ \tilde{p}_j \tilde{q}_j } ] -\frac{1}{4} E[\Sigma^{-1}_{ \tilde{p}_i ,\tilde{q}_i }   [\Sigma^{-1}(y-\mu)(y-\mu)^T\Sigma^{-1}]_{ \tilde{p}_j ,\tilde{q}_j }   ]  -\frac{1}{4} E[\Sigma^{-1}(y-\mu)(y-\mu)^T\Sigma^{-1}]_{ \tilde{p}_i \tilde{q}_i }   \Sigma^{-1}_{ \tilde{p}_j \tilde{q}_j }  ] \nonumber \\
& & +\frac{1}{4} E\{[\Sigma^{-1}(y-\mu)(y-\mu)^T\Sigma^{-1}]_{ \tilde{p}_i \tilde{q}_i } [\Sigma^{-1}(y-\mu)(y-\mu)^T\Sigma^{-1}]_{ \tilde{p}_j \tilde{q}_j }\}  \nonumber \\
&=&  -\frac{1}{4}\Sigma^{-1}_{ \tilde{p}_i \tilde{q}_i  } \Sigma^{-1}_{ \tilde{p}_j \tilde{q}_j  }   + \frac{1}{4}\Sigma^{-1}_{\tilde{p}_i m} \Sigma^{-1}_{n \tilde{q}_i } \Sigma^{-1}_{\tilde{p}_j  r} \Sigma^{-1}_{s \tilde{q}_j } E \{(y-\mu)_m(y-\mu)_n (y-\mu)_r(y-\mu)_s \}  \nonumber 
\end{eqnarray}
The central k-order moments of the variable X are gives as follows; see \cite{triantafyllopoulos2002moments}: \\
(a) If k is odd, $\mu_{1,...,k}(X-\xi) = 0$ \\
(b) If k is even with $k=2\lambda$, then it is $\mu_{1,2,...,2\lambda}(X-\xi) = \sum c_{ij}c_{kl}...c_{xz}$, where the sum is taken over all permutations of $\{1,2,...2\lambda \}$ giving $(2\lambda-1)!/2^(\lambda-1)(\lambda-1)!$ terms in the sum, each being the product of $\lambda$ covariance.
Therefore, the 4-order moments are given by \\
\begin{eqnarray}
E(X_i^4)=3c_{ii} \label{4:moment} \\
E(X_i^3X_j)=3c_{ii}c_{jj} \nonumber \\
E(X_i^2X_j^2)=c_{ii} c_{jj} + 2c_{ij}^2 \nonumber \\
E(X_i^2X_jX_p)=c_{ii}c_{jp} + 2c_{ij}c_{ip} \nonumber \\
E(X_iX_jX_pX_q)=c_{ij}c_{pq} + c_{ip}c_{jq} + c_{iq}c_{jp} \nonumber
\end{eqnarray}
Now, consider the last term on the RHS of equation \ref{I:lower}, 
\begin{equation}
\phi(\tilde{p}_i,\tilde{p}_j,\tilde{q}_i,\tilde{q}_j)=\Sigma^{-1}_{\tilde{p}_i m} \Sigma^{-1}_{n \tilde{p}_j} \Sigma^{-1}_{ \tilde{q}_i r} \Sigma^{-1}_{s \tilde{q}_j} E \{(y-\mu)_m(y-\mu)_n (y-\mu)_r(y-\mu)_s \}
\end{equation}
From the last equation in \ref{4:moment}, we have
\begin{eqnarray}
\phi(\tilde{p}_i,\tilde{p}_j,\tilde{q}_i,\tilde{q}_j) &=&\Sigma^{-1}_{\tilde{p}_i m} \Sigma^{-1}_{n \tilde{q}_i} \Sigma^{-1}_{\tilde{p}_j r} \Sigma^{-1}_{s \tilde{q}_j} (\Sigma_{mn}\Sigma_{rs} + \Sigma_{mr}\Sigma_{ns} + \Sigma_{ms}\Sigma_{nr}) \\
&=& \Sigma^{-1}_{\tilde{p}_i m} \Sigma^{-1}_{n \tilde{q}_i} \Sigma^{-1}_{\tilde{p}_j r} \Sigma^{-1}_{s \tilde{q}_j} \Sigma_{mn}\Sigma_{rs}  + \Sigma^{-1}_{\tilde{p}_i m} \Sigma^{-1}_{n \tilde{q}_i} \Sigma^{-1}_{\tilde{p}_j r} \Sigma^{-1}_{s \tilde{q}_j}\Sigma_{mr}\Sigma_{ns}  + \Sigma^{-1}_{\tilde{p}_i m} \Sigma^{-1}_{n \tilde{q}_i} \Sigma^{-1}_{\tilde{p}_j r} \Sigma^{-1}_{s \tilde{q}_j} \Sigma_{ms}\Sigma_{nr} \nonumber \\
&=& \Sigma^{-1}_{\tilde{p}_i m} \Sigma_{mn} \Sigma^{-1}_{n \tilde{q}_i} \Sigma^{-1}_{\tilde{p}_j r}\Sigma_{rs}\Sigma^{-1}_{s \tilde{q}_j}  + \Sigma^{-1}_{\tilde{p}_i m}\Sigma_{mr} \Sigma^{-1}_{n \tilde{q}_i} \Sigma^{-1}_{ \tilde{p}_j r} \Sigma_{ns} \Sigma^{-1}_{s \tilde{q}_j}  + \Sigma^{-1}_{\tilde{p}_i m}\Sigma_{ms} \Sigma^{-1}_{n \tilde{q}_i} \Sigma^{-1}_{ \tilde{p}_j r} \Sigma^{-1}_{s \tilde{q}_j} \Sigma_{nr} \nonumber \\
&=& \delta_{\tilde{p}_i n} \Sigma^{-1}_{n \tilde{q}_i} \delta_{ \tilde{p}_j s}\Sigma^{-1}_{s \tilde{q}_j}   +\delta_{ \tilde{p}_i r}\Sigma^{-1}_{\tilde{p}_j r} \Sigma^{-1}_{n \tilde{q}_i}  \delta_{n \tilde{q}_j} +\delta_{\tilde{p}_i s}\Sigma^{-1}_{s \tilde{q}_j} \Sigma^{-1}_{n \tilde{q}_i} \delta_{\tilde{p}_j n} \nonumber \\
&=&  \Sigma^{-1}_{\tilde{p}_i \tilde{q}_i} \Sigma^{-1}_{\tilde{p}_j \tilde{q}_j}   +\Sigma^{-1}_{\tilde{p}_j \tilde{p}_i} \Sigma^{-1}_{\tilde{q}_j\tilde{q}_i}   +\Sigma^{-1}_{\tilde{p}_i \tilde{q}_j } \Sigma^{-1}_{\tilde{p}_j\tilde{q}_i} \nonumber 
\end{eqnarray}
\textbf{Note, the assumption that sigma is symmetric has been used in the above equation.}
Substituting above result in equation \ref{I:lower}, we get
\begin{eqnarray}
G_{ij} % &=&  -\frac{1}{4}\Sigma^{-1}_{ab} \Sigma^{-1}_{cd}  +  \frac{1}{4}[\Sigma^{-1}_{ab} \Sigma^{-1}_{cd}  + \Sigma^{-1}_{ca} \Sigma^{-1}_{db}  +\Sigma^{-1}_{ad} \Sigma^{-1}_{cb}] \nonumber \\
& =&\frac{1}{4} [\Sigma^{-1}_{\tilde{p}_j\tilde{p}_i} \Sigma^{-1}_{\tilde{q}_j \tilde{q}_i}  +\Sigma^{-1}_{\tilde{p}_i \tilde{q}_j} \Sigma^{-1}_{\tilde{p}_j \tilde{q}_i}] \label{G:lower}
\end{eqnarray}
%Substituting back the values of $a,b,c,d$ from \ref{abcd}, we have
%\begin{equation}\label{G:lower}
%[I_{\xi}(\xi)]_{ij} = \frac{1}{4}[\Sigma^{-1}_{p(j-d)p(i-d)} \Sigma^{-1}_{q(j-d)q(i-d)}  +\Sigma^{-1}_{p(i-d)q(j-d)} \Sigma^{-1}_{p(j-d)q(i-d)}] \\
%\end{equation}
\textbf{Derivative of G:}\\
From equation \ref{G:upper} we have 
\begin{eqnarray}
G_{ij} &=& E[\: \frac{\partial L}{\partial \mu_i} \times \frac{\partial L}{\partial \mu_j} \: ] \\
&=& \frac{1}{4}E[\{[\Sigma^{-1} + (\Sigma^{-1})^T ](y-\mu)\}_i \times \{[\Sigma^{-1} + (\Sigma^{-1})^T ](y-\mu)\}_j] \nonumber \\
&=& \frac{1}{4}E[\{[\Sigma^{-1} + (\Sigma^{-1})^T ]_{im}(y-\mu)_m\} \times \{[\Sigma^{-1} + (\Sigma^{-1})^T ]_{jn}(y-\mu)_n\}] \nonumber \\
&=& \frac{1}{4}E[(y-\mu)_m(y-\mu)_n] [\Sigma^{-1} + (\Sigma^{-1})^T ]_{im} [\Sigma^{-1} + (\Sigma^{-1})^T ]_{jn} \nonumber \\
&=& \frac{1}{4}\Sigma_{mn}[\Sigma^{-1} + (\Sigma^{-1})^T ]_{im} [\Sigma^{-1} + (\Sigma^{-1})^T ]_{jn} \nonumber \\
&=& \frac{1}{4}[\Sigma^{-1} + (\Sigma^{-1})^T ]_{im} \Sigma_{mn} [\Sigma^{-1} + (\Sigma^{-1})^T ]^T_{nj} \nonumber \\
&=& \frac{1}{4}[\Sigma^{-1} + (\Sigma^{-1})^T ]_{im} \Sigma_{mn} [\Sigma^{-1} + (\Sigma^{-1})^T ]_{nj}  \nonumber \\
&=& \frac{1}{4}\{[\Sigma^{-1} + (\Sigma^{-1})^T ]\Sigma [\Sigma^{-1} + (\Sigma^{-1})^T ]\}_{ij}  \nonumber \\
&=& \frac{1}{4}[ \Sigma^{-1} \Sigma  \Sigma^{-1}   +  \Sigma^{-1} \Sigma  (\Sigma^{-1})^T + (\Sigma^{-1})^T\Sigma   \Sigma^{-1} +  (\Sigma^{-1})^T\Sigma   (\Sigma^{-1})^T ]_{ij} \nonumber \\
&=& \frac{1}{4}[ \Sigma^{-1} + 2(\Sigma^{-1})^T +  (\Sigma^{-1})^T\Sigma   (\Sigma^{-1})^T ]_{ij}   \nonumber 
\end{eqnarray}
Since, we have shown earlier, G is a block diagonal matrix. Let us refer to the upper and lower blocks as $G_{upper}$ and $G_{lower}$
Thus,
\begin{equation}
G_{upper} = \frac{1}{4}\{ \Sigma^{-1} + 2(\Sigma^{-1})^T +  (\Sigma^{-1})^T\Sigma   (\Sigma^{-1})^T \} 
\end{equation}
\begin{eqnarray}
\frac{\partial (G_{upper})_{ij}}{\partial \Sigma_{kl}}  &=& \frac{1}{4}   \frac{\partial  [\Sigma^{-1} + 2(\Sigma^{-1})^T +  (\Sigma^{-1})^T\Sigma   (\Sigma^{-1})^T ]_{ij}  }{\partial \Sigma_{kl}} \\
&=& \frac{1}{4} \frac{\partial  [\Sigma^{-1} + 2(\Sigma^{-1})^T]_{ij}}{\partial \Sigma_{kl}} +  \frac{1}{4} \frac{\partial  [(\Sigma^{-1})^T\Sigma   (\Sigma^{-1})^T]_{ij}}{\partial \Sigma_{kl}} \nonumber \\
&=& -\frac{1}{4} [\Sigma^{-1}_{ik}\Sigma^{-1}_{lj} + 2\Sigma^{-1}_{jk}\Sigma^{-1}_{li} ]  +\frac{1}{4} \frac{\partial  [(\Sigma^{-1})^T_{im} \Sigma_{mn}  (\Sigma^{-1})^T_{nj}]}{\partial \Sigma_{kl}} \nonumber \\
&=& -\frac{1}{4} [\Sigma^{-1}_{ik}\Sigma^{-1}_{lj} + 2\Sigma^{-1}_{jk}\Sigma^{-1}_{li} ] +\frac{1}{4} \frac{\partial (\Sigma^{-1})^T_{im}} {\partial \Sigma_{kl}} \Sigma_{mn}  (\Sigma^{-1})^T_{nj}] \nonumber \\
&&  +\frac{1}{4} (\Sigma^{-1})^T_{im} \frac{\partial  \Sigma_{mn}}{\partial \Sigma_{kl}}  (\Sigma^{-1})^T_{nj} +\frac{1}{4} (\Sigma^{-1})^T_{im} \Sigma_{mn} \frac{\partial   (\Sigma^{-1})^T_{nj}}{\partial \Sigma_{kl}} \nonumber \\ 
&=& -\frac{1}{4} [\Sigma^{-1}_{ik}\Sigma^{-1}_{lj} + 2\Sigma^{-1}_{jk}\Sigma^{-1}_{li} ] -\frac{1}{4}\Sigma^{-1}_{mk}\Sigma^{-1}_{li}\Sigma_{mn}  (\Sigma^{-1})^T_{nj}   \nonumber \\
&&+ \frac{1}{4} (\Sigma^{-1})^T_{im}\delta_{mk}\delta_{nl} (\Sigma^{-1})^T_{nj} - \frac{1}{4} (\Sigma^{-1})^T_{im} \Sigma_{mn}\Sigma^{-1}_{jk}\Sigma^{-1}_{ln} \nonumber \\
&=& -\frac{1}{4} [\Sigma^{-1}_{ik}\Sigma^{-1}_{lj} + 2\Sigma^{-1}_{jk}\Sigma^{-1}_{li} ] -\frac{1}{4}(\Sigma^{-1})^T_{km}\Sigma_{mn}  (\Sigma^{-1})^T_{nj} \Sigma^{-1}_{li}  \nonumber \\
&&+ \frac{1}{4} (\Sigma^{-1})^T_{ik}(\Sigma^{-1})^T_{lj}  - \frac{1}{4} (\Sigma^{-1})^T_{im} \Sigma_{mn}(\Sigma^{-1})^T_{nl}\Sigma^{-1}_{jk} \nonumber \\
&=& -\frac{1}{4} [\Sigma^{-1}_{ik}\Sigma^{-1}_{lj} + 2\Sigma^{-1}_{jk}\Sigma^{-1}_{li} ]  -\frac{1}{4}[(\Sigma^{-1})^T\Sigma(\Sigma^{-1})^T]_{kj} \Sigma^{-1}_{li}   \nonumber \\
&&+ \frac{1}{4} (\Sigma^{-1})^T_{ik}(\Sigma^{-1})^T_{lj} -  \frac{1}{4}[(\Sigma^{-1})^T\Sigma(\Sigma^{-1})^T]_{il} \Sigma^{-1}_{jk} \nonumber \\
&=& - \frac{1}{2}\Sigma^{-1}_{jk}\Sigma^{-1}_{li}  -\frac{1}{4} \Sigma^{-1}_{ik}\Sigma^{-1}_{lj} + \frac{1}{4}  (\Sigma^{-1})^T_{ik}(\Sigma^{-1})^T_{lj}  \nonumber \\
&& -\frac{1}{4}[(\Sigma^{-1})^T\Sigma(\Sigma^{-1})^T]_{kj} \Sigma^{-1}_{li}   -  \frac{1}{4}[(\Sigma^{-1})^T\Sigma(\Sigma^{-1})^T]_{il} \Sigma^{-1}_{jk} \nonumber 
\end{eqnarray}
Now, assuming $\Sigma$ is symmetric, we have 
\begin{equation}
\frac{\partial (G_{upper})_{ij}}{\partial \Sigma_{kl}} = -\Sigma^{-1}_{jk}\Sigma^{-1}_{li} 
\end{equation}
From equation \ref{G:lower},and \text{without assuming $\Sigma$ to be symmetric}, we have
\begin{equation}
G_{ij} = \frac{1}{4}[\Sigma^{-1}_{\tilde{p}_j\tilde{p}_i}  \Sigma^{-1}_{\tilde{q}_j\tilde{q}_i}  +\Sigma^{-1}_{\tilde{p}_i \tilde{q}_j}    \{ (\Sigma^{-1})^T\Sigma (\Sigma^{-1})^T\}_{\tilde{q}_i\tilde{p}_j} ] \\
\end{equation}
%%Re-writing above equation in terms of $a,b,c,d$ where they are functions of $i,j$.
%%\begin{equation}
%%(G_{lower})_{ij} = \frac{1}{4}[\Sigma^{-1}_{ca} \Sigma^{-1}_{db}  +\Sigma^{-1}_{ad}    \{ (\Sigma^{-1})^T\Sigma (\Sigma^{-1})^T\}_{bc}] \\
%%\end{equation}
%a= \tilde{p}_i
%b= \tilde{q}_i
%c=\tilde{p}_j
%d=\tilde{q}_j

Therefore, for $(k,l)$  pairs of indices of lower triangular matrix, we have
\begin{eqnarray}
\frac{\partial (G_{lower})_{ij}}{\partial \Sigma_{kl}} &=& \frac{1}{4}\frac{\partial \{ [\Sigma^{-1}_{\tilde{p}_j\tilde{p}_i} \Sigma^{-1}_{\tilde{q}_j \tilde{q}_i}  + \Sigma^{-1}_{\tilde{p}_i \tilde{q}_j} \{ (\Sigma^{-1})^T\Sigma (\Sigma^{-1})^T\}_{\tilde{q}_i \tilde{p}_j}] \}}{\partial \Sigma_{kl}} \\
&=&\frac{1}{4} \frac{\partial [\Sigma^{-1}_{\tilde{p}_j \tilde{p}_i} \Sigma^{-1}_{\tilde{q}_j \tilde{q}_i}] }{\partial \Sigma_{kl}} + \frac{1}{4} \frac{\partial  [\Sigma^{-1}_{\tilde{p}_i \tilde{q}_j} \{ (\Sigma^{-1})^T\Sigma (\Sigma^{-1})^T\}_{\tilde{q}_i \tilde{p}_j}]   }{\partial \Sigma_{kl}} \nonumber \\
&=&\frac{1}{4} \frac{\partial \Sigma^{-1}_{\tilde{p}_j \tilde{p}_i} }{\partial \Sigma_{kl}}\Sigma^{-1}_{\tilde{q}_j \tilde{q}_i}  + \frac{1}{4} \Sigma^{-1}_{\tilde{p}_j \tilde{p}_i} \frac{\partial \Sigma^{-1}_{\tilde{q}_j \tilde{q}_i} }{\partial \Sigma_{kl}}  + \frac{1}{4} \frac{\partial  \Sigma^{-1}_{ \tilde{p}_i \tilde{q}_j} }{\partial \Sigma_{kl}}\{ (\Sigma^{-1})^T\Sigma (\Sigma^{-1})^T\}_{\tilde{q}_i \tilde{p}_j}  \nonumber \\
&&+ \frac{1}{4}\Sigma^{-1}_{\tilde{p}_i \tilde{q}_j} \frac{\partial  \{ (\Sigma^{-1})^T\Sigma (\Sigma^{-1})^T\}_{ \tilde{q}_i \tilde{p}_j}}{\partial \Sigma_{kl}} \nonumber \\
&=& - \frac{1}{4} \Sigma^{-1}_{\tilde{p}_j k}\Sigma^{-1}_{l\tilde{p}_i}\Sigma^{-1}_{\tilde{q}_j \tilde{q}_i} - \frac{1}{4} \Sigma^{-1}_{\tilde{p}_j \tilde{p}_i} \Sigma^{-1}_{\tilde{q}_j k} \Sigma^{-1}_{l \tilde{q}_i}  -\frac{1}{4}\Sigma^{-1}_{\tilde{p}_i k}\Sigma^{-1}_{l \tilde{q}_j}\{ (\Sigma^{-1})^T\Sigma (\Sigma^{-1})^T\}_{\tilde{q}_i \tilde{p}_j }  \nonumber \\
& &+ \frac{1}{4}\Sigma^{-1}_{\tilde{p}_i \tilde{q}_j } \{ -[(\Sigma^{-1})^T\Sigma (\Sigma^{-1})^T]_{k \tilde{p}_j} \Sigma^{-1}_{l \tilde{q}_i}  + (\Sigma^{-1})^T_{\tilde{q}_i k} (\Sigma^{-1})^T_{l \tilde{p}_j}  - [(\Sigma^{-1})^T\Sigma (\Sigma^{-1})^T]_{\tilde{q}_i l} \Sigma^{-1}_{\tilde{p}_j k} \}  \nonumber \\
&=& - \frac{1}{4} \Sigma^{-1}_{\tilde{p}_j k}\Sigma^{-1}_{l\tilde{p}_i}\Sigma^{-1}_{\tilde{q}_j \tilde{q}_i} - \frac{1}{4} \Sigma^{-1}_{\tilde{p}_j \tilde{p}_i} \Sigma^{-1}_{\tilde{q}_j k} \Sigma^{-1}_{l\tilde{q}_i }  -\frac{1}{4}\Sigma^{-1}_{\tilde{p}_i k}\Sigma^{-1}_{l\tilde{q}_j} \Sigma^{-1}_{\tilde{q}_i \tilde{p}_j}  \nonumber \\
&&+ \frac{1}{4}\Sigma^{-1}_{\tilde{p}_i \tilde{q}_j} \{ -\Sigma^{-1}_{k \tilde{p}_j} \Sigma^{-1}_{l \tilde{q}_i}  + (\Sigma^{-1})^T_{\tilde{q}_i k} (\Sigma^{-1})^T_{l \tilde{p}_j}  - \Sigma^{-1}_{\tilde{q}_i l} \Sigma^{-1}_{\tilde{p}_j k} \}  \nonumber \\
&=& - \frac{1}{4} \Sigma^{-1}_{\tilde{p}_j k}\Sigma^{-1}_{l \tilde{p}_i}\Sigma^{-1}_{\tilde{q}_j \tilde{q}_i} - \frac{1}{4} \Sigma^{-1}_{\tilde{p}_j \tilde{p}_i} \Sigma^{-1}_{\tilde{q}_jk} \Sigma^{-1}_{l \tilde{q}_i}  -\frac{1}{4}\Sigma^{-1}_{\tilde{p}_i k}\Sigma^{-1}_{l \tilde{q}_j} \Sigma^{-1}_{\tilde{q}_i \tilde{p}_j}  \nonumber \\
&& - \frac{1}{4}\Sigma^{-1}_{\tilde{p}_i \tilde{q}_j} \Sigma^{-1}_{k \tilde{p}_j} \Sigma^{-1}_{l \tilde{q}_i}  +\frac{1}{4}\Sigma^{-1}_{\tilde{p}_i \tilde{q}_j} \Sigma^{-1}_{k \tilde{q}_i} \Sigma^{-1}_{\tilde{p}_j l}  - \frac{1}{4}\Sigma^{-1}_{\tilde{p}_i \tilde{q}_j}\Sigma^{-1}_{\tilde{q}_i l} \Sigma^{-1}_{\tilde{p}_j k}  \nonumber 
\end{eqnarray}
\textbf{Connection: }
%From equation 5.4 in the book by John Lee, we have
\begin{eqnarray}
\gamma_{pq}^r %&=& \frac{1}{2}g^{rl}(\partial_pg_{ql} + \partial_qg_{pl} - \partial_lg_{pq}) \\
&=& \frac{1}{2}g^{rl}(\frac{\partial g_{ql}}{\partial \theta_p} + \frac{\partial g_{pl}}{\partial \theta_q} - \frac{ \partial g_{pq}}{\partial \theta_l}) \;\; , \;\; p,q,r,l \in [1,D] \nonumber
\end{eqnarray}
%Where p,q,r, and l vary from  1 to the total number of parameters in a multivariate Gaussian random variable. \\
Assuming the dimension of the problem is d, the total number of parameters is as follows: \\
(a) $\mu_i$  for $i  \in  [1...d]$ \\
(b) $\Sigma_{ij}$ for (i,j) pairs of lower triangular part of the covariance matrix \\%$d \times d$ which is $= 1+2+ ... + d = \frac{d(d+1)}{2}$ \\
%Thus the total number of parameters is $$ D = d +  \frac{d(d+1)}{2}$$
Now, since G is a function of $\Sigma$ alone, and the first d parameters in $\theta$  correspond to $\mu$, $$ \frac{\partial G}{\partial \theta_i} = 0,  i \in [1...d]$$
For $i > d$, we have the following possibilities 
\begin{enumerate}
\item $ q,l \in [1...d]$\\
$\frac{\partial g_{ql}}{\partial \theta_i}  = $  (from the equation  for $\frac{\partial G_{upper}}{\partial \Sigma}$) \\
\item one of q,l $\in [1...d]$ and the other is $>d$ \\
$\frac{\partial g_{ql}}{\partial \theta_i}  =  0 $ \\
\item both q and l are $>d$ \\
$\frac{\partial g_{ql}}{\partial \theta_i}  =  $ (from the equation for  for $\frac{\partial G_{lower}}{\partial \Sigma}$)  
\end{enumerate}

%
%(a)  $ q,l \in [1...d]$\\
%$\frac{\partial g_{ql}}{\partial \theta_i}  = $  (from the equation  for $\frac{\partial G_{upper}}{\partial \Sigma}$) \\
%(b) one of q,l $\in [1...d]$ and the other is $>d$\\
%$\frac{\partial g_{ql}}{\partial \theta_i}  =  0 $ \\
%(c) both q and l are $>d$ \\
%$\frac{\partial g_{ql}}{\partial \theta_i}  =  $ (from the equation for  for $\frac{\partial G_{lower}}{\partial \Sigma}$)  
%%%%%%%%%%%%%%%%%%%%%%%%%%%%%%%%%%%%%%%%%%
\section{Logistic regression problem}\label{app:LR}
$ln(\frac{p}{1-p}) = \beta_0 + \sum_{i=1}^D{\beta_i X_i} \;\; $ where $p$ is the sucess probability. Therefore\\
Probability $(t=1)=p(X)=\frac{1}{1+\exp[-(\beta_0 +  ( \beta^T X) )]}$\\
For a data set $X$ consisting of $N$ points and the corresponding values of $t$, the likelihood of the given vector $t$ for a some $\beta$ is \\
%\begin{equation} \nonumber
$p(t|X,\beta) = \prod_{i=1}^N (\frac{1}{1+\exp[-(\beta_0 +   \beta^T X^{(i)} )]} )^{t_i}  \times  (1- \frac{1}{1+\exp[-(\beta_0 +   \beta^T X^{(i)} )]} )^{(1-t_i)} $\\
%\end{equation}
Log- likelihood: $L = \log(p(t|X,\beta)) = \sum_{i=1}^N [t_i \log(\frac{1}{1+\exp[-(\beta_0 +   \beta^T X^{(i)} )]} )  + (1-t_i)\log(1- \frac{1}{1+\exp[-(\beta_0 +   \beta^T X^{(i)} )]})]$\\
Append an extra element  $(=1)$ at the beginning of every vector $X^{(i)}$ for convenience, call it $\bar{X}$  such that $\bar{X}^{(i)}$ is now $D+1$ dimensional, hence we have
\begin{equation}
L = \sum_{i=1}^N [t_i \log(\frac{1}{1+\exp[-  \beta^T \bar{X}^{(i)} ]} )  + (1-t_i)\log(1- \frac{1}{1+\exp[-\beta^T \bar{X}^{(i)} ]})]
\end{equation}
Gradient of log-likelihood:
\begin{eqnarray}\label{gradL}
\frac{\partial L}{\partial \beta_p} &=& \sum_{i=1}^N [t_i \frac{\partial (\log(\frac{1}{1+\exp[-  \beta^T \bar{X}^{(i)} ]} ) ) }{\partial \beta_p}  +  (1-t_i) \frac{\partial (\log(1- \frac{1}{1+\exp[-\beta^T \bar{X}^{(i)} ]})])}{\partial \beta_p}] \\
&=&  \sum_{i=1}^N [-t_i \frac{\partial (\log(1+\exp[-  \beta^T \bar{X}^{(i)} ] ) ) }{\partial \beta_p} 
+ (1-t_i) \frac{\partial (\log( \exp[-\beta^T \bar{X}^{(i)})])}{\partial \beta_p} \nonumber \\
&&-  (1-t_i)  \frac{\partial (\log( 1+\exp[-\beta^T \bar{X}^{(i)})])}{\partial \beta_p} ]  \nonumber \\
&=& \sum_{i=1}^N [-t_i \frac{\partial (\log(1+\exp[-  \beta^T \bar{X}^{(i)} ] ) ) }{\partial \beta_p} 
+ (1-t_i) \frac{\partial ([-\beta^T \bar{X}^{(i)})])}{\partial \beta_p} \nonumber \\
&&-  (1-t_i)  \frac{\partial (\log( 1+\exp[-\beta^T \bar{X}^{(i)})])}{\partial \beta_p} ]  \nonumber \\
&=& \sum_{i=1}^N [- \frac{\partial (\log(1+\exp[-  \beta^T \bar{X}^{(i)} ] ) ) }{\partial \beta_p} 
- (1-t_i) \frac{\partial (\beta^T \bar{X}^{(i)})}{\partial \beta_p}] \nonumber \\
&=& \sum_{i=1}^N[ \left( -\frac{1}{ 1+\exp(-  \beta^T \bar{X}^{(i)}) }\right) \frac{\partial (1+\exp(-  \beta^T \bar{X}^{(i)} ))}{\partial \beta_p}  -  (1-t_i) \frac{\partial (\beta^T \bar{X}^{(i)})}{\partial \beta_p}] \nonumber 
\end{eqnarray}
 Now,
 \begin{eqnarray}
 \frac{\partial \{1+\exp(-\beta^T \bar{X}^{(j)})\} }{\partial \beta_p} &=& \frac{\partial [\exp(-\beta^T \bar{X}^{(j)})]}{\partial \beta_p} \nonumber \\
 &=& \frac{\partial [\exp(-\beta_i \bar{X}^{(j)}_i)]}{\partial \beta_p} \nonumber \\
 &=& [\exp(-\beta_i \bar{X}^{(j)}_i)] \frac{\partial (-\beta_i \bar{X}^{(j)}_i) }{\partial \beta_p} \nonumber \\
 &=& -  [\exp(-\beta_i \bar{X}^{(j)}_i)] \bar{X}^{(j)}_p \nonumber \\
 &=& -\exp(-\beta^T \bar{X}^{(j)}) \bar{X}^{(j)}_p \nonumber 
 \end{eqnarray}
 Substituting in \ref{gradL}, we get
 \begin{eqnarray}
\frac{\partial L}{\partial \beta_p} &=&  \sum_{i=1}^N[  \frac{\exp(-\beta^T \bar{X}^{(i)}) \bar{X}^{(i)}_p}{ 1+\exp(-  \beta^T \bar{X}^{(i)})} -  (1-t_i) \bar{X}^{(i)}_p ]
 \end{eqnarray}
\textbf{Fisher-Information matrix:}\\
For a given $\{X^{i},t\}$ pair, we have
\begin{eqnarray}\label{FIM}
G &=& E\left(\frac{\partial L}{\partial \beta_p} \frac{\partial L}{\partial \beta_q}\right) \label{FIM} \\
&=& E\left([  \frac{\exp(-\beta^T \bar{X}^{(i)}) \bar{X}^{(i)}_p}{1+\exp(-\beta^T \bar{X}^{(i)}) } -  (1-t_i) \bar{X}^{(i)}_p ]  [  \frac{\exp(-\beta^T \bar{X}^{(i)}) \bar{X}^{(i)}_q}{1+\exp(-\beta^T \bar{X}^{(i)}) } -  (1-t_i) \bar{X}^{(i)}_q ]\right) \nonumber \\ 
&=& \frac{(\exp(-\beta^T \bar{X}^{(i)}) )^2 \bar{X}^{(i)}_p \bar{X}^{(i)}_q }{(1+\exp(-\beta^T \bar{X}^{(i)}) )^2} - \frac{2 \exp(-\beta^T \bar{X}^{(i)}) \bar{X}^{(i)}_p \bar{X}^{(i)}_q}{(1+\exp(-\beta^T \bar{X}^{(i)}) )}  E(1-t) \nonumber \\
&& + \bar{X}^{(i)}_p \bar{X}^{(i)}_q E[1-2t+t^2] \nonumber 
\end{eqnarray}
$t$ is a Bernoulli random variable, therefore
\begin{eqnarray}
E(t) &=& 1 \times p(t=1) + 0 \times p(t=0) \nonumber \\
&=& p(t=1)  \nonumber \\
&=& \frac{1}{1+\exp(-\beta^T \bar{X}^{i})} \nonumber\\
E(t^2) &=& 1^2 \times p(t=1) + 0 \times p(t=0) \nonumber \\
&=& p(t=1)  \nonumber \\
&=& \frac{1}{1+\exp(-\beta^T \bar{X}^{i})} \nonumber
\end{eqnarray}
Substituting in equation (\ref{FIM}), we have
\begin{eqnarray}
G(bar{X}^{(i)},\beta) &=& \frac{(\exp(-\beta^T \bar{X}^{(i)}) )^2 \bar{X}^{(i)}_p \bar{X}^{(i)}_q }{(1+\exp(-\beta^T \bar{X}^{(i)}) )^2} - \frac{2 \exp(-\beta^T \bar{X}^{(i)}) \bar{X}^{(i)}_p \bar{X}^{(i)}_q}{(1+\exp(-\beta^T \bar{X}^{(i)}) )}  \left(1- \frac{1}{1+\exp(-\beta^T \bar{X}^{i})}\right) \nonumber \\
&& + \bar{X}^{(i)}_p \bar{X}^{(i)}_q \left(1-\frac{2}{1+\exp(-\beta^T \bar{X}^{i})}+\frac{1}{1+\exp(-\beta^T \bar{X}^{i})} \right)\nonumber \\
&=& \frac{(\exp(-\beta^T \bar{X}^{(i)}) )^2 \bar{X}^{(i)}_p \bar{X}^{(i)}_q }{(1+\exp(-\beta^T \bar{X}^{(i)}) )^2} 
- \frac{2 (\exp(-\beta^T \bar{X}^{(i)}) )^2\bar{X}^{(i)}_p \bar{X}^{(i)}_q}{(1+\exp(-\beta^T \bar{X}^{(i)}) )^2}  \nonumber \\
&& + \bar{X}^{(i)}_p \bar{X}^{(i)}_q \frac{\exp(-\beta^T \bar{X}^{i})}{1+\exp(-\beta^T \bar{X}^{i})} \nonumber \\
&=& - \frac{(\exp(-\beta^T \bar{X}^{(i)}) )^2 \bar{X}^{(i)}_p \bar{X}^{(i)}_q }{(1+\exp(-\beta^T \bar{X}^{(i)}) )^2}   +  \frac{\exp(-\beta^T \bar{X}^{i})(1+\exp(-\beta^T \bar{X}^{i})) \bar{X}^{(i)}_p \bar{X}^{(i)}_q}{(1+\exp(-\beta^T \bar{X}^{i}))^2} \nonumber \\
&=&  \frac{\exp(-\beta^T \bar{X}^{i}) \bar{X}^{(i)}_p \bar{X}^{(i)}_q}{(1+\exp(-\beta^T \bar{X}^{i}))^2} \nonumber 
\end{eqnarray}
Therefore, G for the product of likelihoods is, $G(\bar{X}) =\sum_{i=1}^N G(\bar{X}^{(i)},\beta) $\\
Derivative of Fisher-Information matrix:
\begin{eqnarray}\label{delG}
\frac{\partial G(\bar{X},\beta) _{pq}}{\partial \beta_r} &= & \sum_{i=1}^N \frac{\partial }{\partial \beta_r} \left(\frac{\exp(-\beta^T \bar{X}^{(i)}) \bar{X}^{(i)}_p \bar{X}^{(i)}_q}{(1+\exp(-\beta^T \bar{X}^{i}))^2} \right)  \\
&=& \sum_{i=1}^N \frac{\partial \exp(-\beta^T \bar{X}^{(i)})}{\partial \beta_r} \left(\frac{ \bar{X}^{(i)}_p \bar{X}^{(i)}_q}{(1+\exp(-\beta^T \bar{X}^{i}))^2} \right) \nonumber \\
&& + \sum_{i=1}^N \frac{\partial }{\partial \beta_r} \left(\frac{1}{(1+\exp(-\beta^T \bar{X}^{i}))^2} \right) \exp(-\beta^T \bar{X}^{i}) \bar{X}^{(i)}_p \bar{X}^{(i)}_q \nonumber \\
\text{Now,}\;\; \frac{\partial [\exp(-\beta^T \bar{X}^{(i)}) ]}{\partial \beta_r}  &=& - \exp(-\beta^T \bar{X}^{(i)}) X^{(i)}_r   \nonumber \\
\text{and} \;\; \frac{\partial }{\partial \beta_r} \frac{1}{\{1+\exp(-\beta^T X^{(j)})\}^2 } &=& - \frac{2}{\{1+\exp(-\beta^T \bar{X}^{(i)})\}^3 } \frac{\partial \{1+\exp(-\beta^T \bar{X}^{(i)})\}}{\partial \beta_r} \nonumber \\
&=& - \frac{2}{\{1+\exp(-\beta^T \bar{X}^{(i)})\}^3 } (-\exp(-\beta^T \bar{X}^{(i)}) \bar{X}^{(i)}_r) \nonumber \\
&=& 2 \frac{\exp(-\beta^T \bar{X}^{(i)})\bar{X}^{(i)}_r}{\{1+\exp(-\beta^T \bar{X}^{(i)})\}^3 }
\end{eqnarray}
Substituting in equation (\ref{delG}), we have
\begin{eqnarray}
\frac{\partial G_{pq}}{\partial \beta_r} &= & - \sum_{i=1}^N  \exp(-\beta^T \bar{X}^{(i)}) X^{(i)}_r \left(\frac{ \bar{X}^{(i)}_p \bar{X}^{(i)}_q}{(1+\exp(-\beta^T \bar{X}^{i}))^2} \right) \nonumber \\
&& + \sum_{i=1}^N  2 \frac{\exp(-\beta^T \bar{X}^{(i)})\bar{X}^{(i)}_r}{\{1+\exp(-\beta^T \bar{X}^{(i)})\}^3 }\exp(-\beta^T \bar{X}^{i}) \bar{X}^{(i)}_p \bar{X}^{(i)}_q \nonumber \\
&=&  - \sum_{i=1}^N   \frac{ \exp(-\beta^T \bar{X}^{(i)})  \bar{X}^{(i)}_p \bar{X}^{(i)}_q X^{(i)}_r}{(1+\exp(-\beta^T \bar{X}^{i}))^2}   + \sum_{i=1}^N  2 \frac{(\exp(-\beta^T \bar{X}^{(i)}))^2  \bar{X}^{(i)}_p \bar{X}^{(i)}_q \bar{X}^{(i)}_r}{\{1+\exp(-\beta^T \bar{X}^{(i)})\}^3 } \nonumber 
\end{eqnarray}
Connection: $\gamma^k_{ij} = \frac{1}{2}g^{kl}(\partial_i g_{jl} +  \partial_j g_{il} - \partial_l g_{ij}) $

\end{document}